\documentclass[aps,nofootinbib,prd,superscriptaddress]{revtex4}
\usepackage{array}
\usepackage{subfigure}
\usepackage{subcaption}
\usepackage{capt-of}
\usepackage{graphicx}
\usepackage{lipsum}
\usepackage{amsmath}
\usepackage{amssymb}

\usepackage{comment}
\usepackage{color,xcolor}
\usepackage[bookmarks=false]{hyperref}
\usepackage{cleveref}
\usepackage{ulem}
\hypersetup{colorlinks=true,citecolor=green,linkcolor=blue,urlcolor=blue,pdfstartview=FitH,bookmarksopen=true}

\newcommand{\eq}{\begin{eqnarray}}
	\newcommand{\en}{\end{eqnarray}}

\newcommand{\bfP}{{\bf P}_{\perp}}
\newcommand{\bfp}{{\bf p}_{\perp}}
\newcommand{\bfk}{{\bf k}_{\perp}}

\newcommand{\mb}[1]{\mathbf{#1}}

\newcommand{\pep}[1]{\mathbf{#1}_{\perp}}

\newcommand{\gdir}[1]{\gamma^{#1}}

\begin{document}
	
	\title{Flavor asymmetry of light sea quarks in proton : A light-front spectator model}
  
    \author{Poonam~Choudhary}
    \email{poonamch@iitk.ac.in} 
    \affiliation{Department of Physics, Indian Institute of Technology Kanpur, Kanpur-208016, India}
    
\author{Dipankar~Chakrabarti}
	\email{dipankar@iitk.ac.in} 
	\affiliation{Department of Physics, Indian Institute of Technology Kanpur, Kanpur-208016, India}

 \author{Chandan Mondal}
  \email{mondal@impcas.ac.cn}
  \affiliation{Institute of Modern Physics, Chinese Academy of Sciences, Lanzhou 730000, China}
  \affiliation{School of Nuclear Science and Technology, University of Chinese Academy of Sciences, Beijing 100049, China}

\begin{abstract}
We formulate a light-front spectator model for the proton that incorporates the presence of light sea quarks. In this particular model, the sea quarks are seen as active partons, whereas the remaining components of the proton are treated as spectators. The proposed model relies on the formulation of the light-front wave function constructed by the soft wall AdS/QCD. The model wave functions are parameterized by fitting the unpolarized parton distribution functions of light sea quarks from the CTEQ18 global analysis. We then
employ the light-front wave functions to
obtain the sea quarks generalized parton distribution functions,  transverse momentum dependent parton distributions, and their asymmetries, which are accessible in the upcoming Electron-Ion-Colliders. We investigate sea quarks' spin and orbital angular momentum contributions to the proton spin.
\end{abstract}
\maketitle
%==================================	
\section{Introduction}\label{Sec1}
%==================================
Quantum chromodynamics (QCD) is the well-established theory for the strong interactions, where the proton is viewed as a strongly
coupled and correlated 
confined system of quarks and gluons. The original quark model, where the proton is described by
two up quarks ($u$) and one down quark ($d$), has an appealing simplicity. However, various experiments that measure the parton distribution functions (PDFs) have revealed a substantial structure with additional quarks, antiquarks, and gluons beyond the leading three-quark Fock component. These
additional quarks and antiquarks are known  as sea quarks.
It is not possible
to distinguish any individual down or up quark as a valence or sea quark,
but antiquarks must belong to the sea and so their study promises to illuminate new insights into the structure of the proton. Going beyond the leading Fock component $|uud\rangle$, the concept of the five-particle Fock state with charm sea quark, $|uudc\bar{c}\rangle$, in the proton was first suggested by Brodsky {\it et. al.}~\cite{Brodsky:1980pb,Brodsky:1981se}.
The sea-quark distributions in the proton, together with the valence quark and gluon distributions, have been investigated in numerous experiments employing the deep inelastic scattering  (DIS), the Drell-Yan, and other hard processes~\cite{Drell:1970yt,Chang:2014jba}.
In the study of sea quarks in the proton, several asymmetries such as $\bar{d}-\bar{u}$, $ \bar{d}/\bar{u}$ and $(\bar{d}-\bar{u})/(\bar{u}+\bar{d})$  originating from the flavor asymmetry have been observed~\cite{Buccella:1992zs,Kumano:1997cy}. It states that the anti-up quark and the anti-down quark distributions are not the same within the proton. 
The initial naive expectation was that the sea quark distributions of up and down quark are the same as the sea was generated predominantly by gluons splitting into quark-antiquark pairs and
the splitting is flavor independent and the masses of up and down quarks are also almost degenerate.
However, the deviation of this assumption can be evaluated by the 
following expression
\begin{eqnarray}
S_G	&=&\int_{0}^{1}\left[F_{2}^{p}(x)-F_{2}^{n}(x)\right]
	\frac{{\rm d}x}{x} = 
	\frac{1}{3}\int_0^1 \left[u_v(x)-d_v(x)\right]{\rm d}x - \frac{2}{3} \int_{0}^{1}\left[\bar{d}_{p}(x)-
	\bar{u}_{p}(x)\right]{\text dx}\,\nonumber\\
&=&\frac{1}{3} - \frac{2}{3} \int_{0}^{1}\left[\bar{d}_{p}(x)-
	\bar{u}_{p}(x)\right]{\text dx}\,,
 \label{eq:1}
\end{eqnarray}
where $F_{2}^{p}(x)$ and $F_{2}^{n}(x)$ are the proton and the neutron
inelastic structure functions, respectively; $u_v(x)$ and $d_v(x)$ are the valence quark distributions in the proton; and $\bar{d}_{p}(x)$ and
$\bar{u}_{p}(x)$ are the anti-down and the anti-up quark distributions in
the proton sea as a function of Bjorken-$x$.  
Equation~\eqref{eq:1}
requires the assumption of charge symmetry between the proton and the
neutron (i.e. $u_p = d_n$, $\bar{u}_p = \bar{d}_n$, etc.,).  If the
nucleon sea is flavor symmetric in light quarks, then the value of $S_G=1/3$, a result referred as the Gottfried Sum Rule (GSR)~\cite{Gottfried:1967kk}.
 In the early 1990’s, the New Muon Collaboration (NMC) \cite{NewMuon:1991hlj,NewMuon:1993oys} announced a precise DIS measurement indicating a violation of the GSR~\cite{Gottfried:1967kk}, which leads to an asymmetries between the $\bar{u}$ and $\bar{d}$ distributions in the proton. 
The NMC Collaboration reported a  value of
 $S_G =
 0.235 \pm 0.026$~\cite{Kumano:1990mj}, which implies that
 \begin{equation}
 	\int_{0}^{1}\left[\bar{d}_{p}(x)-\bar{u}_{p}(x)\right]{\rm d}x
 	= 0.147\pm 0.039,
 	\label{eq:2}
 \end{equation}
 a considerable excess of $\bar{d}$ relative to $\bar{u}$ in the proton. 

The discovery of the flavor asymmetry in the nucleon sea triggered numerous dedicated experimental and theoretical efforts~\cite{Chang:2014jba,Buccella:1992zs,Kumano:1997cy,Wakamatsu:2014asa,Lin:2014zya,Song:2011fc, Brodsky:2022kef,Garvey:2001yq,SeaQuest:2022vwp,Accardi:2019ofk,Kotlorz:2021goc,NuSea:1998kqi,NuSea:1998tww,Dove:2020iuk,Nagai:2019lqo,McGaughey:1992kz,Choudhary:2023bap,Dorokhov:1991pv,Dorokhov:1993fc,Trevisan:2008zz,He:2022leb}. The study of sea quarks' properties within the nucleon becomes one of the major goals at the Large Hadron Collider (LHC)~\cite{Rojo:2015acz, Harland-Lang:2014zoa} and the future Electron-Ion-Colliders (EICs)~\cite{Accardi:2012qut,AbdulKhalek:2021gbh,Anderle:2021wcy}. The sea quark asymmetry is not only observed in the light flavors but also in the strange~\cite{Cao:2016uva,He:2017viu} and the heavy flavors~\cite{Brodsky:2022kef,Lyubovitskij:2022wjt,Sufian:2020coz}.  In this work, we  focus on the light sea quarks.  The light flavor asymmetries have been measured experimentally by several collaborations~\cite{SeaQuest:2022vwp,Accardi:2019ofk,Kotlorz:2021goc,NuSea:1998kqi,NuSea:1998tww,Dove:2020iuk,Nagai:2019lqo,McGaughey:1992kz,NuSea:2001idv,SeaQuest:2021zxb,HERMES:1998uvc}. The results from the NuSea/E866 Collaboration~\cite{NuSea:2001idv} demonstrate  a rapid growth of $\bar{d} (x)/\bar{u}(x)$ up to $x\lesssim 0.2$ and then a sharp fall of this ratio in the region $x>0.2$. However, the recent measurement by the SeaQuest/E906 Collaboration~\cite{SeaQuest:2021zxb,SeaQuest:2022vwp} shows  a modest growth of $\bar{d} (x)/\bar{u}(x)$ at the region $0.13<x<0.45$. The trends between these two experiments at higher $x$ region are quite different. No explanation has been found yet for the differing results. The experimental data is not available for the entire $x$ region. The data of $\bar{d} (x)-\bar{u}(x)$ from the HERMES Collaboration are more or less consistent with the prominent features as shown by the NuSea/E866~\cite{NuSea:2001idv} and the SeaQuest/E906~\cite{SeaQuest:2021zxb,SeaQuest:2022vwp} Collaboration, where the asymmetry is expected to fall monotonously as $x$ increases. 
A precise knowledge of the sea quark distributions is required for the analysis and interpretation of the scattering experiments in
the LHC era.  Global fitting collaborations such as  CTEQ~\cite{Dulat:2015mca}, NNPDF~\cite{NNPDF:2017mvq}, HERAPDF~\cite{Alekhin:2017kpj},
MSTW~\cite{Martin:2009iq},  and MMHT~\cite{Harland-Lang:2014zoa} have made considerable efforts to determine sea quark PDFs and their uncertainties.
The proton sea has also been investigated using different theoretical approaches such as the chiral quark model~\cite{Song:2011fc,Choudhary:2023bap}, instanton model~\cite{Dorokhov:1991pv,Dorokhov:1993fc}, statistical model~\cite{Trevisan:2008zz}, chiral soliton model~\cite{ Thomas:1983fh}, chiral light-front perturbation theory~\cite{Alberg:2017ijg}, chiral effective theory~\cite{Wang:2022bxo}, lattice QCD~\cite{Lin:2014zya,Alexandrou:2021oih}, BHPS model \cite{Chang:2011vx}, etc.  In Ref.~\cite{Henley:1990kw} the asymmetry is explained in terms of pion cloud effects and by imposing constraints on the pion-nucleon vertex function. A
common feature of  most of these theoretical approaches is a growth of $\bar{d} (x)/\bar{u}(x)$ with increasing $x$, consistent with the characteristic shown by the NuSea/E866 data at lower-$x$ and the SeaQuest/E906 data. An overview of the nucleon sea and their asymmetries can be found in Refs.~ \cite{Kumano:1997cy,Geesaman:2018ixo,Chang:2014jba}.  
The spin structure of sea quarks is an additional captivating element to consider. The sea quark helicity distributions~\cite{Geesaman:2018ixo,Chang:2014jba} could give us important clues about how to solve the nucleon spin puzzle~\cite{Ji:2020ena}.

 Recently, there are numerous  experiments and theoretical investigations~\cite{ZEUS:2003pwh,COMPASS:2008isr,HERMES:2009lmz} aimed to understand the generalized parton distributions (GPDs) and the transverse momentum parton dependent distributions (TMD), which encode the three-dimensional (3D) information of the quark in the proton. The GPDs~\cite{Diehl:2003ny,Belitsky:2005qn,Goeke:2001tz,Ji:1998pc,Burkardt:2002hr} are required for exclusive processes such as deeply virtual Compton scattering (DVCS)~\cite{Ji:1996nm,Goeke:2001tz} or vector meson productions~\cite{Goloskokov:2007nt,Collins:1996fb}, while the TMDs~\cite{Collins:1984kg,Boer:1997nt,Angeles-Martinez:2015sea} are necessary to explain the Semi-Inclusive Deep Inelastic Scattering (SIDIS)~\cite{Bacchetta:2006tn,Ji:2004wu} or Drell-Yan processes~\cite{Tangerman:1994eh,Collins:2002kn,Zhou:2009jm}. The GPDs give us with valuable information about the spatial distributions, spin and orbital motion of quarks inside the proton. The TMDs, on the other hand, are the extended version of collinear PDFs, capturing the 3D structural information of the proton in momentum space. The TMDs also encode the knowledge about the correlations between spins of the proton and momenta of the quarks.  While the valence quark GPDs
 and TMDs have been
investigated for decades and are  determined from various theoretical approaches~\cite{Alexandrou:2020zbe,Lin:2021brq,Ji:1997gm,Scopetta:2002xq,Petrov:1998kf,Penttinen:1999th,Boffi:2002yy,Boffi:2003yj,Vega:2010ns,Chakrabarti:2013gra,Mondal:2015uha,Chakrabarti:2015ama,Mondal:2017wbf,deTeramond:2018ecg,Xu:2021wwj,Liu:2022fvl,Kaur:2023lun,Avakian:2010br,Efremov:2009ze,Bastami:2020rxn,Bacchetta:2008af,Maji:2017bcz,Pasquini:2008ax,Musch:2010ka,Ji:2014hxa,Hu:2022ctr}, much
less information is available on the sea quark distributions. Recently, the sea quark GPDs in the proton have  been investigated with the nonlocal covariant chiral effective theory~\cite{He:2022leb} and using a light-cone baryon-meson fluctuation model~\cite{Luan:2023lmt}. To our knowledge, the study of sea quark TMDs in the proton not yet available in the literature. Precise determination of 3D structure of the proton, specifically the sea quark and gluon distributions, are two of the major scientific objectives of the upcoming EICs~\cite{Accardi:2012qut,AbdulKhalek:2021gbh,Anderle:2021wcy}.

In this work, we construct a light-front spectator model for the proton  embodying the light sea quarks, which are seen as active partons and the remaining components of the proton are considered as spectators. We model the light-front wave functions using the wave functions predicted by the soft-wall AdS/QCD and  parameterize them by fitting the unpolarized PDFs of light sea quarks from the CTEQ global analysis at the scale $Q = 2$ GeV~\cite{Hou:2019efy}. We then investigate the asymmetries of sea quarks in this model by employing  the resulting light-front wave functions. We compare our model predictions for the $\bar{d}(x)-\bar{u}(x)$ and $\bar{d}(x)/\bar{u}(x)$ with the available experimental data from the NuSea/E866~\cite{NuSea:2001idv}, HERMES \cite{HERMES:1998uvc}, and SeaQuest/E906~\cite{SeaQuest:2022vwp} Collaborations as well as with the CTEQ18 global fit~\cite{Hou:2019efy}. We further study the sea quarks asymmetries through their GPDs and TMDs  in this model and also investigate sea quarks spin and orbital angular momentum (OAM) contributions to the proton spin.

The paper is organized as follows: In section~\ref{sec:model}, we discuss about our model construction and determine the value of model parameters by fitting the unpolarized sea quark PDFs to the CT18NNLO global analyses~\cite{Hou:2019efy}. We present the comparison between our model predictions for the flavor asymmetries and the experimental data in this section. The sea quark helicity and transversity PDFs are also presented in this section. We evaluate the sea quark GPDs and the OAM in section~\ref{GPDs}. Flavor asymmetries for sea quarks at finite momentum transfer are obtained from the calculated GPDs and the results are discussed in this section. Section~\ref{tmdsasym} presents the sea quark TMDs and describes the flavor asymmetries for sea quarks in the transverse  momentum plane. We discuss a different phenomenological model and reevaluate all the sea quark distributions and compare them with the previous results in section~\ref{sec5}.
  We provide a summary in section~\ref{SUMMARY}. 
%==============================
\section{Model description}\label{sec:model}
%==============================
QCD predicts the existence of both 
perturbative extrinsic and nonperturbative intrinsic quark contributions to the structure of hadrons. We analyze the sea quark content in the proton, induced either directly by the nonperturbative $|qqq q\bar{q}\rangle$ Fock component or by the $|qqq g\rangle$ Fock component, where the gluon splits into a quark-antiquark pair. The complete distribution of sharing the longitudinal momentum could be written as sum of the intrinsic and the extrinsic distributions. 
The intrinsic quarks are observed in the low energy region, while the extrinsic sea quarks resulting from gluon splitting appear in relatively high energy. 

The light-front formalism provides a convenient way to study the  hadronic system. Here we adopt the generic ansatz for the light-front quark-spectator model for the proton~\cite{Brodsky:2022kef}.  In this model, one contemplates the proton as an effective system
composed of an active sea quark (fermion) and a composite state of
spectator (boson), where the spin of the spectator is assumed to be zero (scalar) only. We model the light-front wave functions from the solution of soft-wall AdS/QCD. The $2$-particle Fock-state expansion for proton with spin components: $J^z = \pm 1/2$, in a frame, where the transverse momentum of proton vanishes, i.e., $P \equiv \big(P^+,\textbf{0}_\perp,{M^2}/{P^+}\big)$, 
can be expressed as
 \begin{align}\label{state}
  |P;\uparrow(\downarrow)\rangle 
 &= \sum_{\bar{q}} \int \frac{{\rm d}x\, {\rm d}^2\textbf{k}_{\perp}}{2(2\pi)^3\sqrt{x(1-x)}}\Big[ \psi^{\uparrow(\downarrow)}_{\bar{q};+\frac{1}{2}}(x,\textbf{k}_{\perp})|+\frac{1}{2},0; xP^+,\textbf{k}_{\perp}\rangle + \psi^{\uparrow(\downarrow)}_{\bar{q};-\frac{1}{2}}(x,\textbf{k}_{\perp})|-\frac{1}{2},0; xP^+,\textbf{k}_{\perp}\rangle\Big].
  \end{align}
For nonzero transverse momentum of the proton, i.e., $\bfP\ne0$, the physical transverse momenta of the active quark and the spectator are $\bfp^{\bar{q}}=x\bfP+\bfk$ and $\bfp^S=(1-x)\bfP-\bfk$, respectively, where $\bfk$ corresponds to the relative transverse momentum of the constituents. $\psi_{\bar{q};\lambda_{\bar{q}}}^{\lambda_p}(x,\bfk)$ represent the light-front wave functions with helicities of the proton $\lambda_p=\uparrow(\downarrow)$ and the sea quark $\lambda_{\bar{q}}=\pm \frac{1}{2}$. The light-front wave functions at an initial scale $Q_0=2$ GeV are given by
\eq\label{wfs}
\psi^{\uparrow}_{\bar{q};+\frac{1}{2}}(x,\bfk) &=& \varphi(x,\bfk^2) 
\nonumber\\
\psi^{\uparrow}_{\bar{q};-\frac{1}{2}}(x,\bfk) &=&
- \dfrac{k^1 + i k^2}{\kappa} \, \varphi(x,\bfk^2) 
\,, \nonumber\\
\psi^{\downarrow}_{\bar{q};+\frac{1}{2}}(x,\bfk) &=&
- \Big[\psi^{\uparrow}_{\bar{q};-\frac{1}{2}}(x,\bfk)\Big]^\dagger =
\dfrac{k^1 - i k^2}{\kappa} \, \varphi(x,\bfk^2) 
\,, \nonumber\\
\psi^{\downarrow}_{\bar{q};-\frac{1}{2}}(x,\bfk) &=&
\Big[\psi^{\uparrow}_{\bar{q};+\frac{1}{2}}(x,\bfk)\Big]^\dagger =
\varphi(x,\bfk^2) \,,
\label{LFWFs}
\en
where $\varphi(x,\bfk^2)$ is the modified soft wall AdS/QCD wave function~\cite{Chakrabarti:2023djs},
\eq
\varphi(x,\bfk)=\sqrt{A} \frac{4 \pi }{\kappa}\sqrt{\frac{\log[1/(1-x)]}{x}}x^{\alpha}(1-x)^{3+\beta} \exp{\left[-\frac{\log[1/(1-x)]}{2\kappa^{2}x (1-x)}\bfk^{2}\right]},
\label{WaveFun}
\en
with $\alpha$ and $\beta$ are our model parameters. We consider the AdS/QCD scale parameter $\kappa = 0.4$ GeV~\cite{Chakrabarti:2013gra}. The values of the model parameters $\alpha$ and $\beta$, and the normalization constant $A$ are fixed by fitting the light sea quarks unpolarized PDFs at the scale $Q_0=2$ GeV with CTEQ18 next-to-next-leading order (NNLO) data.

The PDF, the probability that a parton carries a certain
fraction of the total light-front longitudinal momentum of the proton, provides us information about the nonperturbative structure of the proton. The sea quark PDF of the proton,
which encodes the distribution of longitudinal momentum and polarization carried by the sea quark in the proton, can be expressed in terms of the light-front wave functions. The unpolarized sea quark PDF at the model scale reads
\eq 
f_{1}(x) &=&
\int \frac{{\rm d}^2\bfk}{16\pi^3} \,
\left[ |\psi^{\uparrow}_{\bar{q}; + \frac{1}{2}}(x,\bfk)|^2
+ |\psi^{\uparrow}_{\bar{q};- \frac{1}{2}}(x,\bfk)|^2
\right] \nonumber\\
&=&
\int \frac{{\rm d}^2\bfk}{16\pi^3} \,
\left[ |\psi^{\downarrow}_{\bar{q}; + \frac{1}{2}}(x,\bfk)|^2
+ |\psi^{\downarrow}_{\bar{q}; - \frac{1}{2}}(x,\bfk)|^2
\right] \, \\
&=&  A x^{2 \alpha}(1-x)^{7+2 \beta} \left(1+ \frac{x (1-x)}{\log\frac{1}{1-x}} \right )\,,\nonumber
\en
while the helicity PDF is expressed as
\eq 
g_{1L}(x) &=&
\int \frac{{\rm d}^2\bfk}{16\pi^3} \,
\left[ |\psi^{\uparrow}_{\bar{q}; + \frac{1}{2}}(x,\bfk)|^2
- |\psi^{\uparrow}_{\bar{q};- \frac{1}{2}}(x,\bfk)|^2
\right] \nonumber\\
&=&
\int \frac{{\rm d}^2\bfk}{16\pi^3} \,
\left[ |\psi^{\downarrow}_{\bar{q}; + \frac{1}{2}}(x,\bfk)|^2
- |\psi^{\downarrow}_{\bar{q}; - \frac{1}{2}}(x,\bfk)|^2
\right] \, \\
&=&  A x^{2 \alpha}(1-x)^{7+2 \beta} \left(1- \frac{x (1-x)}{\log\frac{1}{1-x}} \right )\,,\nonumber
\en
and the Transversity PDF is given by 
\eq 
h_{1}(x) &=&\int \frac{{\rm d}^2\bfk}{16\pi^3} \psi ^{\uparrow \dagger}_{\bar{q};+\frac{1}{2}}(x,\textbf{k}_{\perp})\psi^{\downarrow }_{\bar{q};-\frac{1}{2}}(x,\textbf{k}_{\perp})\,, \nonumber \\
&=&  A x^{2 \alpha}(1-x)^{7+2 \beta}\,.
\en
%=============================
 \begin{table}[htp]
	\caption{Fitted model parameters at $Q_0=2$ GeV. }
	\centering
	\begin{tabular}{ |c|c|c|c|c|c|c|c|} 
		\hline\hline
		& A  & $\alpha $  & $\beta$   \\
		\hline
		%	\multirow{3}{4em}{Multiple row} & cell2 & cell3 \\ 
		$\overline{u}$ \quad \quad	&  \quad $0.055_{-0.002}^{0.003} $ &  \quad $-0.611_{0.001}^{0.001} $ & \quad $-1.33_{-0.181}^{-0.131} $  \\
		$\overline{d}$	\quad  \quad &  \quad $ 0.082_{0.007}^{-0.006} $ &  \quad $-0.572_{0.016}^{-0.014} $ & \quad $ -1.33_{0.163}^{-0.132}$ \\
		\hline \hline
	\end{tabular}
	\label{Table-1}
\end{table}
%==================================
%===================================
\subsection{Numerical fitting and model parameters}\label{sec:fitting}
%===================================
There are three parameters $\alpha$, $\beta$, and $A$ in our model that manifest the goodness of the model. 
The parameter $A$ is a  free parameter, which decides the amplitude of the sea quark PDF in the $x$ close to zero region. The parameters $\alpha$ and $\beta$ decide the behavior of the distributions in extreme limits of $x$. The parameter $\alpha$ is crucial to lower $x$, while $\beta$ controls large $x$ behaviour of the distributions. We determine these  parameters by fitting the model unpolarized light sea quark distributions with the latest available data at NNLO of the sea quark distributions $ x {\bar u}(x)$ and $ x {\bar d}(x)$ from the global analysis by the CTEQ Collaboration \cite{Hou:2019efy}.  We particularly fit CTEQ18 NNLO  sea quark PDFs at the scale $Q_0=2 $ GeV. We choose 201 number of data points distributed logarithmically within the interval $0.001 < x < 1 $. There are $58$ CTEQ18 NNLO replicas accessible along the centre line. The zeroth replicas represent the central lines of the distributions. We determine the central values of our model parameters by fitting those zeroth replicas of the CTEQ18 NNLO data. We choose replicas number $4$ and $8$ for the $\bar d$ and the $\bar u $ PDFs, respectively from the  data to estimate the uncertainties in our model parameters. 
The value of $\chi^2$ per degree of freedom (d.o.f.) for the fit of the $ x {\bar d}(x)$  is $ 0.271 $, whereas for the $ x {\bar u}(x)$ the value of $\chi^2$/d.o.f. is $0.0409$ in the region $0.001< x< 0.1$. The value of $\chi^2$ enlarges in the high $x$ region particularly for the $\bar u$ sea quarks. 
%The value of $\chi^2$ per d.o.f in the entire region of $0.001< x< 1$ for the fit of the $ x {\bar d}(x)$ PDF is $ 6.35 $ while it is $1876.26$ for the $ x f_1^{\bar u}(x)$ PDF fit. 
The fitted model parameters with their uncertainties are listed in Table~\ref{Table-1}. 

%===============================
\begin{figure}
	\centering
	\includegraphics[scale=0.32]{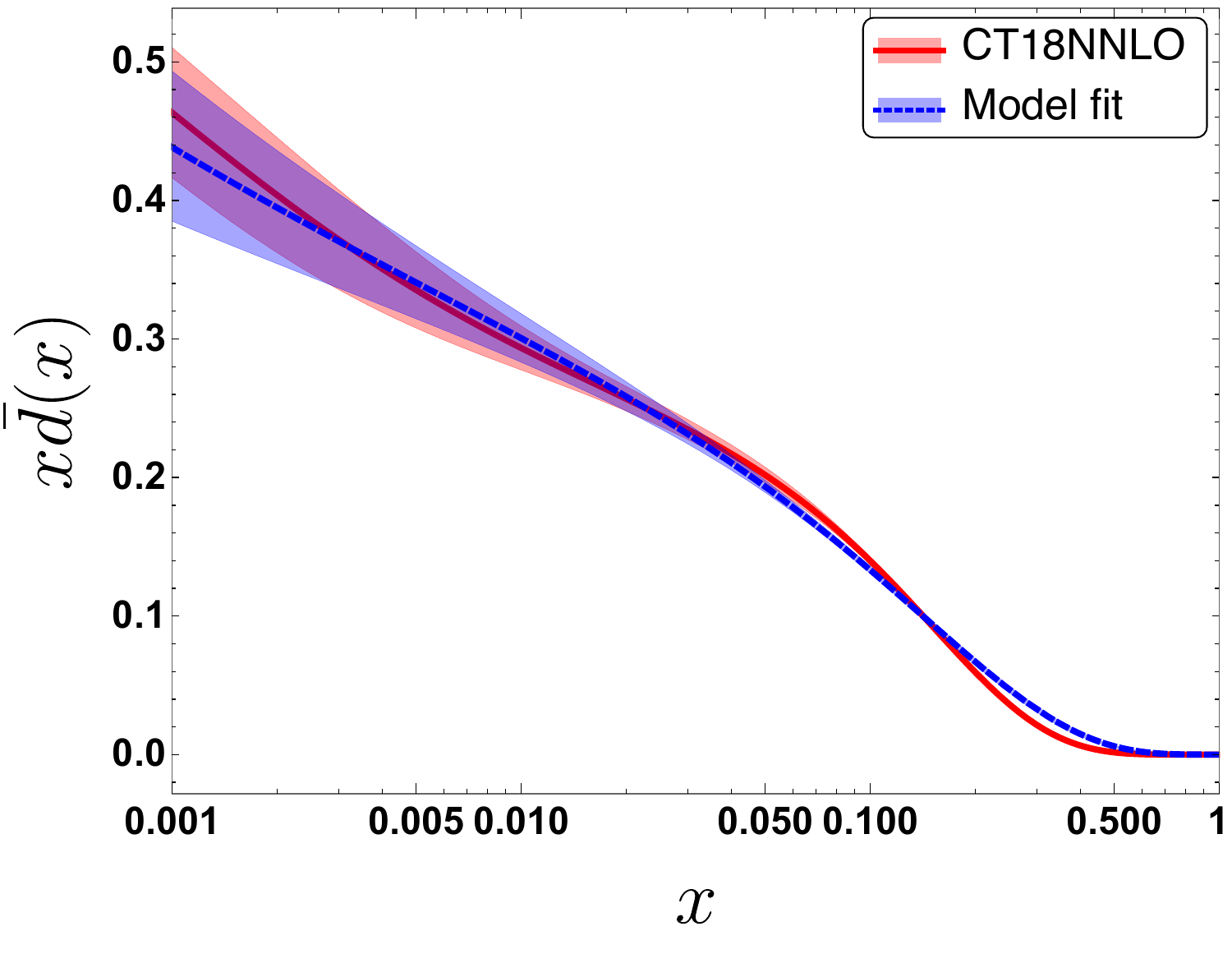}\hspace{0.5cm}
	\includegraphics[scale=0.32]{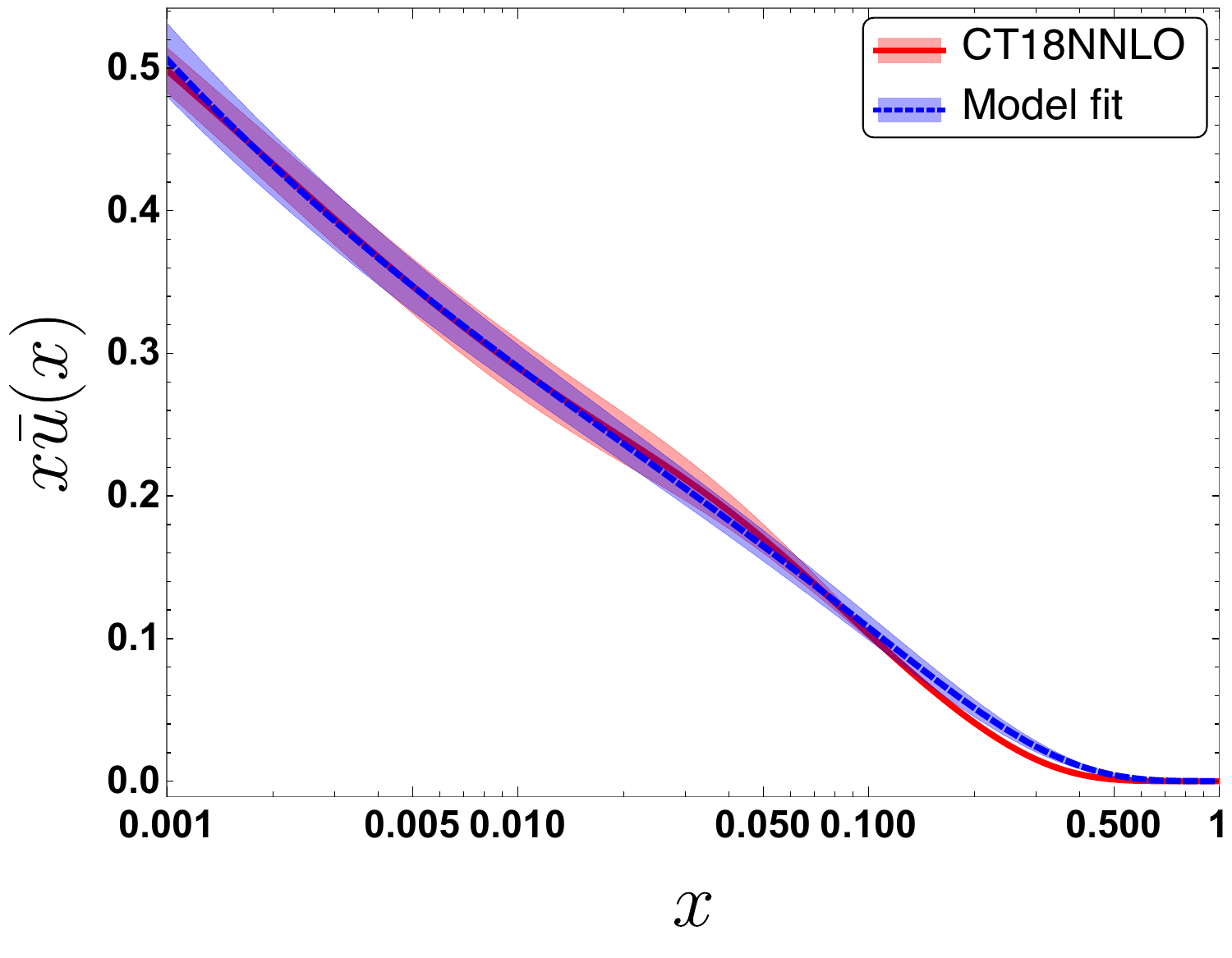}
	\caption{Sea quark unpolarized distributions in the proton, left panel:  $ x {\bar d}(x)$ and right panel: $ x {\bar u}(x)$ as a function of longitudinal momentum fraction $x$ in the kinematics region $0.001 \leq x \leq 1 $ at $Q_0=2$ GeV. Our model results (dashed-blue lines with blue bands) are compared with the CTEQ18 NNLO dataset (solid-red lines with red bands)~\cite{Hou:2019efy}}.
	\label{model fit}
\end{figure}
%==================================
\begin{figure}
	\centering
	\includegraphics[scale=0.32]{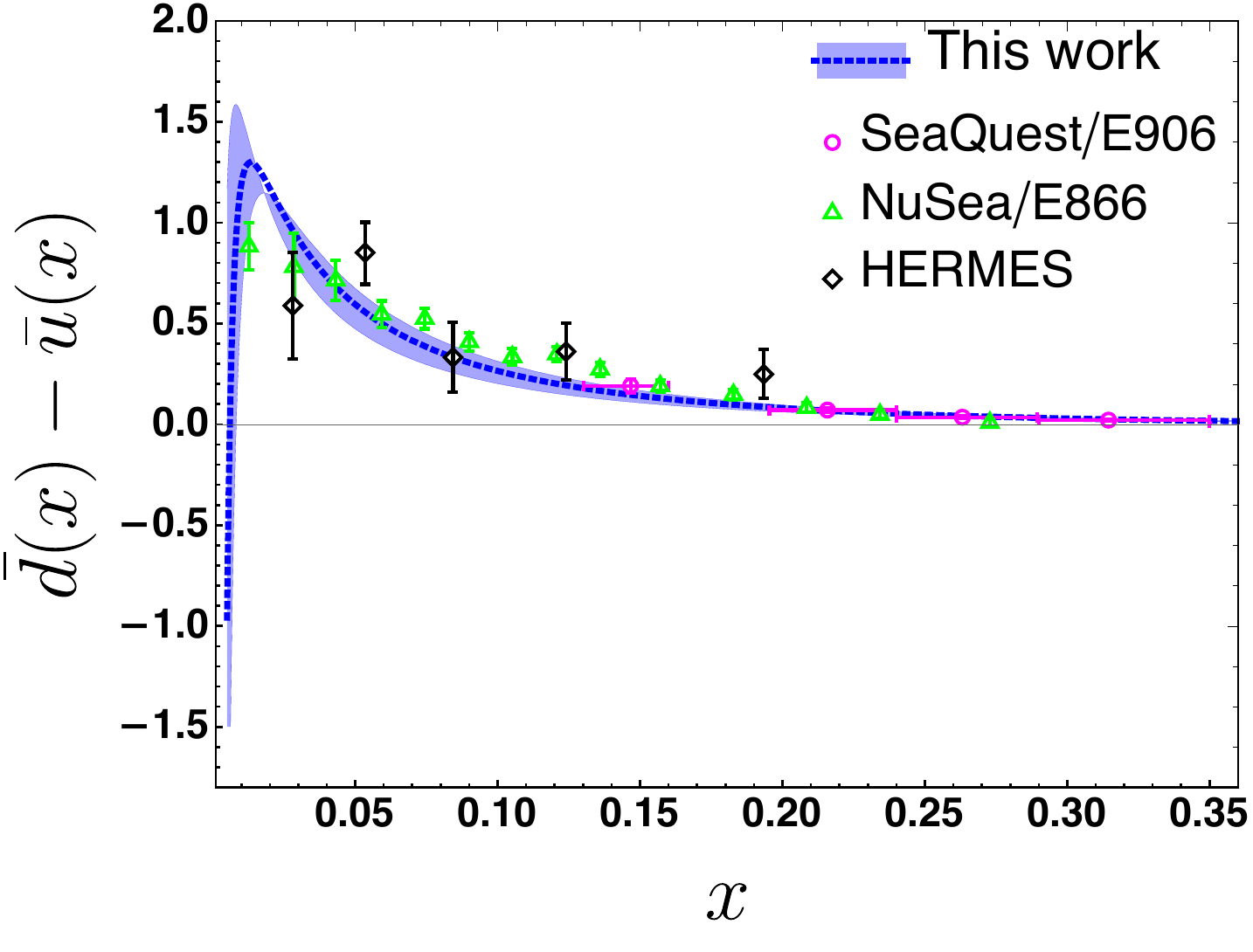}\hspace{0.5cm}
	\includegraphics[scale=0.32]{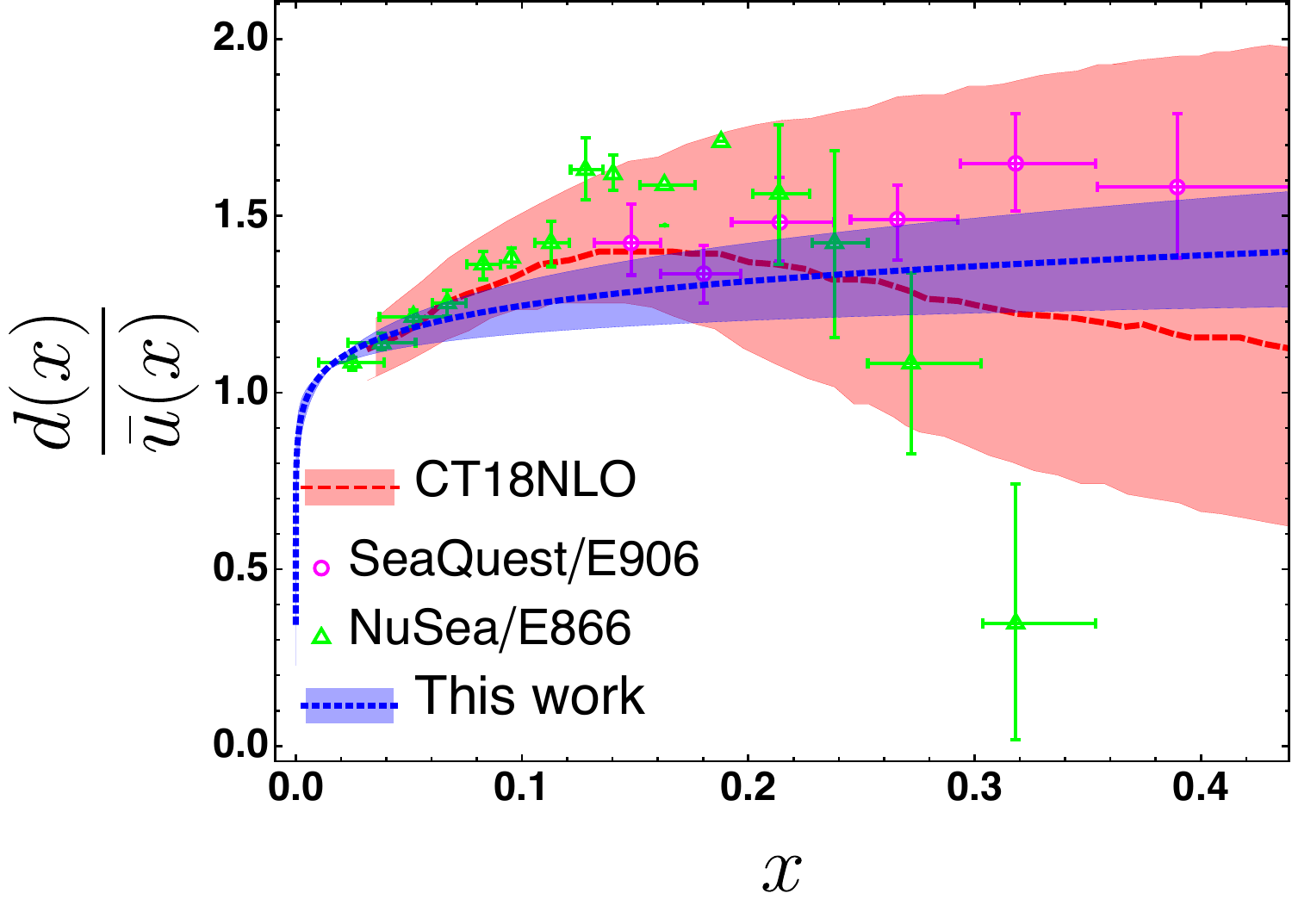}
	\caption{The light flavor sea asymmetries in the proton, left panel: $\bar{d}(x)-\bar{u}(x)$ and right panel: $\bar{d}(x)/\bar{u}(x)$ as a function of $x$. Our model results
(blue dotted lines with blue bands) are compared with the experimental data from the SeaQuest/E906~\cite{SeaQuest:2022vwp} (magenta open circle), NuSea/E866~\cite{NuSea:2001idv} (green open triangle), and HERMES~\cite{HERMES:1998uvc}(black open rombos) Collaborations. In the right panel, we also include the CTEQ18 NLO global analysis~\cite{Hou:2019efy} (red dashed lines with red bands) for the comparison. }
	\label{pdf asym}
\end{figure}
%===================================
  \begin{table}
 	\caption{Value of $\int  {\rm d}x\left[\bar d(x) - \bar u(x)\right]  $ in different range of $x$. Our results are compared with various experiments.}
 	\label{tab2}
 	\centering
 		\begin{tabular}{|c|c|c|}
           \hline\hline
 			~Model/Experiments   ~&~ $x$-range       ~&~ $\int {\rm d}x \left[ \bar d(x) - \bar u(x)\right]  $\\ \hline 
              This work         & $0.13<x<0.45$  & $0.015 \pm 0.004 $ \\ \hline
              SeaQuest/E906 \cite{SeaQuest:2022vwp} & $0.13<x<0.45$  & $0.015 \pm 0.003 $ \\ \hline
              This work         & $0.015<x<0.35$  & $0.069 \pm 0.015 $ \\ \hline
              NuSea/E866 \cite{NuSea:2001idv}        & $0.015<x<0.35$  & $0.0803 \pm 0.011$ \\ \hline
 			NMC \cite{NewMuon:1993oys}      & $0.004<x<0.80$  & $0.148 \pm 0.039$ \\ \hline
 			HERMES  \cite{HERMES:1998uvc}     & $0.020<x<0.30$  & $0.16  \pm 0.03$  \\ \hline \hline
 		\end{tabular}
 	%\end{center}
 \end{table}
 %=================================

In Fig.~\ref{model fit}, we show the results of our fits for the sea quark distributions, $ x {\bar u}(x)$ and $ x{\bar d}(x)$, at the scale $Q_0=2 $ GeV. The solid red lines with the red bands identify the CTEQ analyses of those distributions~\cite{Hou:2019efy} and the dashed blue lines with the blue bands represent our results computed with the effective parameters.
The second moments of the sea quark PDFs, which describe the average longitudinal momenta carried by the sea quarks are given by
 \begin{align}
\langle x\rangle_{\bar {d}}&=\int_{0.001}^{1}{\rm d}x\, x {\bar {d}}(x)=0.039\pm 0.002\,, \\ \nonumber
\langle x\rangle_{\bar {u}}&=\int_{0.001}^{1}{\rm d}x\, x {\bar {u}}(x)=0.032\pm 0.003\,.
\end{align} 

 We present the flavor asymmetries, ${\bar{d}}(x)-{\bar{u}}(x)$ and ${\bar{d}}(x)/{\bar{u}}(x)$, computed using our model wave functions in Fig.~\ref{pdf asym}. 
 We compare our  prediction for ${\bar{d}}(x)-{\bar{u}}(x)$  with the available data from the SeaQuest/E906 \cite{SeaQuest:2022vwp}, HERMES~\cite{HERMES:1998uvc} and NuSea/E866~\cite{NuSea:2001idv} Collaborations in Fig.~\ref{pdf asym} (left plot).   We find that below $x<0.01$, ${\bar{u}}(x)$ is larger than ${\bar{d}}(x)$ in our model and the $\bar{d}(x)-\bar{u}(x)$ asymmetry is opposite to what generally observed in the literature. However, this pattern has also been observed in the global analyses made by CT14~\cite{Dulat:2015mca} and the latest NNPDF4.0 \cite{NNPDF:2021njg} Collaborations. 
 On the other hand, in the region $x>0.01$, we find more or less consistency between our result and the available experimental data~\cite{NuSea:2001idv,HERMES:1998uvc,SeaQuest:2022vwp}. Note that the value of the quantity $ \int_{0}^{1}\left[\bar{d}(x)-\bar{u}(x)\right]dx $ depends on the energy scale of the measurement.
The numerical results for the quantity at 
 the scale $Q_0=2$ GeV for selected $x$ domains are given in Table~\ref{tab2}. We compare our results with results from various experiments~\cite{NewMuon:1993oys,HERMES:1998uvc,NuSea:2001idv}. We find that our predictions are in good agreement with the SeaQuest/E906 results~\cite{SeaQuest:2022vwp} as shown in Table~\ref{tab2}. Additionally, the SeaQuest/E906~\cite{SeaQuest:2022vwp} result for the difference of second moments: $ \int_{0.13}^{0.45} x \left[\bar{d}(x)-\bar{u}(x)\right]dx=0.00318 ^{+0.0005}_{-0.0006}$ is in excellent agreement with our model prediction: $ \int_{0.13}^{0.45} x \left[\bar{d}(x)-\bar{u}(x)\right]dx=0.003 \pm 0.001$.

 Figure~\ref{pdf asym}  (right plot) compares our result for the ${{\bar{d}}(x)}/{{\bar{u}}(x)}$ with the experimental data  from the NuSea/E866~\cite{NuSea:2001idv}, SeaQuest/E906~\cite{SeaQuest:2022vwp} Collaborations as well as with the CTEQ18 NLO global analysis~\cite{Hou:2019efy}.  It can be noticed that there is a 
 difference in qualitative behaviour between the SeaQuest/E906 data, which exhibits more or less flat behavior of the $\bar{d}(x)/\bar{u}(x)$ asymmetry, and the NuSea/E866 measurement showing a rapid fall of the ${{\bar{d}}(x)}/{{\bar{u}}(x)}$  in the  region $x>0.2$. The trends between the two experiments at higher $x$ domain are quite different. As can be seen, our
prediction for this asymmetry agrees well with the global fit and the trend of the latest experimental data from the SeaQuest/E906 Collaboration~\cite{SeaQuest:2022vwp} available in the range $0.015<x<0.35$. Meanwhile, Our result deviates from the overall behavior of portrayed by NuSea/E866 data~\cite{NuSea:2001idv}. 
In a Drell-Yan measurement at CERN, the NA51 Collaboration ~\cite{NA51:1994xrz} confirmed that $\bar{d}(x)$ is larger than
$\bar{u}(x)$ at an average $x$ value of $0.18$. They reported $\bar{d}(x)/\bar{u}(x)|_{x=0.18}=1.96 \pm 0.15  \pm 0.19$, while in our model, we obtain $\bar{d}(x)/\bar{u}(x)|_{x=0.18}=1.30\pm 0.08$.

\begin{figure}
	\centering
	\includegraphics[scale=0.32]{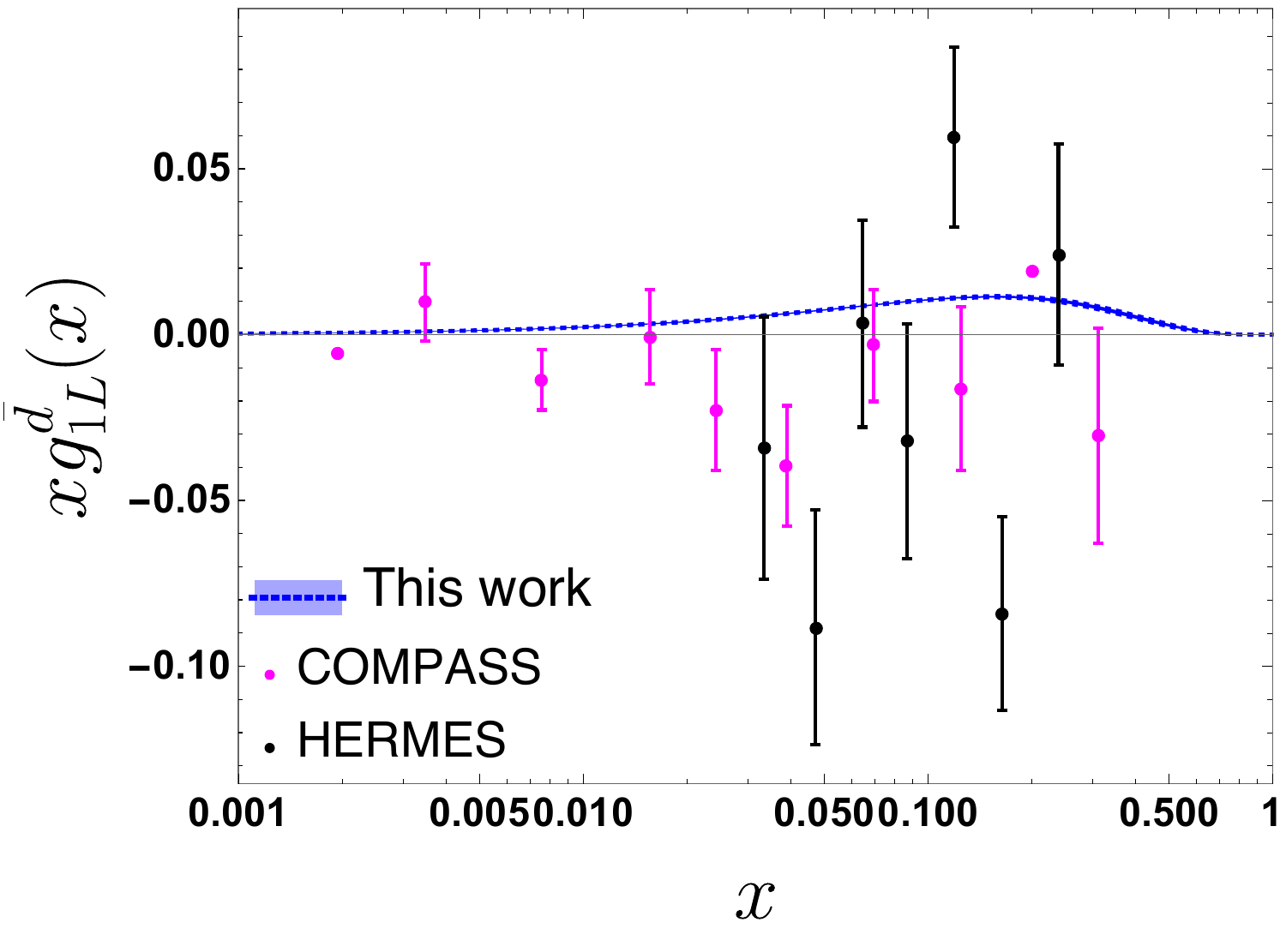}\hspace{0.5cm}
	\includegraphics[scale=0.32]{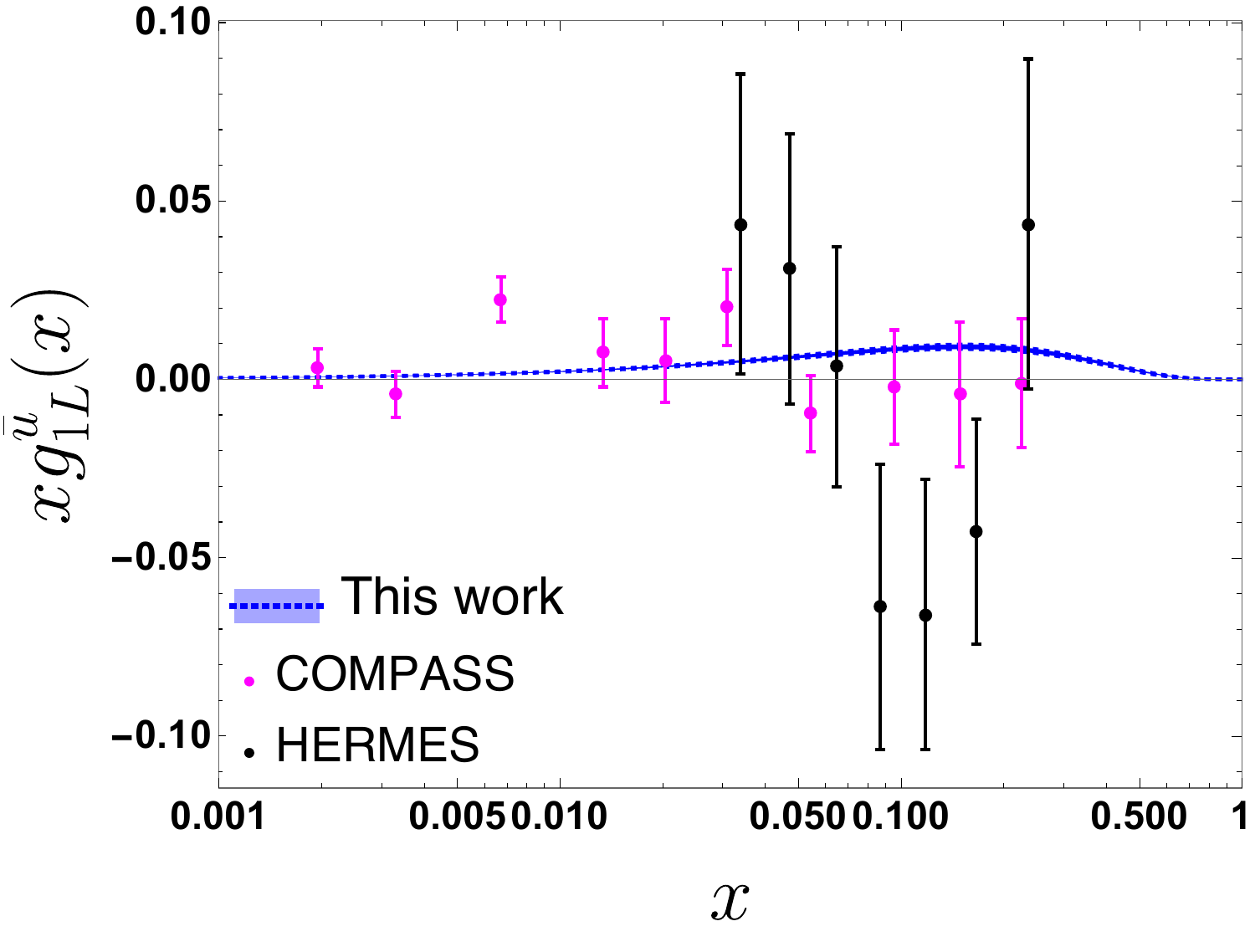}\vspace{0.5cm} \\
	\caption{ The helicity PDFs for $\bar{d}$ (left panel) and $\bar{u}$ (right panel) in the proton. Our model results
(blue lines with blue bands) are compared with the experimental data from the COMPASS \cite{COMPASS:2010hwr} (magenta solid  circle) and HERMES \cite{HERMES:2003gbu} (black solid circle) Collaborations.}
	\label{helicity}
\end{figure}

We show the results for helicity distribution for $\bar{d}$ and $\bar{u}$ in Fig.~\ref{helicity}, where we compare our model predictions with the experimental data from the COMPASS \cite{COMPASS:2010hwr} and HERMES \cite{HERMES:2003gbu} Collaborations. We find acceptable agreement between our sea quark helicity distributions
and the experimental data. Note that the signs of sea quark helicity
PDFs  are not fixed by the experiments.
The spin contribution of the $\bar{d}$ and $\bar{u}$ to the total proton spin can be quantified as the first moment of the helicity PDFs,
 \begin{align}
\Delta \Sigma^{\bar {d}}&=\int_{0.001}^{1} {\text dx} g_{1L}^ {\bar {d}}(x)=0.031\pm 0.001\,, \\ \nonumber
\Delta \Sigma^{\bar {u}}&=\int_{0.001}^{1} {\text dx}  g_{1L}^{\bar {u}}(x)=0.026\pm 0.002\,.
\end{align} 

\begin{figure}
	\centering
	\includegraphics[scale=0.32]{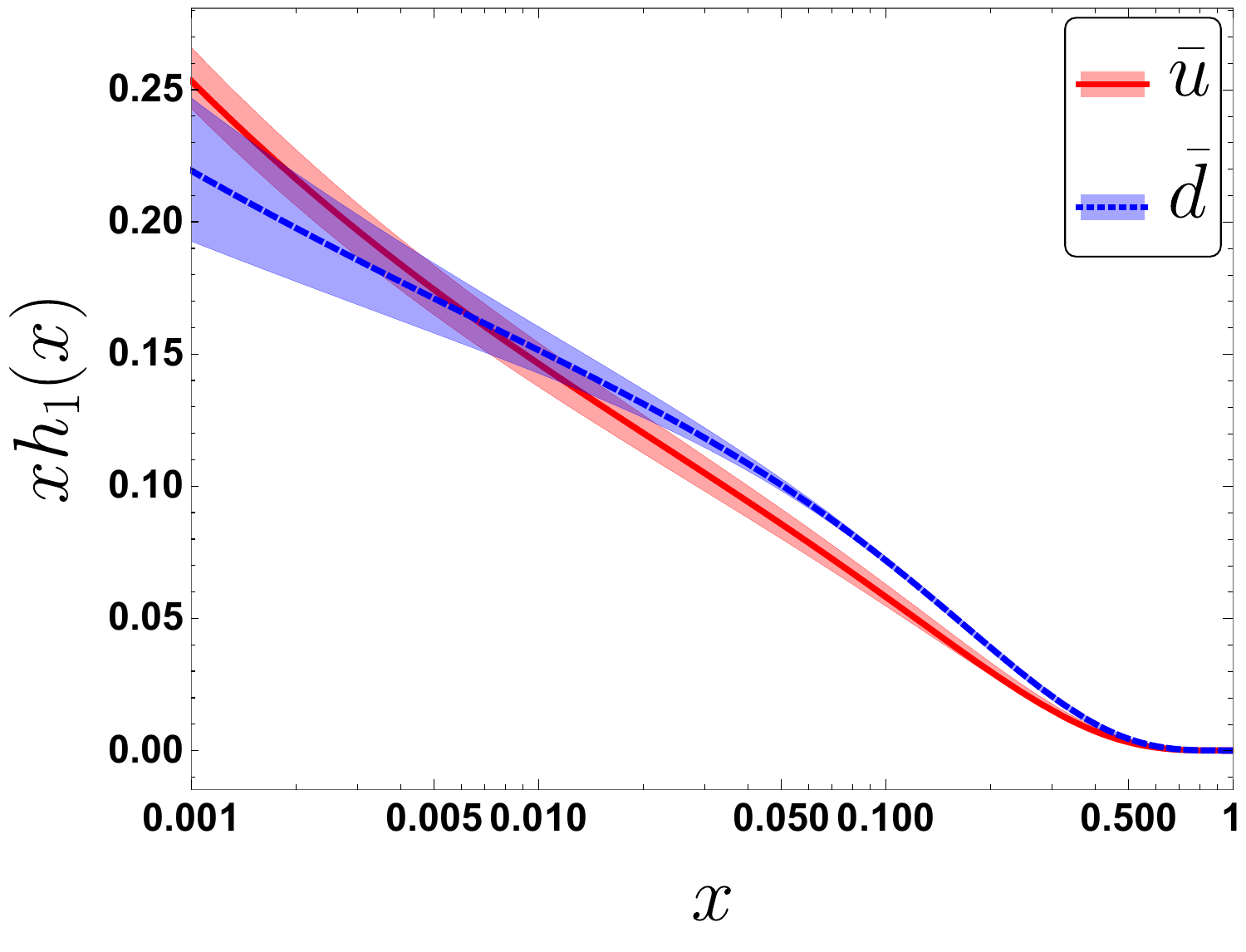}\hspace{0.5cm}
	\caption{The transversity PDFs for $\bar{d}$ (blue line with blue band) and $\bar{u}$ (red line with red band) in the proton evaluated in our model.} 
	\label{Transversity}
\end{figure}

The sea quark's transversity PDFs at the scale $Q_0=2$ GeV are shown in Fig.~\ref{Transversity}. They appear similar to unpolarized PDFs, mainly distributed at small-$x$ region and vanish after $x>0.5$. At small-$x$ ($<0.05$), $h_1^{\bar{u}}(x)$ dominates over $h_1^{\bar{d}}(x)$, whereas $h_1^{\bar{d}}(x)$ is larger than $h_1^{\bar{u}}(x)$ at $x>0.05$. The first moment of the transversity PDFs give the tensor
charges. We find the sea quark tensor charges  at the scale $Q_0=2$ GeV as 
 \begin{align}
\delta {\bar {d}}&=\int_{0.001}^{1}{\text dx} h_{1}^ {\bar {d}}(x)=0.75\pm 0.05\,, \\ \nonumber
\delta {\bar {u}}&=\int_{0.001}^{1} {\text dx}  h_{1}^{\bar {u}}(x)=0.73\pm 0.04\,.
\end{align} 

%================================
%====================================
%====================================

%====================
\section{Sea quark GPDs} \label{GPDs}
%====================
In general, the GPDs are obtained through the off-forward matrix elements of the bilocal operators between hadronic
states. There are four chiral-even  quark GPDs at leading twist, denoted as $H^q$, $E^q$, $\widetilde{H}^q$, and $\widetilde{E}^q$. The unpolarized ($H$ and $E$) and helicity-dependent ($\widetilde{H}$ and $\widetilde{E}$) quark GPDs for the nucleon are parameterized as~\cite{Diehl:2001pm,Diehl:2003ny}, 
\begin{align}
%\label{gpd_eq}
& \frac{1}{2}\int\frac{\text d z^-}{2\pi}e^{i \bar xP^+z^-}\langle
p', \lambda'|\bar{\psi}_q(-{z}/2)\gamma^+
\psi_q({z}/2)|p, \lambda\rangle_{|_{z^+ = 0, \vec z_\perp=
		0}}
=\frac{1}{2P^+}\bar{u}(p',\lambda')\bigg[H^q\gamma^+ + E^q\frac{i}{2m}\sigma^{+\alpha}\Delta_{\alpha}\bigg]u(p,\lambda)\,,\label{gpd_unpol}
\\
& \frac{1}{2}\int\frac{\text dz^-}{2\pi}e^{i \bar xP^+z^-}\langle
p', \lambda'|\bar{\psi}_q(-{z}/2)\gamma^+\gamma_5
\psi_q({z}/2)|p, \lambda\rangle_{|_{z^+ = 0, \vec z_\perp=
		0}}
= \frac{1}{2P^+} \bar{u}(p',\lambda') \left[
\widetilde{H}^q\, \gamma^+ \gamma_5 +
\widetilde{E}^q\, \frac{\gamma_5 \Delta^+}{2m}
\right] u(p,\lambda)\,,\label{gpd_helicity}
\end{align}
where $\psi_q(x)$ represents the quark field in the light-cone gauge. The momenta of the nucleon initial and final states are denoted by $p$ and $p'$, respectively, with their corresponding helicities denoted as $\lambda$ and $\lambda'$ and  $m$ is the nucleon mass. The light-front spinor for the nucleon is  denoted by $u(p,\lambda)$. The momentum transfer in this picture is expressed as $\Delta^{\mu} = p^{\prime\mu}-p^{\mu}$, while the skewness is given by $\zeta = -\frac{\Delta^+}{2\bar{P}^+}$. 
In this work, we consider the zero skewness case ($\zeta =0$) and thus only study the kinematic region corresponding to $\zeta < x < 1 $ also called the DGLAP region. The invariant momentum transfer in the process is denoted by $t=\Delta^2$ and for zero skewness, one has $t=-\Delta_\perp^2$. Note that one has to take into account nonzero skewness to compute $\widetilde{E}^g$. The remaining three chiral-even quark GPDs at leading twist can be written in terms of diagonal overlap of LFWFs as follows:
\begin{align}
H^{\bar{q}}(x, 0, -\Delta_\perp^2) &=  \int \frac{\text {d}^2\bfk}{16 \pi^3} \, 
\biggl[ \psi^{\dagger\uparrow}_{\bar{q}; + \frac{1}{2}}(x,\bfk^{\prime \prime}) \,   
\psi^{\uparrow}_{\bar{q}; + \frac{1}{2}}(x,\bfk^{\prime}) 
+ \psi^{\dagger \uparrow}_{\bar{q}; - \frac{1}{2}}(x,\bfk^{\prime \prime}) \, 
\psi^{\uparrow}_{\bar{q}; - \frac{1}{2}}(x,\bfk^{\prime}) \biggr] \ \label{GPDHq},, \\
E^{\bar{q}}(x, 0, -\Delta_\perp^2) &= - \frac{2 M_N}{q^1-iq^2} \int \frac{\text {d}^2\bfk}{16 \pi^3} \, 
\biggl[ \psi^{\dagger\uparrow}_{\bar{q}; + \frac{1}{2}}(x,\bfk^{\prime \prime}) \,   
\psi^{\downarrow}_{\bar{q}; + \frac{1}{2}}(x,\bfk^{\prime}) + \psi^{\dagger\uparrow}_{\bar{q}; - \frac{1}{2}}(x,\bfk^{\prime \prime}) \,  \psi^{\downarrow}_{\bar{q}; - \frac{1}{2}}(x,\bfk^{\prime}) \biggr]\ \label{GPDEq},, \\
\tilde{H}^{\bar{q}}(x, 0, -\Delta_\perp^2) &= \int \frac{\text {d}^2\bfk}{16 \pi^3} \, 
\biggl[ \psi^{\dagger\uparrow}_{\bar{q}; + \frac{1}{2}}(x,\bfk^{\prime \prime}) \,   
\psi^{\uparrow}_{\bar{q}; + \frac{1}{2}}(x,\bfk^{\prime}) -\psi^{\dagger\uparrow}_{\bar{q}; -\frac{1}{2}}(x,\bfk^{\prime \prime}) \, \psi^{\uparrow}_{\bar{q}; -\frac{1}{2}}(x,\bfk^{\prime}) \biggr] \,\label{GPDHtildeq},
\end{align}
where, the final transverse momentum of the struck quark
\begin{align}
\bfk^{\prime \prime}=\bfk + (1-x) \frac{\Delta_\perp}{2}
\end{align}
and the initial transverse momentum of the struck quark
\begin{align}
\bfk^{\prime}= \bfk - (1-x) \frac{\Delta_\perp}{2}\,.
\end{align}

Using the light-front wave functions in our model given in Eq.~\eqref{wfs}, 
the explicit expressions of the sea quark GPDs at zero skewness read
\eq
H^{\bar{q}}(x, 0, -\Delta_\perp^2)&=&  A x^{2 \alpha+1}(1-x)^{6+2 \beta} \left(1+ \frac{x^2}{\log\frac{1}{1-x}} -\frac{\Delta_\perp^2 (1-x)^2}{4 \kappa^2}\right )  \exp{\bigg[-\frac{\log[1/(1-x)]}{2\kappa^{2}x^2}\Delta_\perp^2\bigg]}\,, \label{GPDH}\\
E^{\bar{q}}(x, 0, -\Delta_\perp^2) &=& \frac{2 M_N}{\kappa}  A x^{2 \alpha+1}(1-x)^{7+2 \beta} \exp{\bigg[-\frac{\log[1/(1-x)]}{2\kappa^{2}x^2}\Delta_\perp^2\bigg]}\,, \\
\tilde{H}^{\bar{q}}(x, 0, -\Delta_\perp^2) &= &  A x^{2 \alpha+1}(1-x)^{6+2 \beta} \left(1- \frac{x^2}{\log\frac{1}{1-x}} -\frac{\Delta_\perp^2 (1-x)^2}{4 \kappa^2}\right )  \exp{\bigg[-\frac{\log[1/(1-x)]}{2\kappa^{2}x^2}\Delta_\perp^2\bigg]}\,.\label{GPDHt}
\en

%===================================
\begin{figure}
	\centering
	
	\includegraphics[scale=0.22]{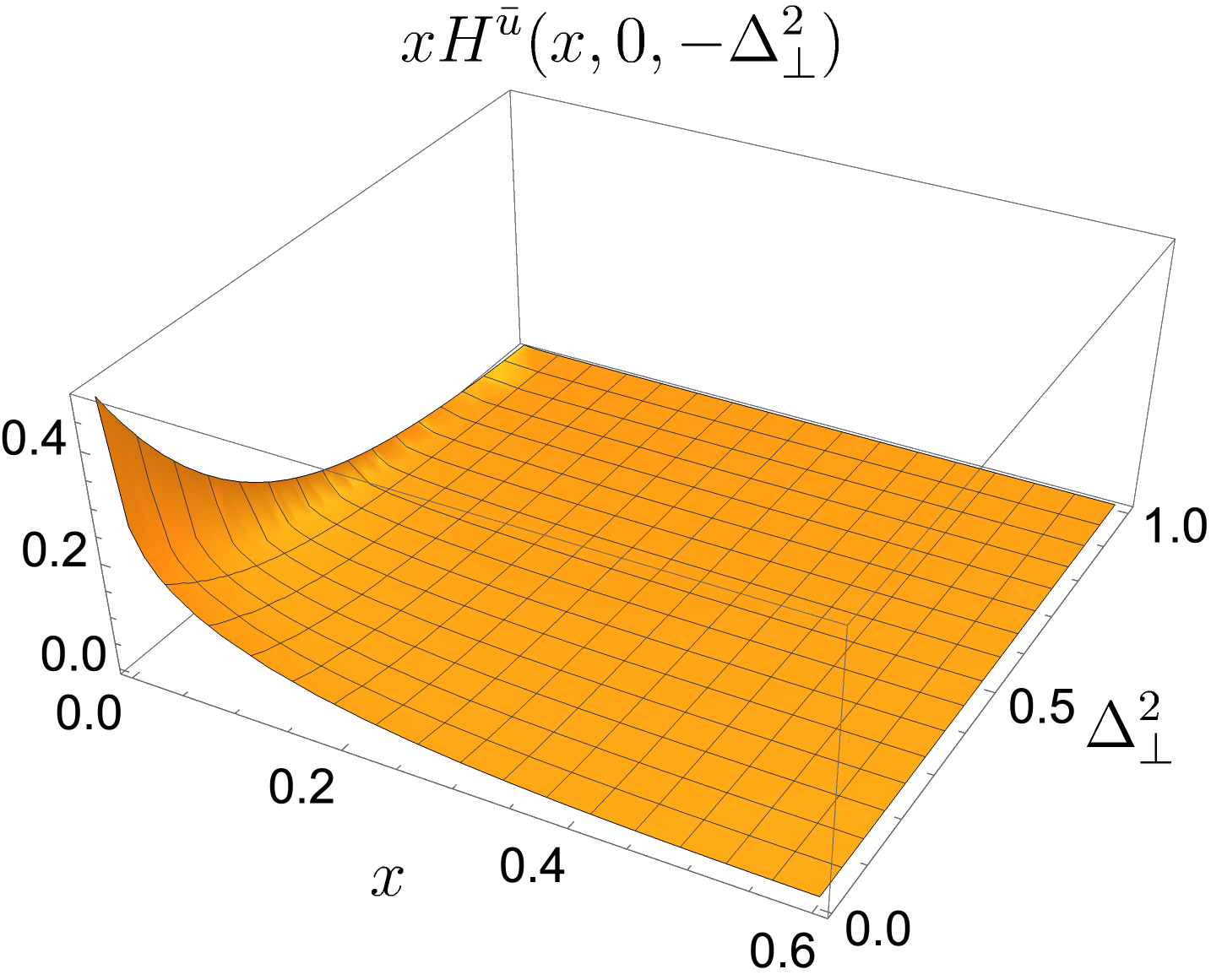}\hspace{0.4cm} 
	\includegraphics[scale=0.22]{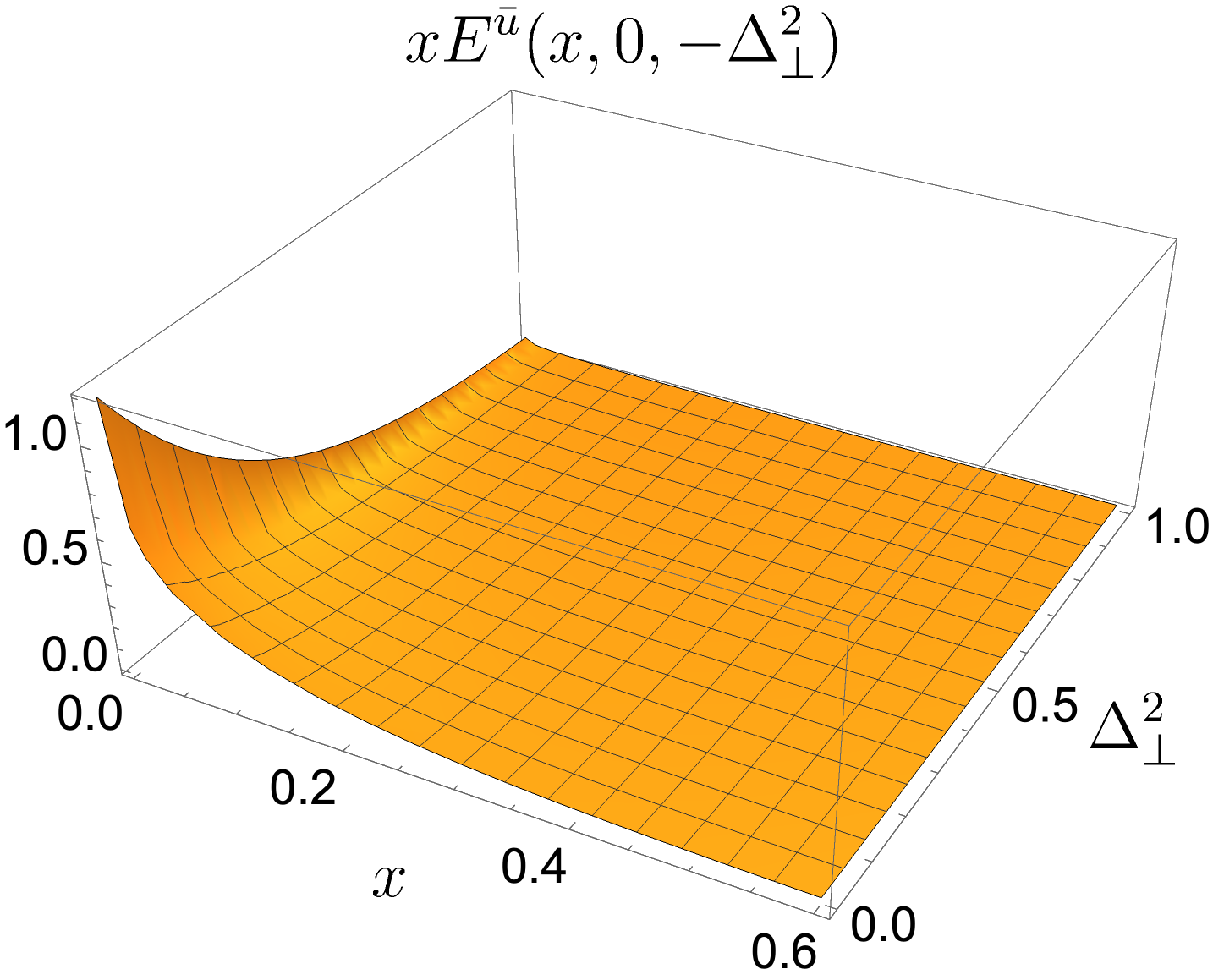}\hspace{0.4cm}
	\includegraphics[scale=0.22]{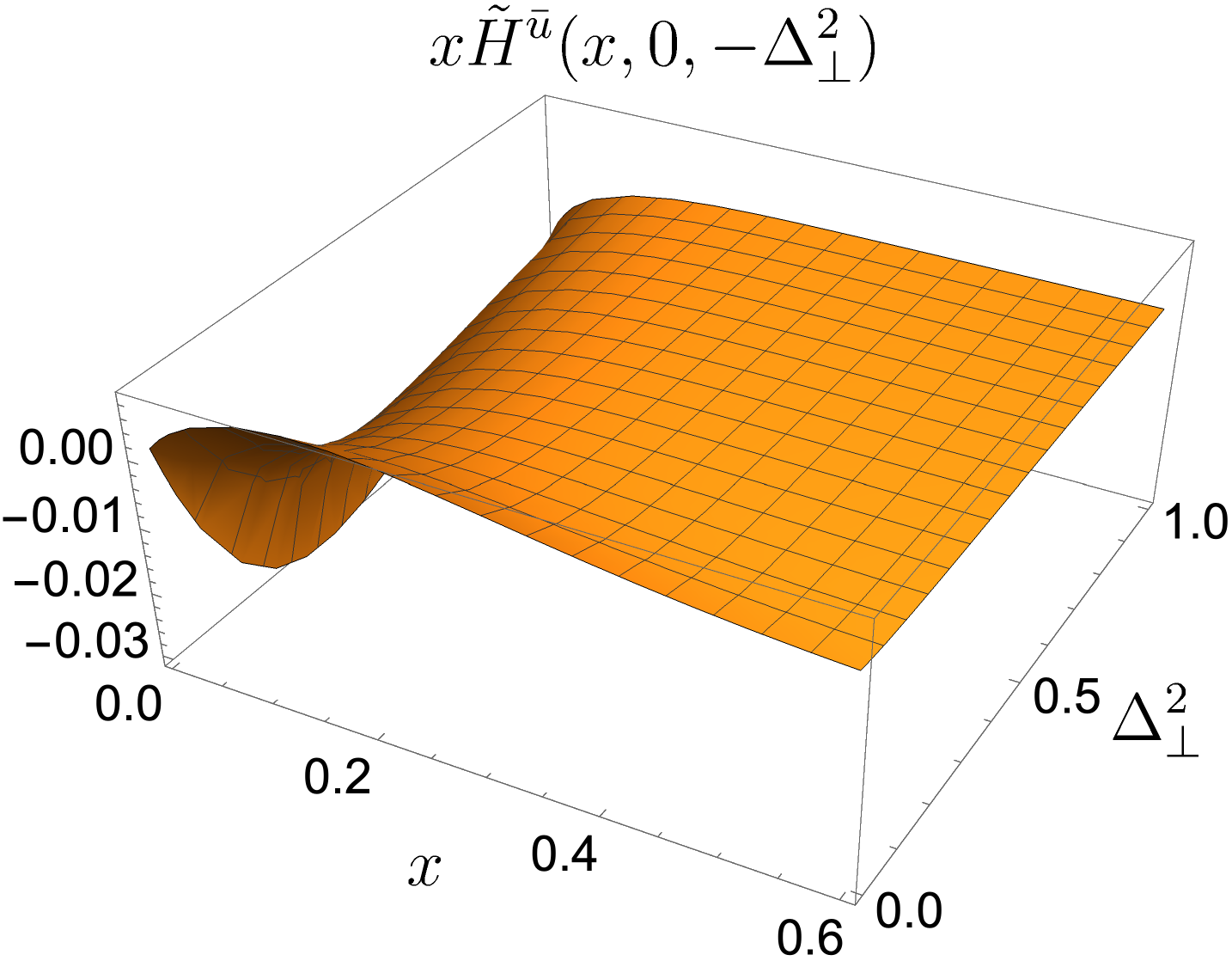}\vspace{0.4cm} \\
        \includegraphics[scale=0.22]{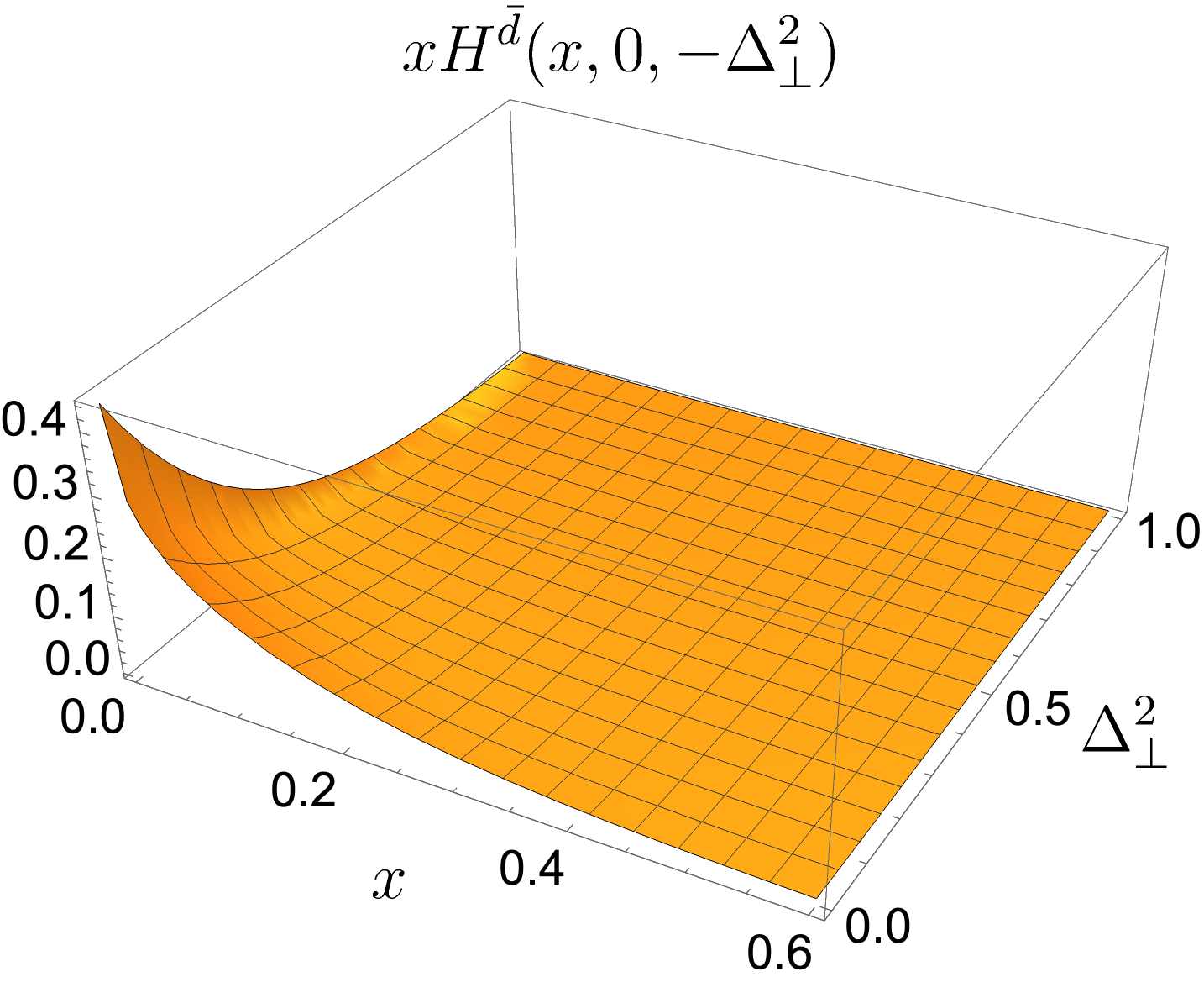}\hspace{0.4cm}
	\includegraphics[scale=0.22]{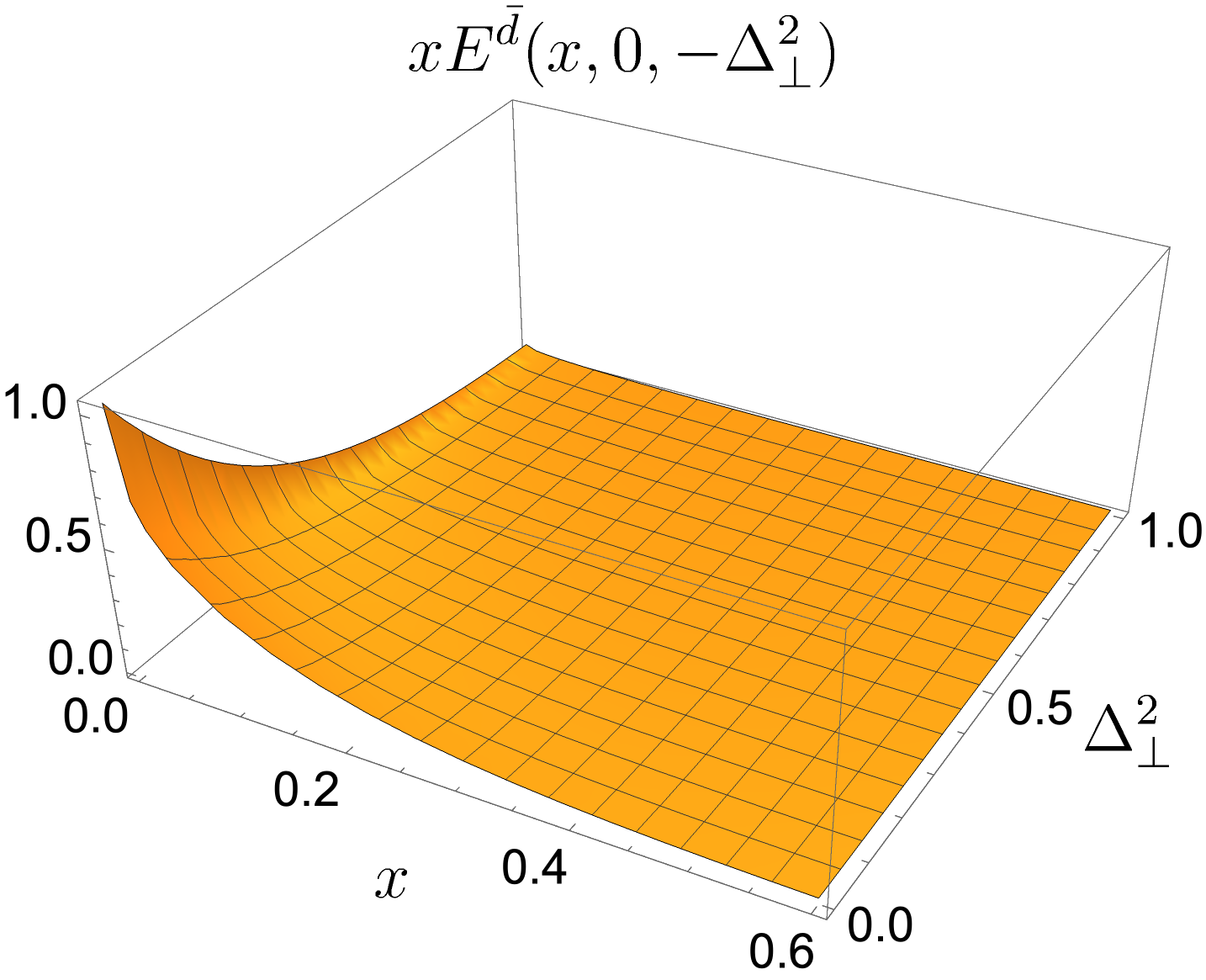}\hspace{0.4cm} 
	\includegraphics[scale=0.22]{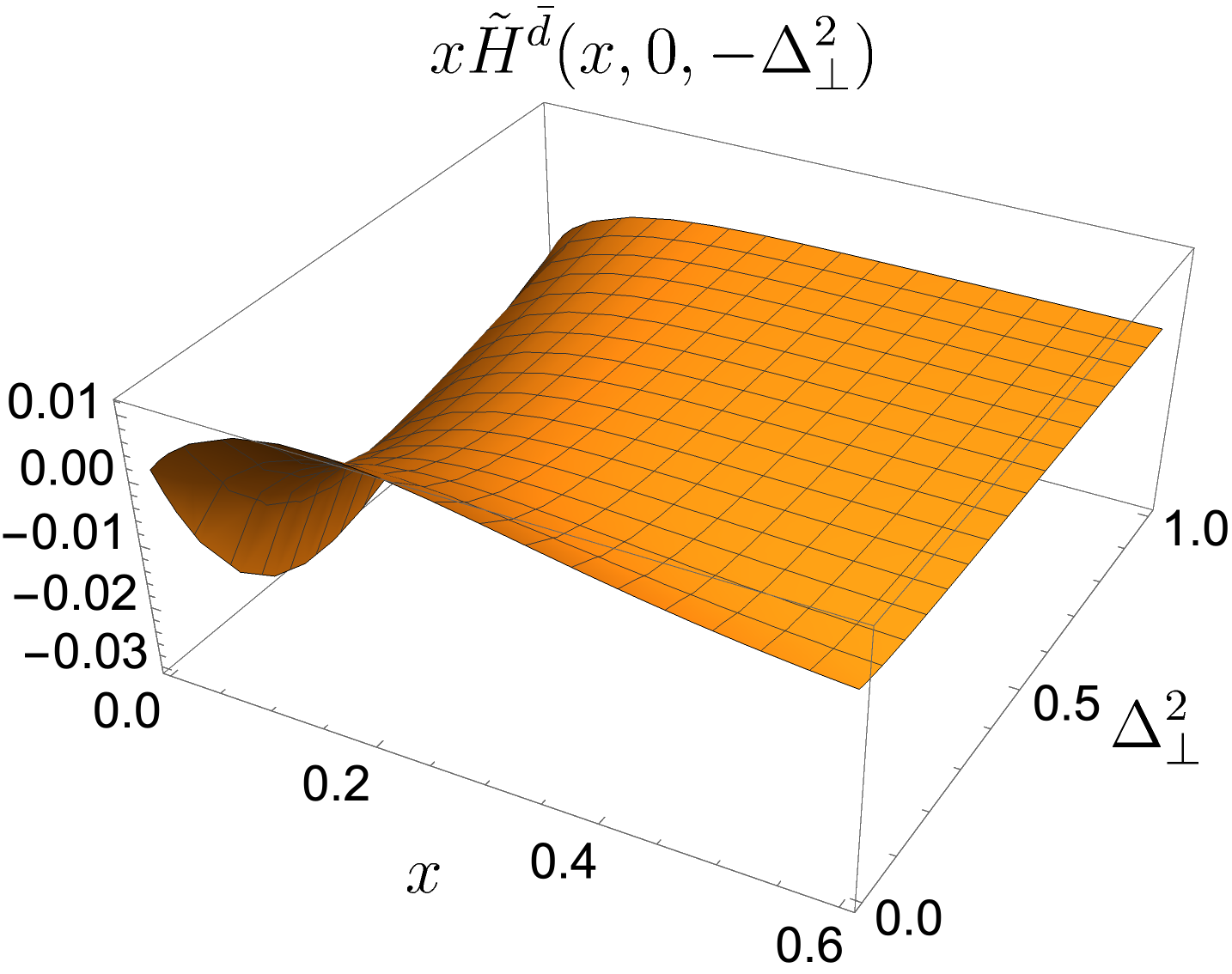}\vspace{0.4cm}\\
	\caption{The sea quark chiral-even GPDs, $ H^{\bar{q}}(x,0,-\Delta_\perp^2)$ (left panel), $ E^{\bar{q}}(x,0,-\Delta_\perp^2)$ (middle panel), and $ \tilde{H}^{\bar{q}} (x,0,-\Delta_\perp^2)$ (right panel) plotted in the range $0.001<x<0.6 $ at zero skewness for the proton calculated within our model. The upper panel is for the $\bar{u}$ quark and the lower panel represents the distributions for the $\bar{d}$ quark.} 
	\label{3DGPDs}
\end{figure}
%===================================

We utilize the obtained
parameters listed in Table~\ref{Table-1} in conjunction with the expressions from Eqs.~\eqref{GPDH} $-$ \eqref{GPDHt} to compute the sea quark GPDs in momentum
space. We present three-dimensional (3D) plots
for the three distributions, $x H^{\bar q} (x,0,-\Delta_\perp^2)$, $x E^{\bar q} (x,0,-\Delta_\perp^2)$, and $x \widetilde{H}^{\bar q} (x,0,-\Delta_\perp^2)$ in Fig.~\ref{3DGPDs}. We observe that $x H^{\bar q} (x,0,-\Delta_\perp^2)$ and $x E^{\bar q} (x,0,-\Delta_\perp^2)$  exhibit similar behavior having positive peaks at low-$x$, whereas the distribution $x \widetilde{H}^{\bar q} (x,0,-\Delta_\perp^2)$ shows distinctly different behavior from the other two GPDs, having negative peaks along the $\Delta_\perp^2$ direction at low-$x$. The maximum
peak value for $x H^{\bar q} (x,0,-\Delta_\perp^2)$ and $x E^{\bar q} (x,0,-\Delta_\perp^2)$ occurs at the forward limit, $\Delta^2=0$. 
Notably, the largest peak for $E^{\bar q}(x,-\Delta_\perp^2)$ is significantly larger than those for $H^{\bar q}(x,-\Delta_\perp^2)$ and $\widetilde{H}^{\bar q}(x,-\Delta_\perp^2)$. The maximum peak magnitude for  $xH^{\bar q}(x,-\Delta_\perp^2)$ is also larger than that of $x\widetilde{H}^{\bar q}(x,-\Delta_\perp^2)$. We also notice that the magnitudes of all the distributions decrease as the momentum transfer $\Delta_\perp^2$ increases similar to that noticed for the quark GPDs in the nucleon~\cite{Ji:1997gm,Scopetta:2002xq,Petrov:1998kf,Penttinen:1999th,Boffi:2002yy,Boffi:2003yj,Vega:2010ns,Mondal:2015uha,Chakrabarti:2015ama,Mondal:2017wbf,deTeramond:2018ecg,Xu:2021wwj,Chakrabarti:2013gra}. However, $x H^{\bar q} (x,0,-\Delta_\perp^2)$ and $x E^{\bar q} (x,0,-\Delta_\perp^2)$ fall faster than $x\widetilde{H}^{\bar q}(x,-\Delta_\perp^2)$.   The forward limits of $H^{\bar q}(x,0)$ and $\widetilde{H}^{\bar q}(x,0)$ correspond to ordinary helicity-independent and helicity-dependent sea quark PDFs, which are shown in Figs.~\ref{model fit} and \ref{helicity}, respectively. 
%We note that $E^{\bar q}(x,-\Delta_\perp^2)$ decouples in the forward limit, so no such limit exists for $E^{\bar q}(x,-\Delta_\perp^2)$ ~\cite{Diehl:2003ny}.

%\sout{When comparing with the results from the nonlocal covariant chiral effective theory~\cite{He:2022leb}, a significant difference can be observed.}
Contrary to our model, where the $E^{\bar q}(x,0,-\Delta_\perp^2)$  represents positive definite quantity for both the anti-up and anti-down quarks, which is consistent with the finding obtained in a light-cone baryon-meson fluctuation model~\cite{Luan:2023lmt}, the covariant chiral effective theory \cite{He:2022leb} displays a negative distribution for the anti-up quark but positive for the anti-down quark. It is anticipated that future investigations will employ theoretical and experimental analyses to examine the underlying causes of these discrepancies. We also notice that the magnitudes of the distributions in our model are comparable to those computed within the baryon-meson fluctuation model~\cite{Luan:2023lmt} but much higher than those results from the nonlocal covariant chiral effective theory~\cite{He:2022leb}. It should also be noted that the scale of the baryon-meson fluctuation model, $Q_0\sim 0.7$ GeV, and the covariant chiral effective theory approach, $Q_0\sim 1.0$ GeV are different from model scale, $Q_0=2$ GeV.    However, the qualitative nature of the distributions as shown in Refs.~\cite{He:2022leb,Luan:2023lmt} aligns with our findings.

%======================
\subsection{Flavor asymmetry at non zero momentum transfer} \label{GPD asymmetry}
\begin{figure}
	\centering
	\includegraphics[scale=0.22]{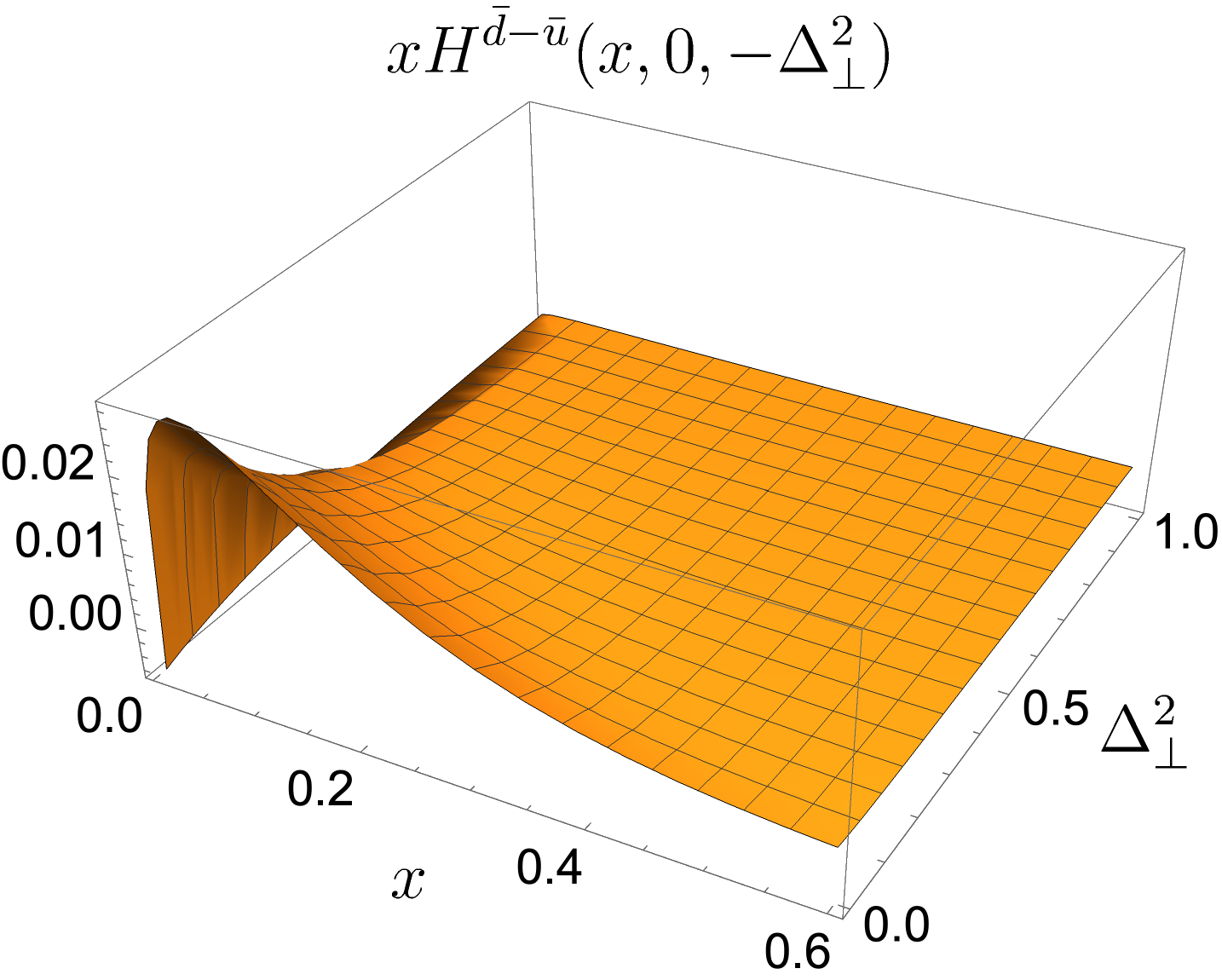}\hspace{0.4cm}
	\includegraphics[scale=0.22]{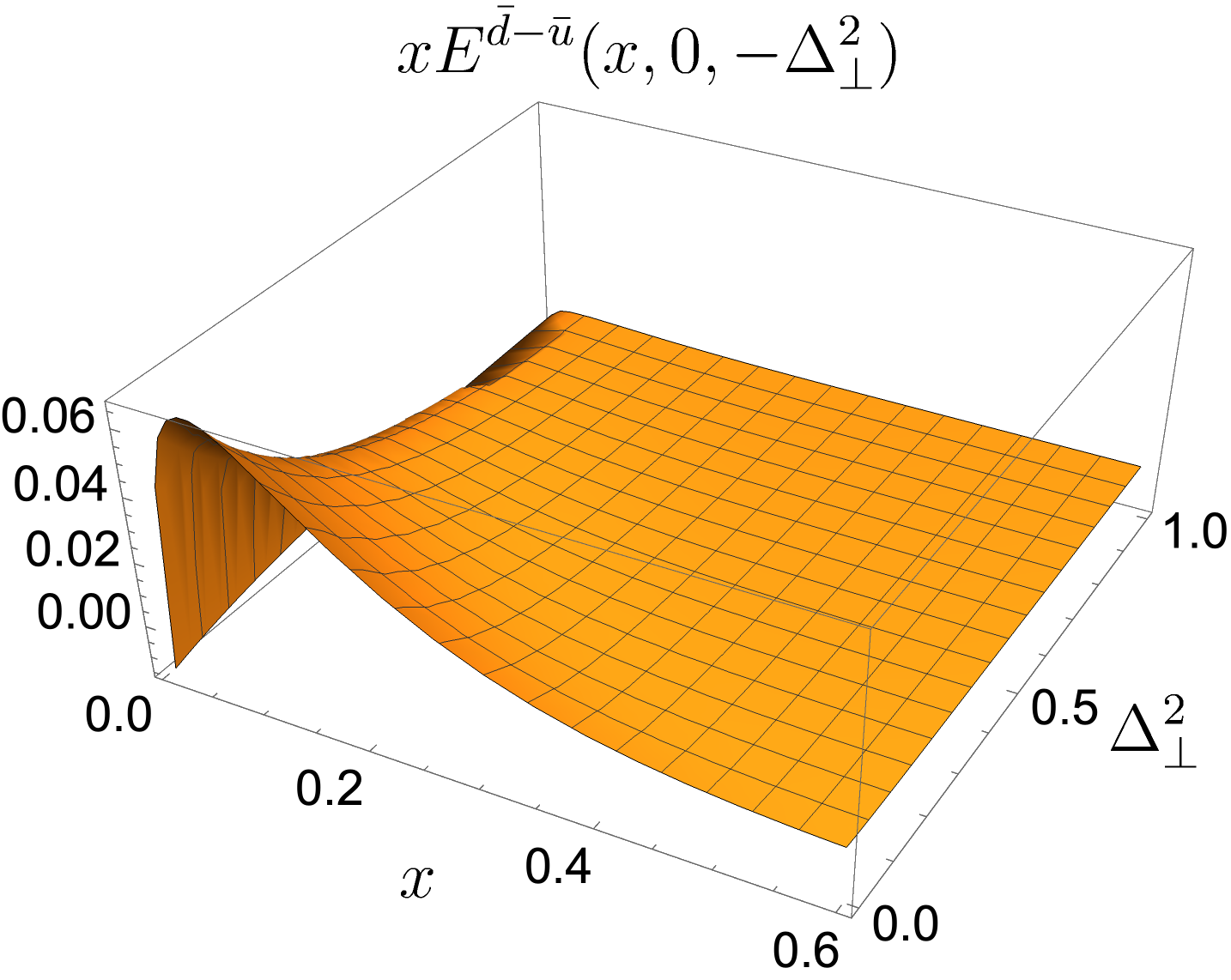}\hspace{0.4cm}
         \includegraphics[scale=0.22]{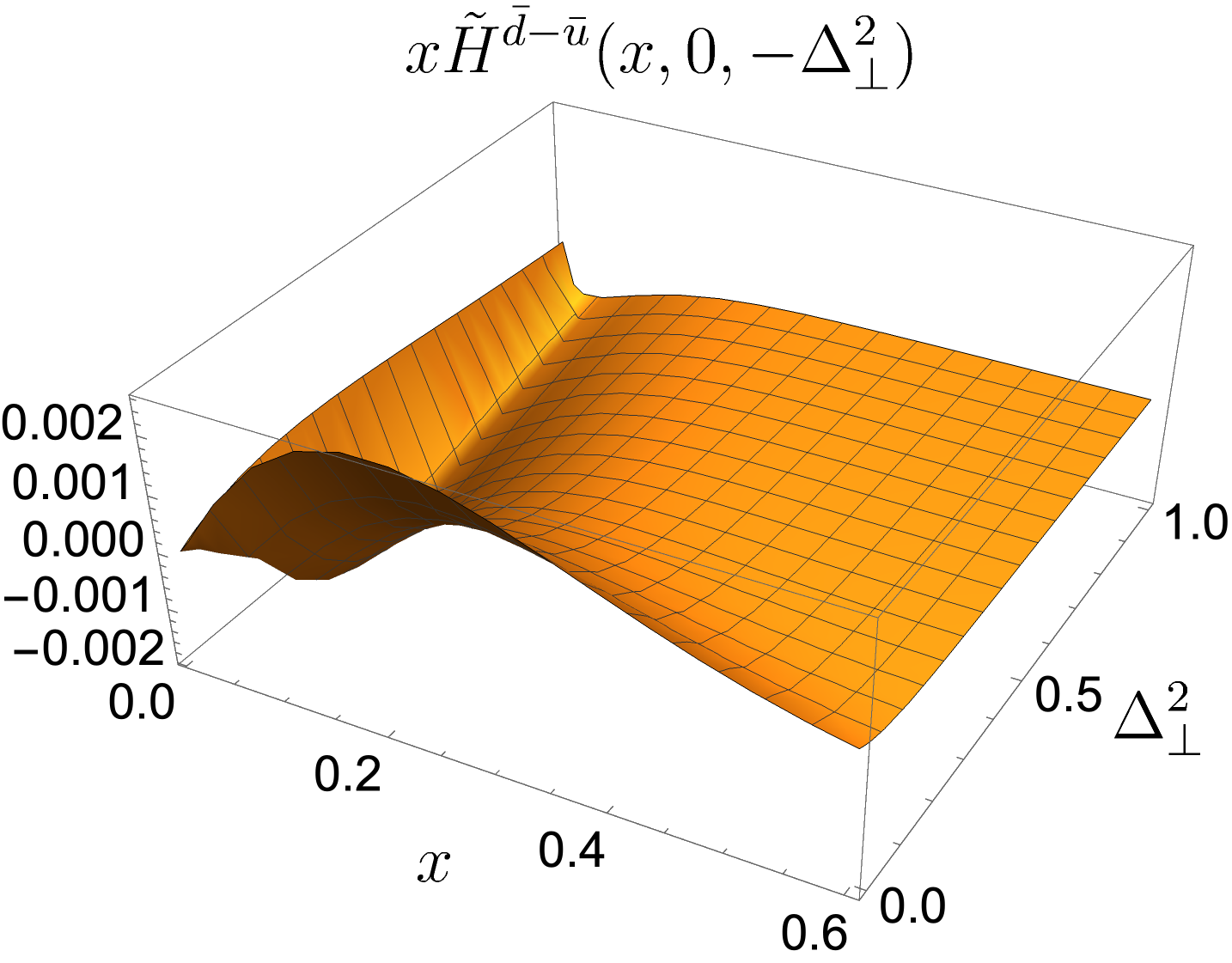}\vspace{0.5cm}
	\caption{Flavor asymmetries for light sea quarks at finite momentum transfer, the electric asymmetry  $ x H^{\bar{d}-\bar{u}} $ (left panel), the magnetic asymmetry $ x E^{\bar{d}-\bar{u}} $ (middle panel), and the helicity asymmetry $ x \widetilde{H}^{\bar{d}-\bar{u}} $ (right panel) in the range $0.005<x<0.6$ and momentum transfer $0< \Delta_{\perp}^2<1 $ GeV$^2$. }
\label{GPDasym}
\end{figure}

\begin{figure}
	\centering
	\includegraphics[scale=0.22]{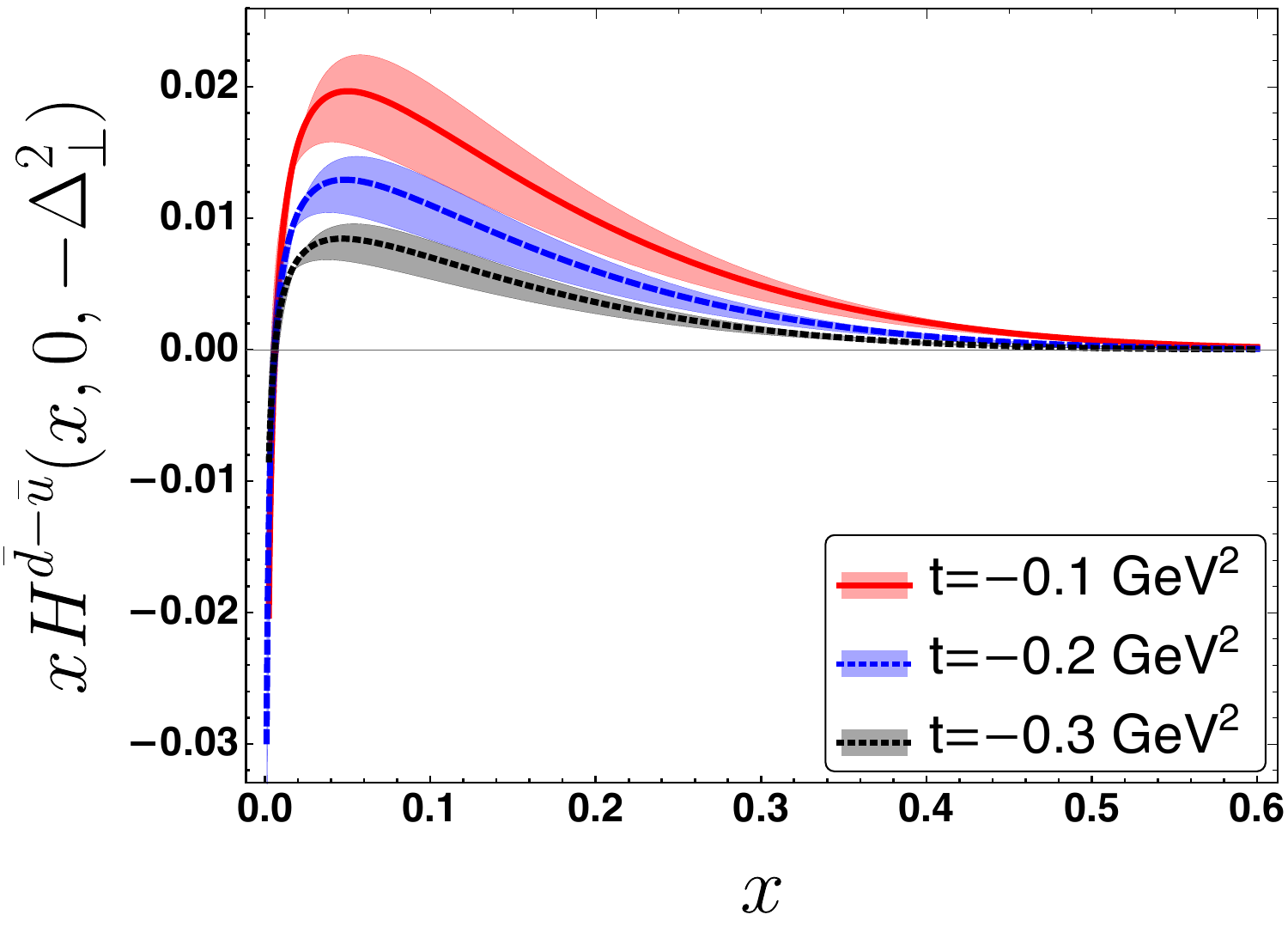}\hspace{0.4cm}
	\includegraphics[scale=0.22]{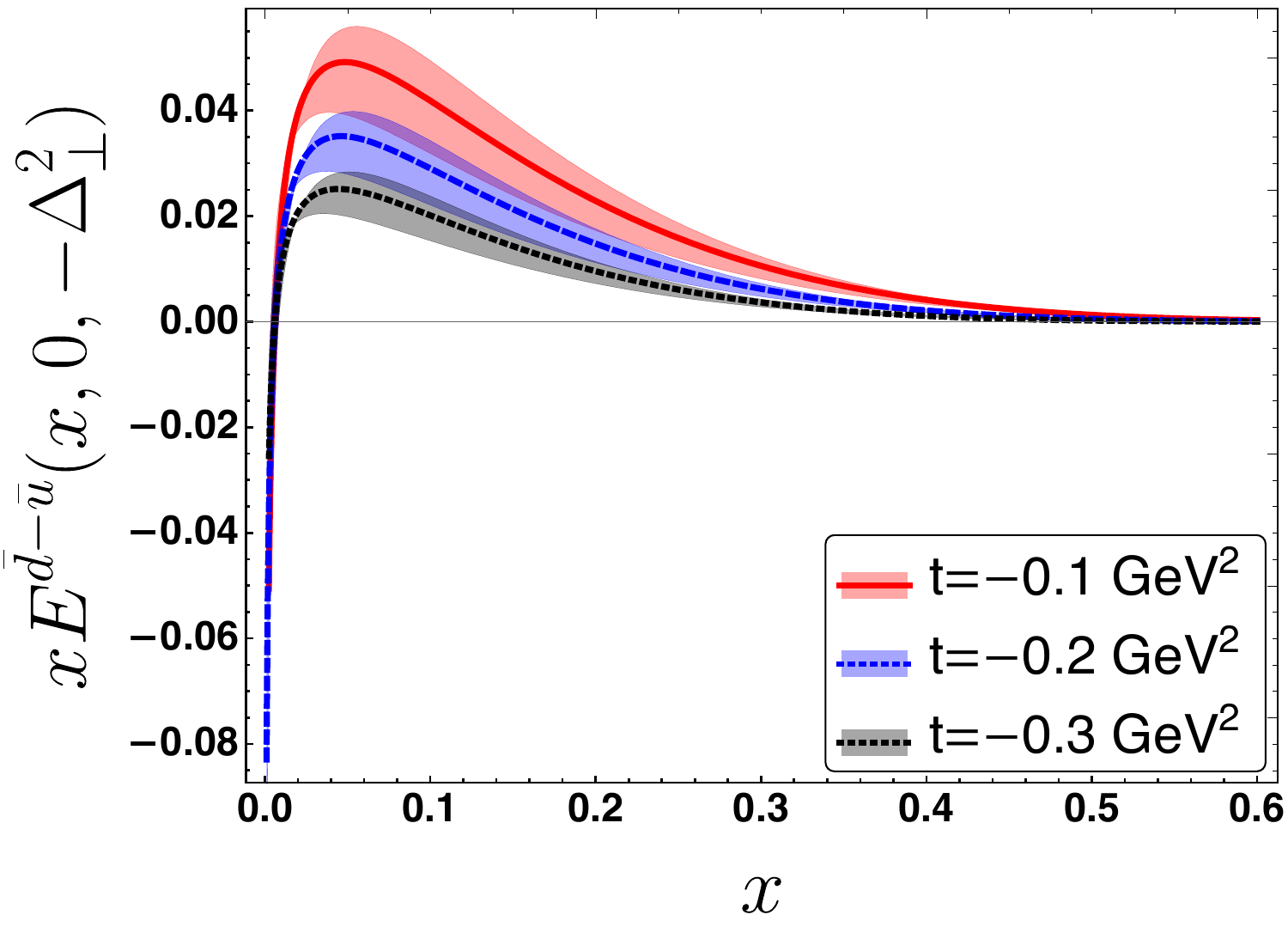}\hspace{0.4cm} 
        \includegraphics[scale=0.22]{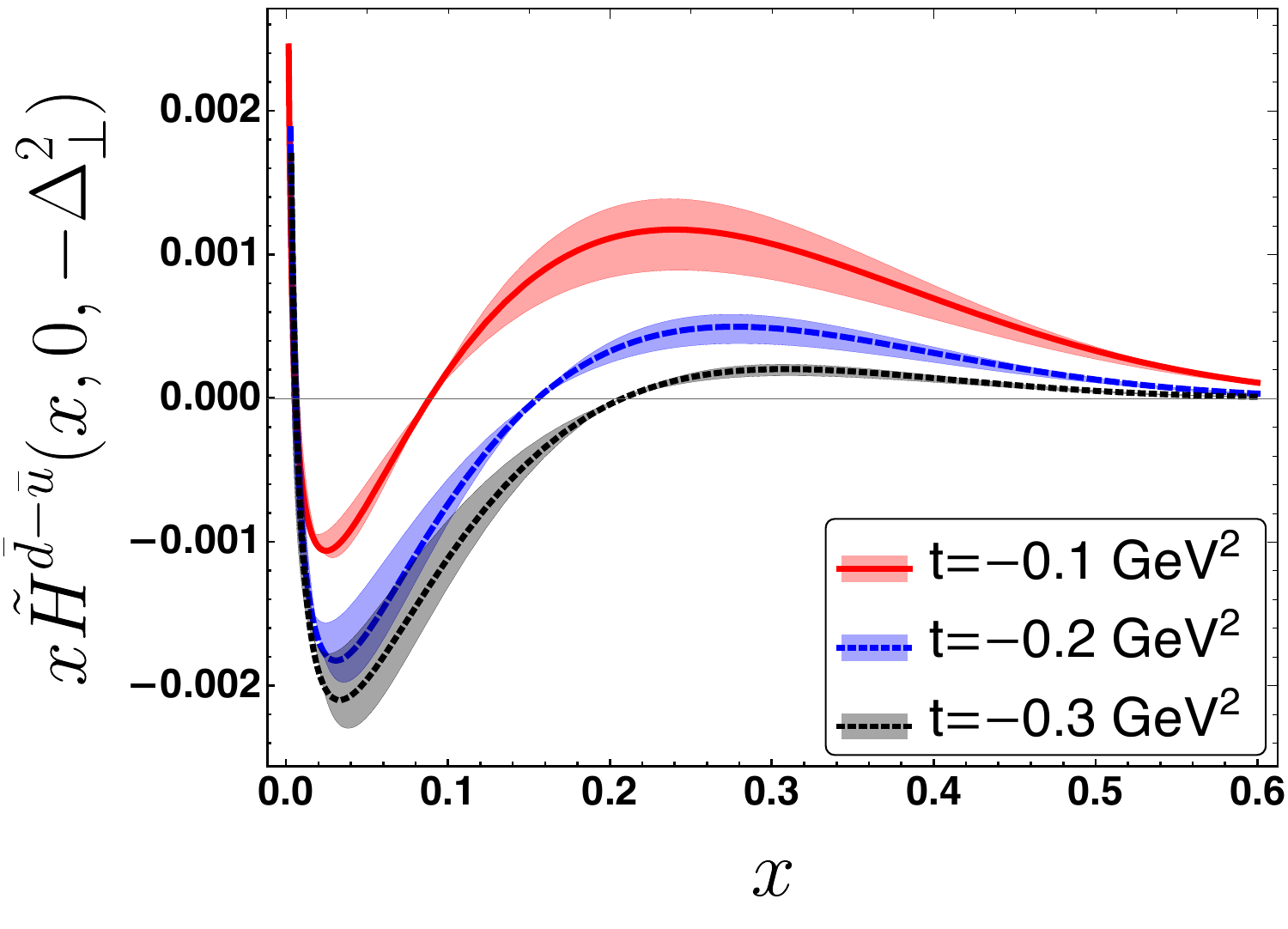}\vspace{0.5cm} 
\caption{Two-dimensional plots of the sea quarks' GPDs asymmetries $xH^{\bar{d}-\bar{u}}(x,0,-\Delta_\perp^2)$ (left panel), $xE^{\bar{d}-\bar{u}}(x,0,-\Delta_\perp^2)$ (middle panel), and $x\widetilde{H}^{\bar{d}-\bar{u}}(x,0,-\Delta_\perp^2)$ (right panel) in the range $0.001<x<0.6$ with different values of momentum transfer $t=-\Delta_\perp^2$.}
	\label{2DGPDs}
\end{figure}

%======================
Turning now to the light sea quark asymmetry of the GPDs, in Figs.~\ref{GPDasym} and \ref{2DGPDs}, we illustrate the distributions $ x H^{\bar{d}-\bar{u}}(x,0,-\Delta_{\perp}^2) $, $ x E^{\bar{d}-\bar{u}}(x,0,-\Delta_{\perp}^2) $, and $ x \widetilde{H}^{\bar{d}-\bar{u}}(x,0,-\Delta_{\perp}^2) $, which are known as the Electric, Magnetic, and Helicity flavor asymmetries, respectively. In Fig.~\ref{GPDasym}, we show the 3D distributions in the region $0.005\le x\le 0.6$ and $0\le \Delta_{\perp}^2\le 1$ GeV$^2$, while in Fig.~\ref{2DGPDs}, we present the 2D sections of the distributions  in the range $0.001\le x\le 0.6$ focusing on small-$x$ region for fixed momentum transfers.
%{\PC {In Fig.~\ref{2DGPDs}, we present 2D asymmetry distributions in the range $0.001\le x\le 0.6$ for the fixed momentum transfer. }}
We observe that both the Electric and Magnetic asymmetries are negative in the range $0.001<x<0.005 $ independent of choice of $\Delta_\perp^2$. 
This characteristic of the GPDs in our model is reassuring from Fig.~\ref{pdf asym}, since the  GPDs in forward limit reduce to PDFs. Both the asymmetries, on the other hand, are positive for $x>0.005$ with a peak at low-$x$  that decreases with increasing $\Delta_{\perp}^2$.  
At the peak, the magnetic GPD asymmetry $ x E^{\bar{d}-\bar{u}} $ has magnitude $0.06$, which is about three times larger than that of the electric asymmetry $ x H^{\bar{d}-\bar{u}} $. Note that these results are model  and scale  dependent. We predict them at the scale $Q_0=2 $ GeV. The asymmetry behaviour may vary as  it is evolved to different energy scales. Except in magnitude, both the asymmetries exhibit more or less similar behavior having their peaks around the same value of $x$
as can be seen in Fig.~\ref{2DGPDs}. These general properties should be nearly model-independent characteristics of the GPDs and, indeed, they are also observed in
other theoretical study of the GPDs~\cite{He:2022leb}. We also notice that in our model, both $ x H^{\bar{d}-\bar{u}} $ and $ x E^{\bar{d}-\bar{u}} $ fall off sharply with $\Delta_{\perp}^2$ and the distributions almost vanish for $x> 0.6 $, whereas the GPDs asymmetries reported in the nonlocal covariant chiral effective theory~\cite{He:2022leb} do not fall that fast. The third plots in Fig.~\ref{GPDasym} and \ref{2DGPDs} correspond to the asymmetry in the helicity distribution at different momentum transfer $\Delta_\perp^2$. We find that the distribution  $ x \widetilde{H}^{\bar{d}-\bar{u}} $ in our model changes its sign with $x$ and $\Delta_\perp^2$. This is similar to the finding reported by the HERMES~\cite{HERMES:2003gbu} when  momentum transfer is zero. 

%================================ 
%================================
%======================
\subsection{Sea quark orbital angular momentum}\label{Sec4}
%====================
By exploring the GPDs associated with sea quarks, it becomes feasible to investigate the OAM contributions of the $\bar{u}$ and $\bar{d}$ quarks within the proton, which hold considerable importance in understanding the underlying characteristics of nucleon spin structure. Using the spin decomposition according to Ji \cite{Ji:1996ek},  the quark contribution can be separated further into the usual quark helicity and the gauge-invariant OAM,
\begin{align}
L^{\bar{q}} =\int \text {d}x\,L^{\bar{q}}(x)\,,
\label{eq:Lq}
\end{align}
where $ L^{\bar{q}}(x)$ can be represented in terms of the chiral-even GPDs by the following relation 
\begin{align}
	L^{\bar{q}}(x) =\frac{1}{2}\left\{x\left[H^{\bar{q}/P}(x,0,0)
	+E^{\bar{q}/P}(x,0,0)\right]-\widetilde{H}^{\bar{q}/P}(x,0,0)\right\}\,.\label{eq:LqX}
\end{align}
%\cite{Xu:2023nqv}

\begin{figure}
	\centering
	\includegraphics[scale=0.32]{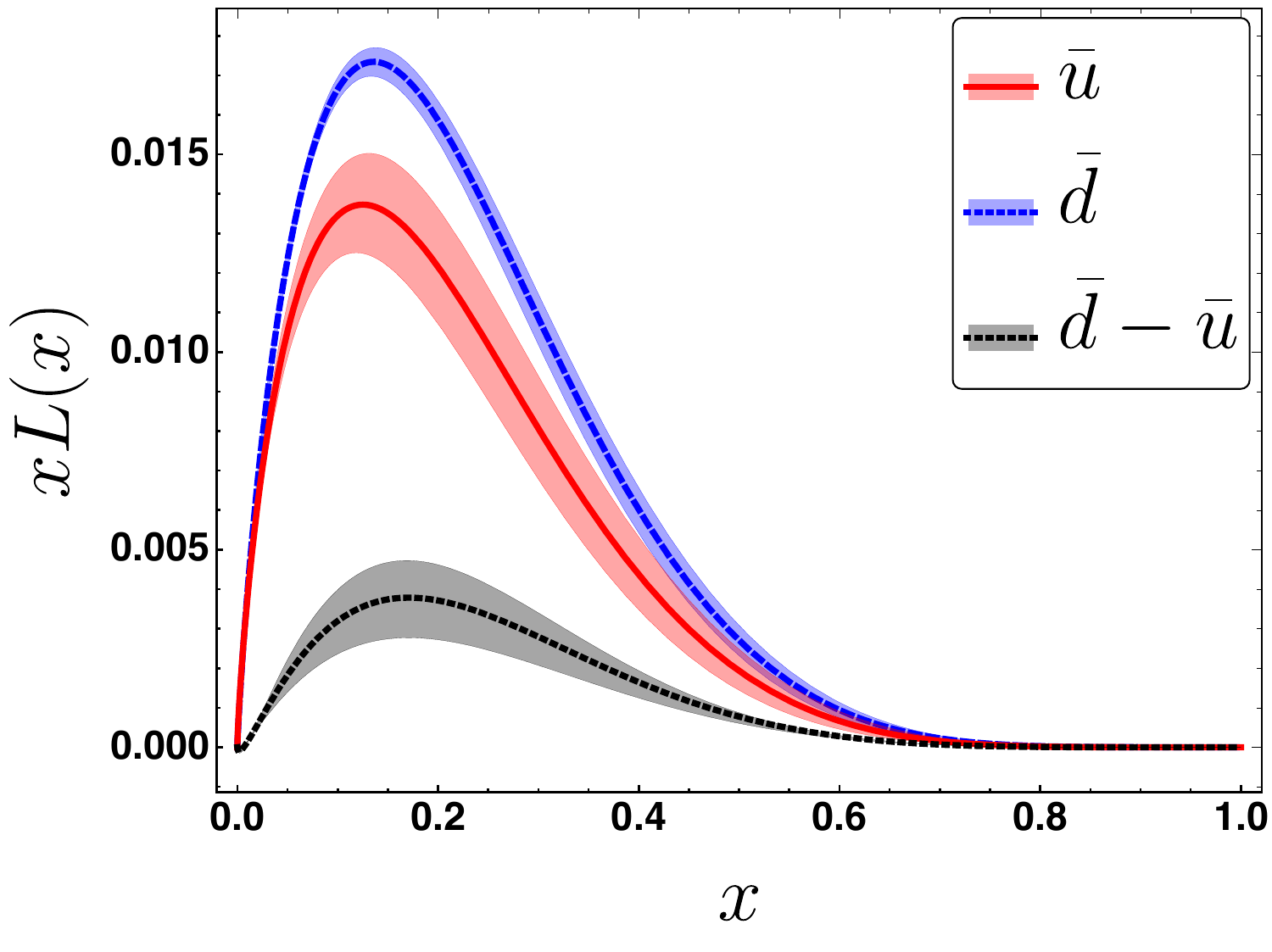}\hspace{0.5cm}
\caption{The sea quark OAM distribution, $xL(x)$, for $\bar{d}$ (blue line with blue band) and $\bar{u}$ (red line with red band) and their asymmetry $xL^{\bar{d}-\bar{u}}(x)$ (black line with black band) in the proton evaluated within our model.}
	\label{OAM asmmetry}
\end{figure}

Figure~\ref{OAM asmmetry} shows the  $xL^{\bar{q}}(x)$ for both the up and down sea quarks. We find that the distributions are positive for all values of $x$, having a peak around $x\sim 0.15$. The distribution for the $\bar{d}$ is slightly larger than that of the $\bar{u}$ quark  making the asymmetry $xL^{\bar{d}-\bar{u}}(x)$ small but positive. It vanishes after $x$ increases over $0.6$. 

It is expected that the sea quarks do not contribute much to the OAM of the proton but it also depends on the region of $x$. Since there are more sea quarks involved in the lower-$x$ region, therefore, the result of the OAM $L^{\bar{q}}$ depends on the limit of $x$ in the Eq.~\eqref{eq:Lq}.  
In the region, $0.001<x<1$, our model provides 
\begin{align}
L^{\bar{u}}=0.040 \pm 0.003 ,~~~~ L^{\bar{d}}= 0.048 \pm 0.002 , 
%~~~~ L^{\bar{d}-\bar{u}}= 0.008,
\end{align}
while in the region $0.01<x<1$, we obtain
\begin{align}
L^{\bar{u}}=0.035 \pm 0.003 ,~~~~ L^{\bar{d}}= 0.044  \pm 0.002 ,
%~~~~ L^{\bar{d}-\bar{u}}= 0.009. 
\end{align}

%=============================
%===============================

%==========================
\section{Sea quark TMDs} \label{tmdsasym}
%==========================
TMDs encode three-dimensional momentum information of a parton inside a hadron. TMDs are able to describe a wide range of phenomena following QCD factorization theorems~\cite{Barone:2010zz,Collins:1989gx}. They are essential in order to %adequately
characterise the SIDIS or the Drell-Yan processes~\cite{Collins:1984kg}, where the collinear picture of the parton model fails. The quark TMDs are parameterization factors of the quark correlation
function~\cite{Goeke:2005hb,Meissner:2007rx}
%We find the transverse-momentum-dependent correlator \cite{Diehl:2015uka, Lorce:2011zta} that is incorporated into the TMD definition:
\begin{equation} 
\Phi^{\Gamma}(x,\bfk ;P,S) = \frac{1}{xP^+}\int\frac{\text {d}\xi^-}{2\pi}\,\frac{\text {d}^2 \mathbf{\xi}{_\perp}}{(2\pi)^2}\,e^{i k\cdot \xi}\,\langle
  P,S\lvert \bar{\psi}(0) \mathcal{W}(0,\,\xi)\Gamma \psi(\xi)\rvert P,S\rangle \bigg \rvert_{\xi^{+}=0}\,,
\label{corr2}
\end{equation}  
where,
$\psi$ represents the quark field, and $\Gamma$ is the Dirac matrix, which in the leading twist, is taken as $\Gamma\equiv\gamma^+$, $\gamma^+\gamma^5$, $i\sigma^{j+}\gamma^5$.
\begin{figure}
	\centering
	\includegraphics[scale=0.22]{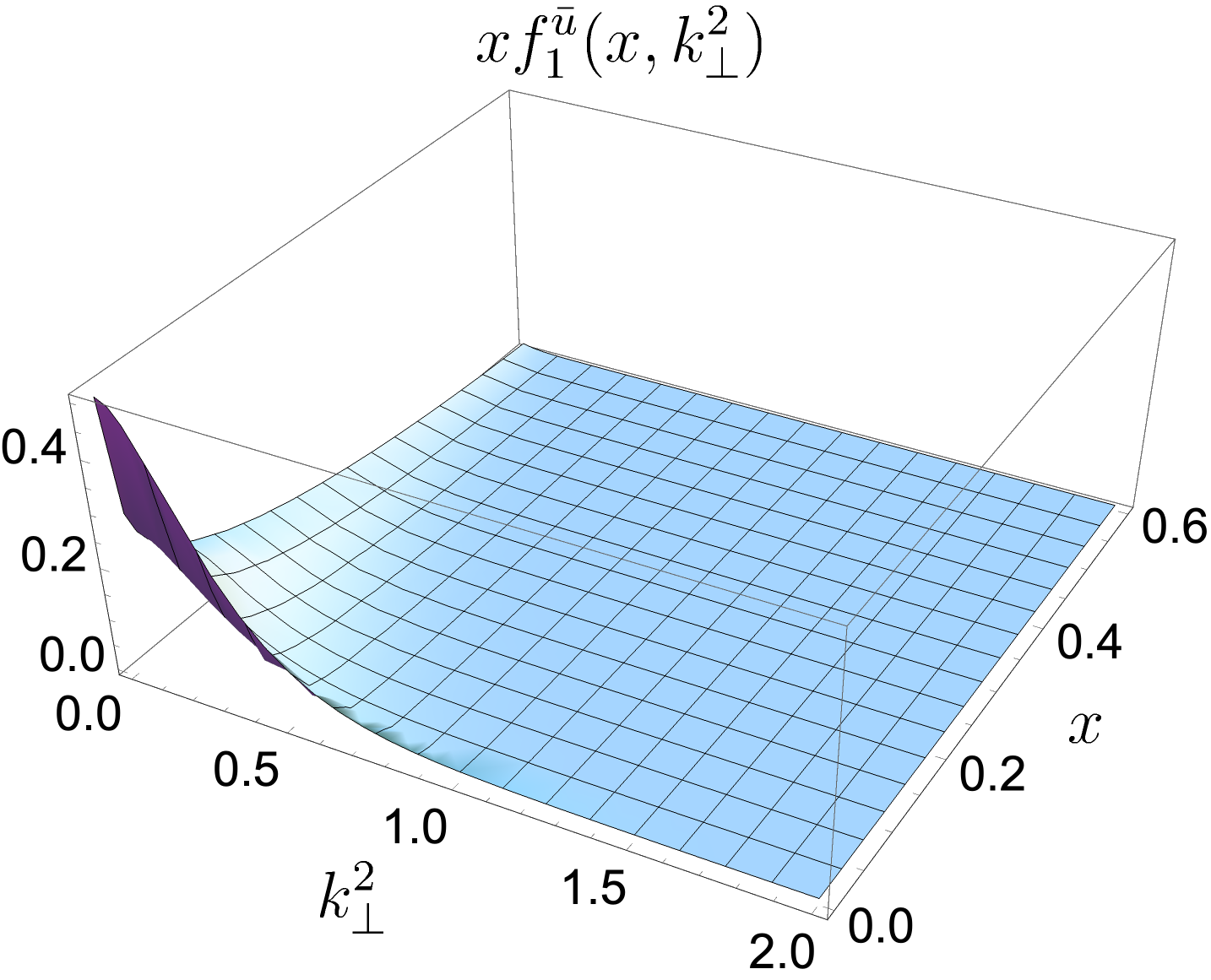}\hspace{0.4cm}
	\includegraphics[scale=0.22]{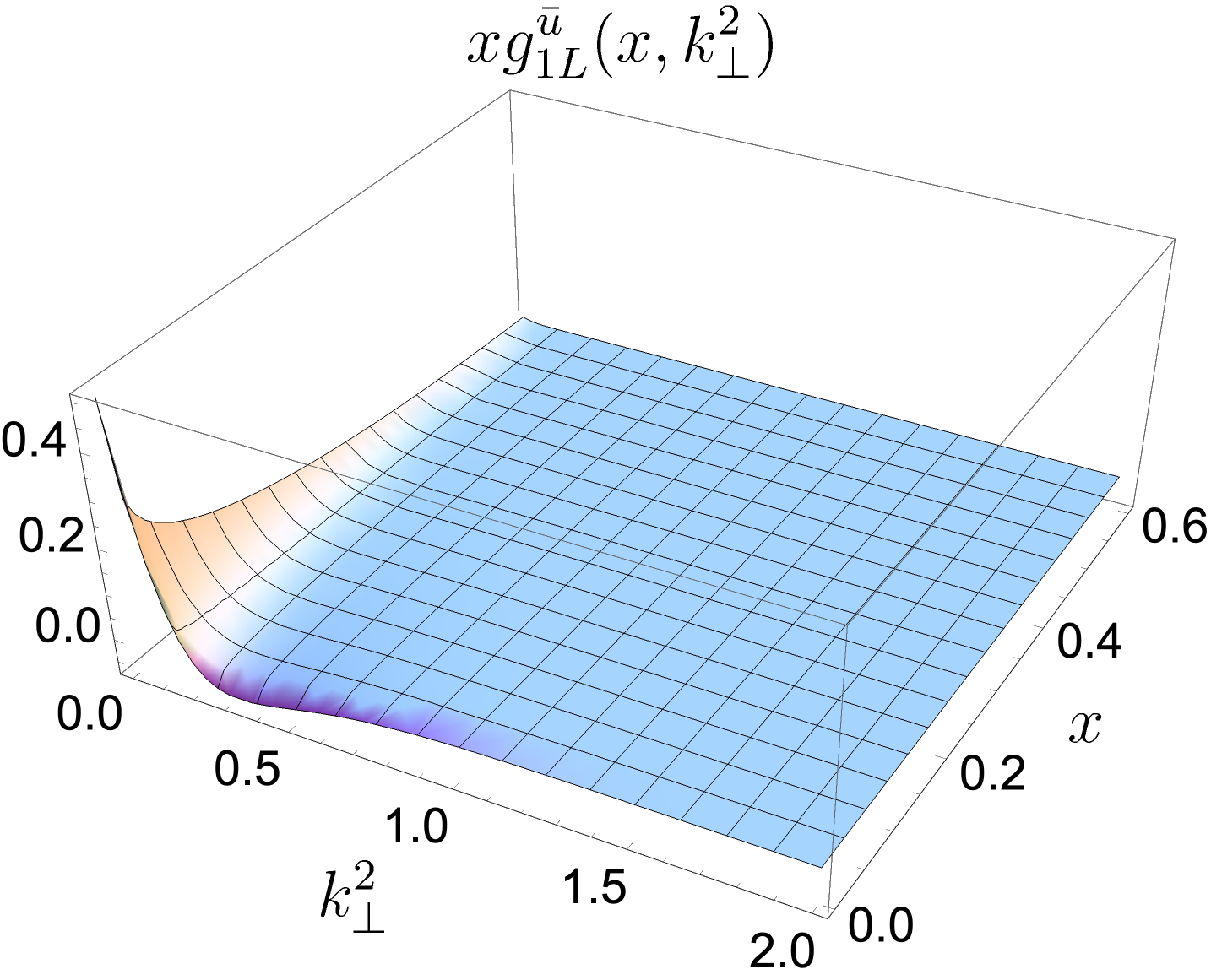}\hspace{0.4cm} 
	\includegraphics[scale=0.22]{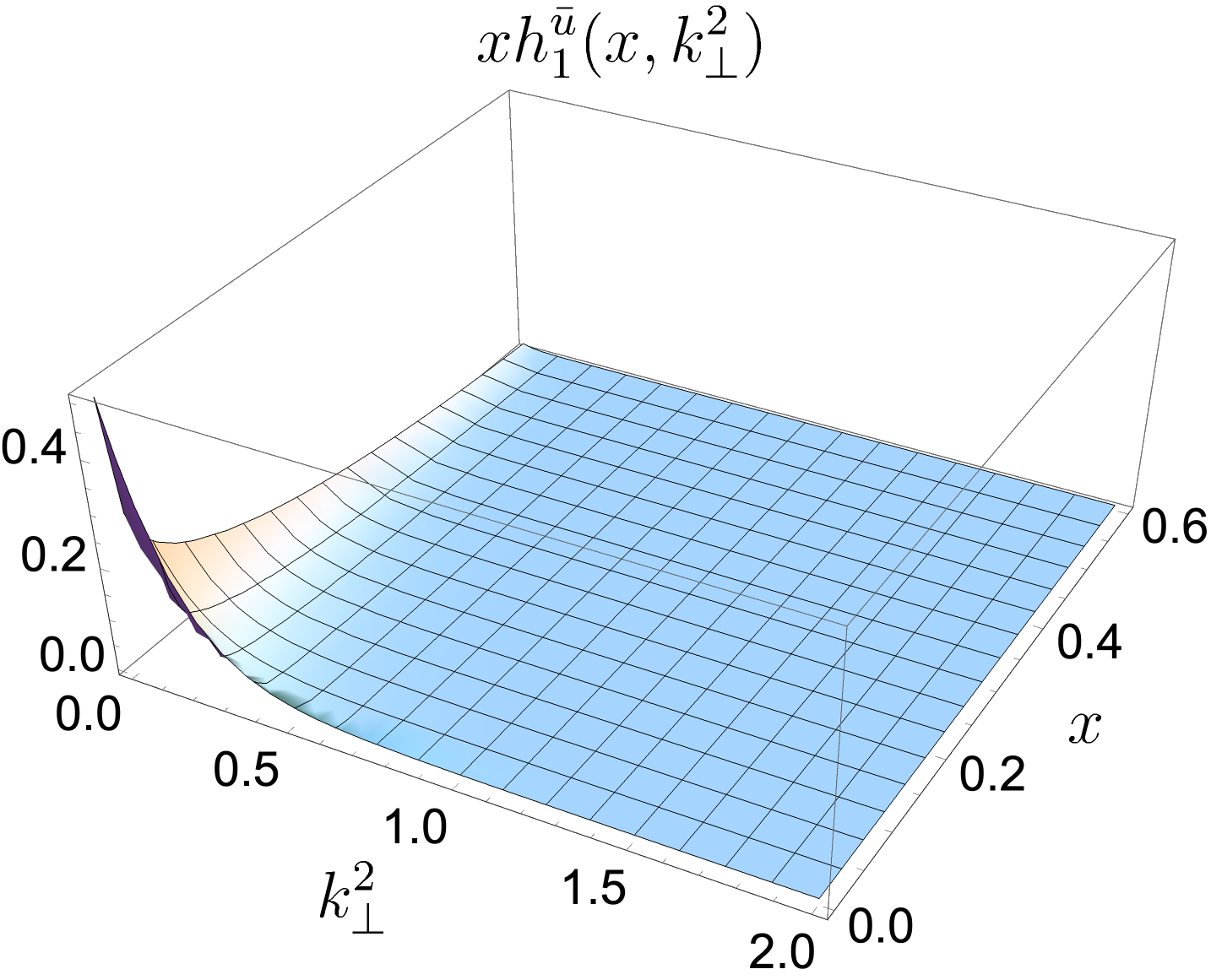}\vspace{0.4cm}\\
	\includegraphics[scale=0.22]{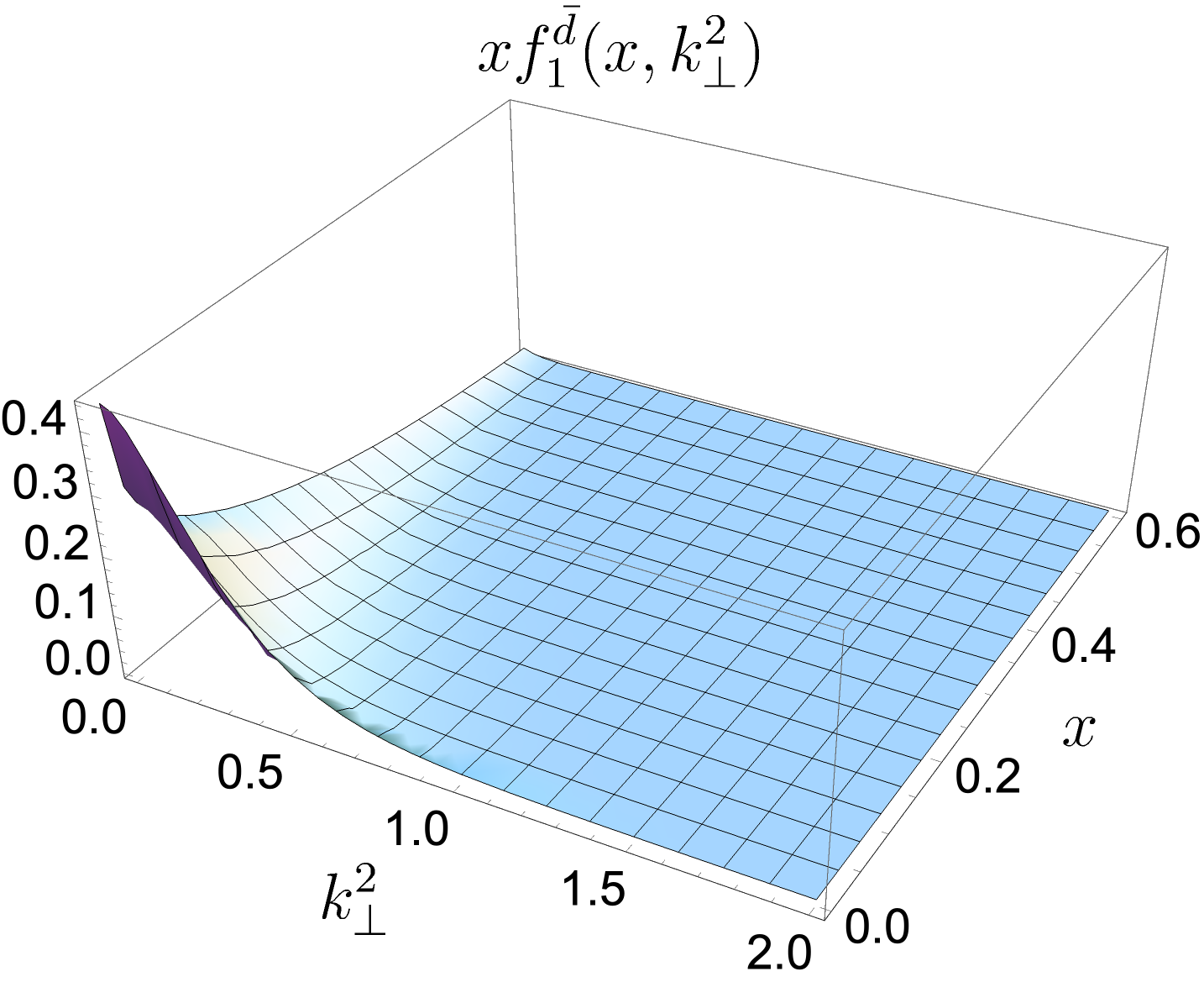}\hspace{0.4cm} 
	\includegraphics[scale=0.22]{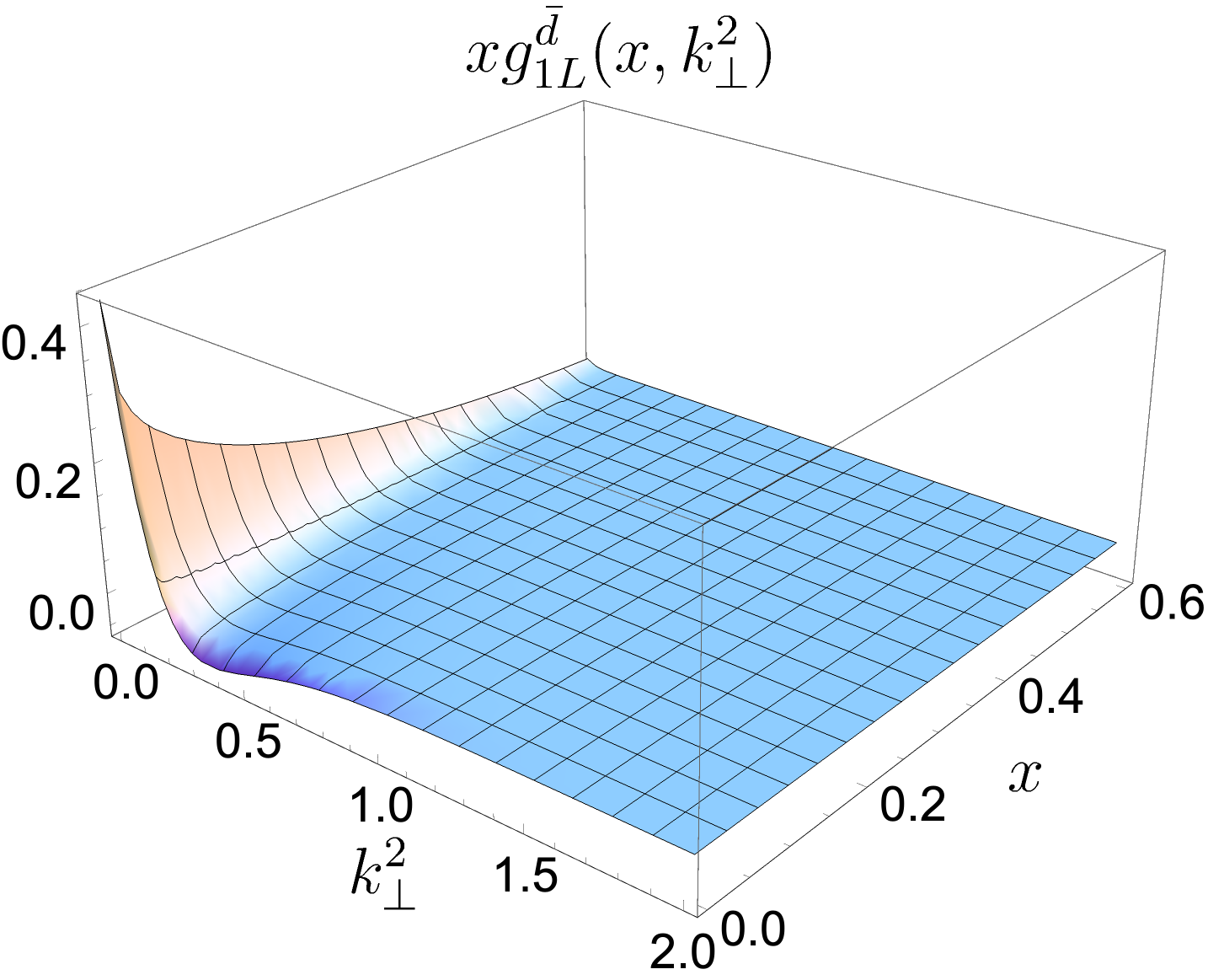}\hspace{0.4cm}
	\includegraphics[scale=0.22]{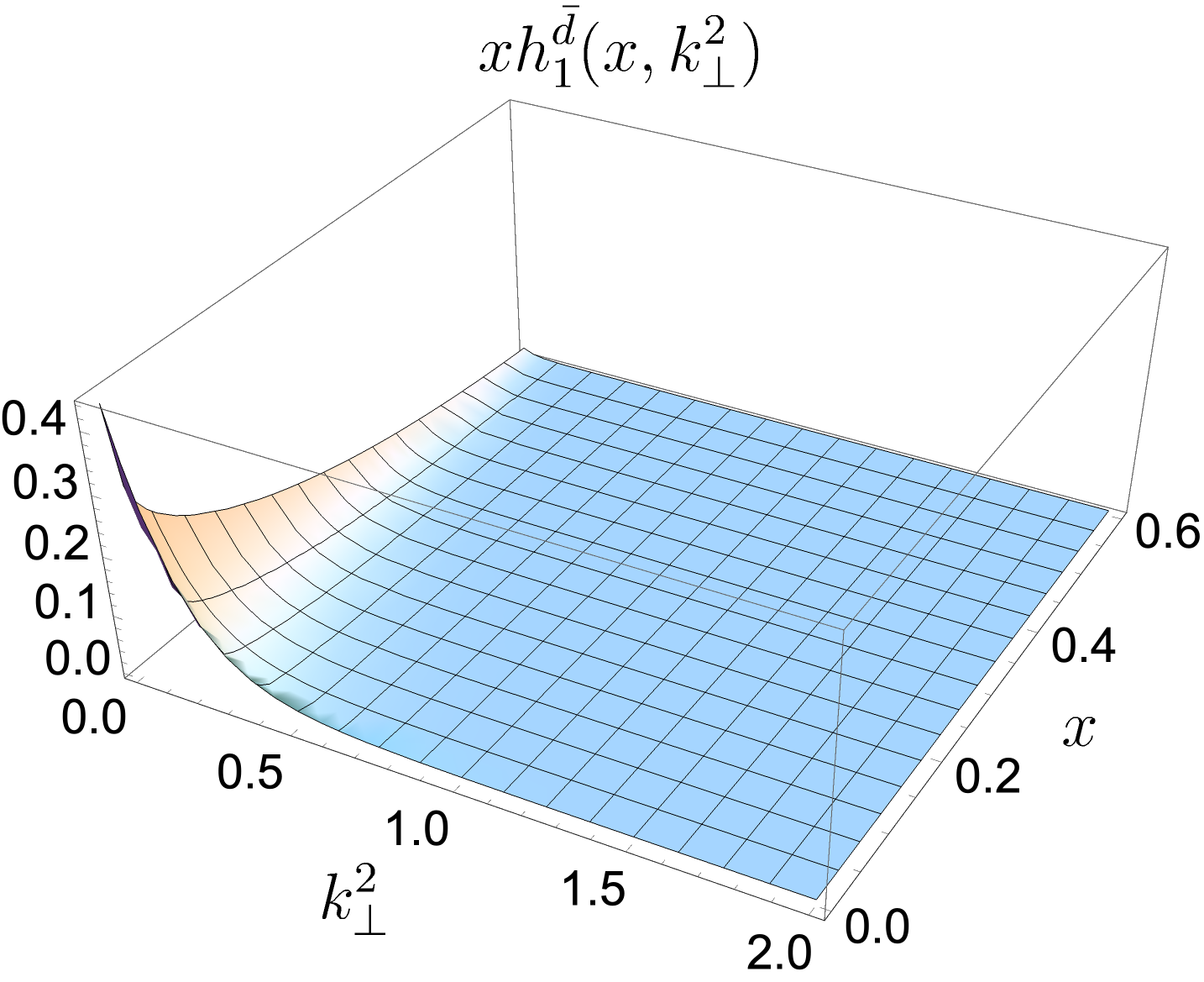}\vspace{0.4cm} \\
	\caption{The sea quark T-even TMDs, $ x{f}^{\bar{q} }_1(x,\bfk^2)$ (left panel), $x {g}^{\bar{q}  }_{1L}(x,\bfk^2)$ (middle panel), and $ xh_{1}^{\bar{q}}(x,\bfk^2)$ (right panel) calculated in the range $0.001<x<0.6$ and momentum transfer $0< \bfk^2<2 $ GeV$^2$ in the proton within our model. The upper panel is for the $\bar{u}$ quark and the lower panel represents the distributions for the $\bar{d}$ quark. $\bfk^2$ is in units of GeV$^2$.}
	\label{3DGTMDs1}
\end{figure}
It turns out that, at the leading twist there are eight quark TMDs for the nucleon. They are parameterized as follows~~\cite{Goeke:2005hb,Meissner:2007rx}:
%They can be projected as the following matrices: 
\begin{align} &\Phi^{[\gdir{+}]}(x,\pep{k};\mb{S})=f_{1}^{q}-\frac{\epsilon_{\perp}^{ij}k_{\perp}^{i}S_{\perp}^{j}}{m}f_{1T}^{\perp q} \; , \label{tmd1}\\
&\Phi^{[\gdir{+}\gdir{5}]}(x,\pep{k};\mb{S})=S_{z}g_{1L}^{q}+\frac{\pep{k}\cdot\pep{S}}{m}g_{1T}^{q} \; , \label{tmd2} \\
&\Phi^{[i\sigma^{j+}\gdir{5}]}(x,\pep{k};\mb{S})=S_{\perp}^{j}h_{1}^{q}+S_{z}\frac{k^{j}_{\perp}}{m}h_{1L}^{\perp q}+S^{i}_{\perp}\frac{2k_{\perp}^{i}k_{\perp}^{j}-\pep{k}^{2}\delta^{ij}}{2m^{2}}h_{1T}^{\perp q} +\frac{\epsilon_{\perp}^{ji}k_{\perp}^{i}}{m}h_{1}^{\perp q}\; , \label{tmd3}\end{align}
where 
$h_{1}^{q}:\,=h_{1T}^{q}+\frac{\pep{k}^{2}}{2m^{2}}h_{1T}^{\perp q}
$
and we use the convention $\epsilon^{12}=1$. 
We choose a frame in which the incoming proton has zero transverse momentum i.e. $P=(P^{+},\pep{0},P^{-})$ and the spin four vector such that $P\cdot S=0$ and $S^{2}=-1$. 
The covariant four-vector $S$  can be represented by the three-vector $\mb{S}:\,=(\perp{S},S_z)$, with $\mb{S}^{2}=1$.

In the current study, we only retain the zeroth-order expansion of the gauge link. Note that $f_{1T}^\perp$ and $h_{1}^\perp$, are T-odd distributions
and vanish under the zeroth-order expansion
of the gauge link, i.e., $\mathcal{W}(0,\,\xi)=1$ in Eq.~\eqref{corr2}.
All six T-even leading-twist TMDs can be expressed in terms of the LFWFs in our model as \cite{Bacchetta:2015qka}, 
\eq
{f}^{\bar{q} }_1(x,\bfk)&=&\frac{1}{16\pi^3} \biggl[ |\psi^{\uparrow}_{\bar{q}; + \frac{1}{2}}(x,\bfk)|^2
+ |\psi^{\uparrow}_{\bar{q};- \frac{1}{2}}(x,\bfk)|^2
\biggr]\,, \\
{g}^{\bar{q}  }_{1L}(x,\bfk)&=&\frac{1}{16\pi^3} \biggl[ |\psi^{\uparrow}_{\bar{q}; + \frac{1}{2}}(x,\bfk)|^2
- |\psi^{\uparrow}_{\bar{q};- \frac{1}{2}}(x,\bfk)|^2
\biggr]\,, \\
 g_{1T}^{\bar{q}}(x,\textbf{k}_{\perp}^2)&=& \frac{M}{2(2\pi)^{3} \bfk^{2}} \bigg[k_{R} \psi ^{\uparrow \dagger}_{\bar{q},+\frac{1}{2}}(x,\textbf{k}_{\perp})\psi^{\downarrow }_{\bar{q},+\frac{1}{2}}(x,\textbf{k}_{\perp})+ k_{L}\psi ^{\downarrow \dagger}_{\bar{q},+\frac{1}{2}}(x,\textbf{k}_{\perp})\psi^{\uparrow }_{\bar{q},+\frac{1}{2}}(x,\textbf{k}_{\perp})
\bigg]\,, \\
 h_{1L}^{\perp \bar{q}}(x,\textbf{k}_{\perp}^2)&=& \frac{M}{2(2\pi)^{3} \bfk^{2}} \bigg[k_{R} \psi ^{\uparrow \dagger}_{\bar{q},-\frac{1}{2}}(x,\textbf{k}_{\perp})\psi^{\uparrow }_{\bar{q},+\frac{1}{2}}(x,\textbf{k}_{\perp})+ k_{L}\psi ^{\uparrow \dagger}_{\bar{q},+\frac{1}{2}}(x,\textbf{k}_{\perp})\psi^{\uparrow }_{\bar{q},-\frac{1}{2}}(x,\textbf{k}_{\perp})
\bigg]\,, \\
 h_{1T}^{\perp \bar{q}}(x,\textbf{k}_{\perp}^2)&=& \frac{M}{2(2\pi)^{3} \bfk^{4}} \bigg[k_{R}^2 \psi ^{\uparrow \dagger}_{\bar{q},-\frac{1}{2}}(x,\textbf{k}_{\perp})\psi^{\downarrow }_{\bar{q},+\frac{1}{2}}(x,\textbf{k}_{\perp})+ k_{L}^2\psi ^{\downarrow \dagger}_{\bar{q},+\frac{1}{2}}(x,\textbf{k}_{\perp})\psi^{\uparrow }_{\bar{q},-\frac{1}{2}}(x,\textbf{k}_{\perp})
\bigg]\,, \\
h_{1}^{\bar{q}}(x,\textbf{k}_{\perp}^2) &=& \frac{1}{2(2\pi)^{3} } \psi ^{\uparrow \dagger}_{\bar{q},+\frac{1}{2}}(x,\textbf{k}_{\perp})\psi^{\downarrow }_{\bar{q},-\frac{1}{2}}(x,\textbf{k}_{\perp})\,,
\en
where $ k_{R/L} = k_1 \pm i k_2 $ are defined in terms of the components of the transverse momentum.
Utilizing the explicit form of the LFWFs, Eq.~\eqref{WaveFun}, we derive the following expressions of the sea quark T-even TMDs within our model,
\eq
{f}^{q  }_1(x,\bfk^2)&=& \frac{1}{\pi \kappa^2}\biggl[1+\frac{\bfk^2}{\kappa^2} 
\biggr]   A x^{2 \alpha-1}(1-x)^{6+2 \beta} \log(1/1-x)  \exp{\bigg[-\frac{\log[1/(1-x)]}{\kappa^{2}x (1-x)}\bfk^{2}\bigg]}\,, \\
{g}^{q  }_{1L}(x,\bfk^2)&=&  \frac{1}{\pi \kappa^2} \biggl[1-\frac{\bfk^2}{\kappa^2} 
\biggr]   A  x^{2 \alpha-1}(1-x)^{6+2 \beta} \log(1/1-x) 
\exp{\bigg[-\frac{\log[1/(1-x)]}{\kappa^{2}x (1-x)}\bfk^{2}\bigg]}\,, \\
{g}^{q  }_{1T}(x,\bfk^2)&=&  \frac{1}{\pi \kappa^2}  \frac{ 2 M}{\kappa} A  x^{2 \alpha-1}(1-x)^{6+2 \beta} \log(1/1-x) 
\exp{\bigg[-\frac{\log[1/(1-x)]}{\kappa^{2}x (1-x)}\bfk^{2}\bigg]} \\
{h}^{q \perp }_{1L}(x,\bfk^2)&=& - \frac{1}{\pi \kappa^2}  \frac{2 M}{ \kappa} A  x^{2 \alpha-1}(1-x)^{6+2 \beta} \log(1/1-x) 
\exp{\bigg[-\frac{\log[1/(1-x)]}{\kappa^{2}x (1-x)}\bfk^{2}\bigg]}\,, \\
{h}^{q \perp }_{1T}(x,\bfk^2)&=& - \frac{1}{\pi \kappa^2}  \frac{2 M^2}{ \kappa^2} A  x^{2 \alpha-1}(1-x)^{6+2 \beta} \log(1/1-x) 
\exp{\bigg[-\frac{\log[1/(1-x)]}{\kappa^{2}x (1-x)}\bfk^{2}\bigg]}\,, \\
{h}^{q  }_1(x,\bfk^2)&=& \frac{1}{\pi \kappa^2}  A x^{2 \alpha-1}(1-x)^{6+2 \beta} \log(1/1-x)  \exp{\bigg[-\frac{\log[1/(1-x)]}{\kappa^{2}x (1-x)}\bfk^{2}\bigg]}\,. 
\en

\begin{figure}
	\centering
	\includegraphics[scale=0.22]{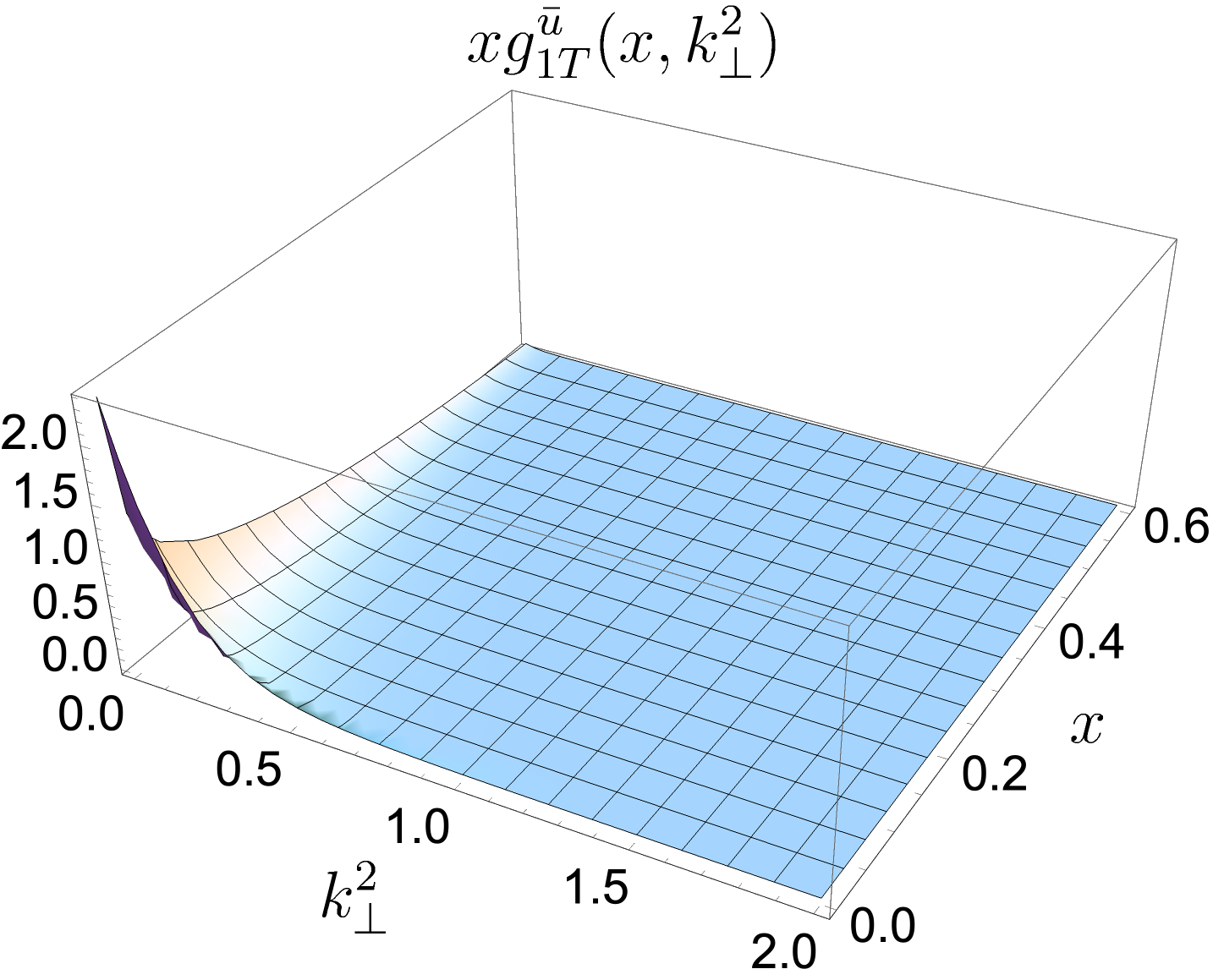}\hspace{0.4cm}
	\includegraphics[scale=0.22]{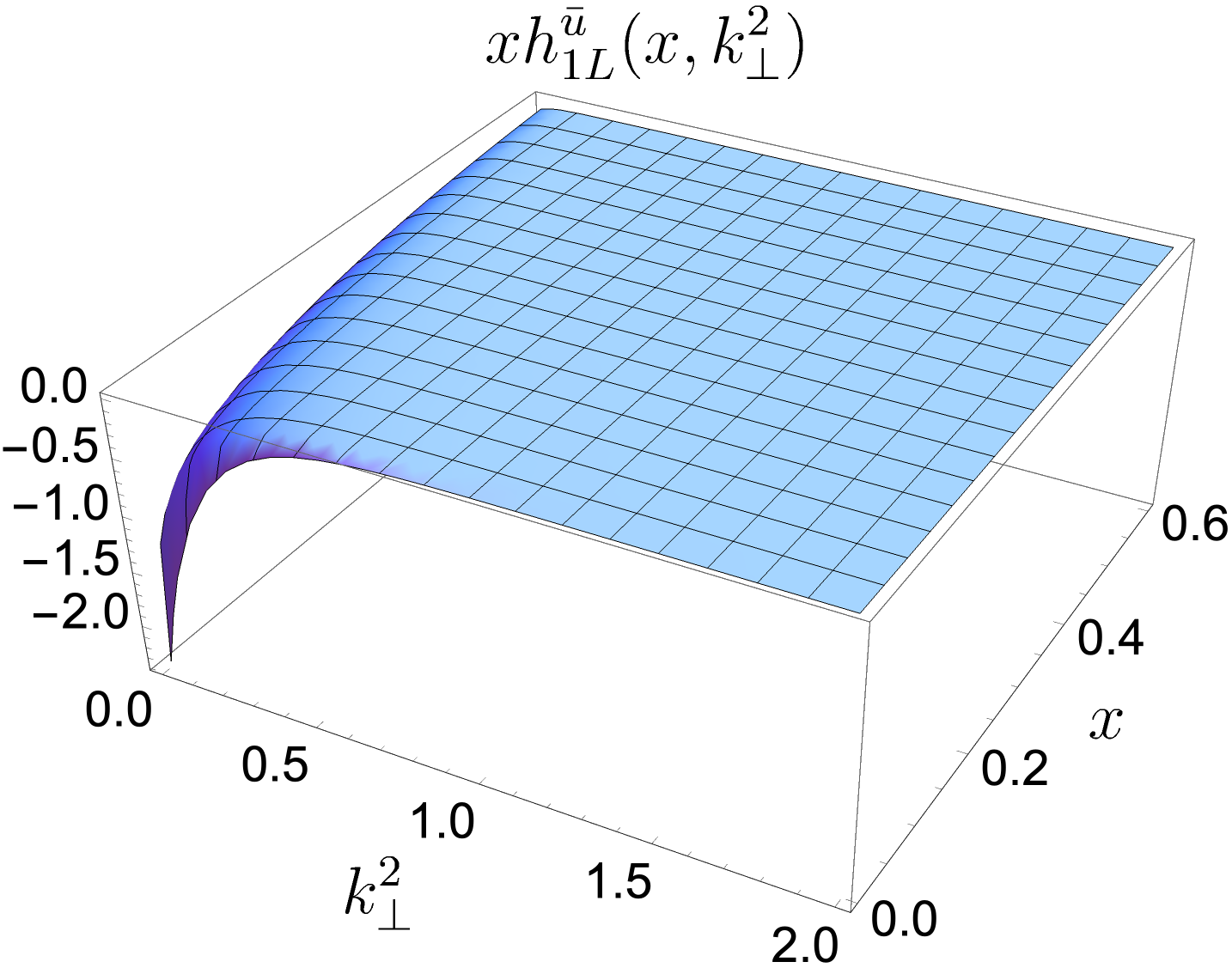}\hspace{0.4cm} 
	\includegraphics[scale=0.22]{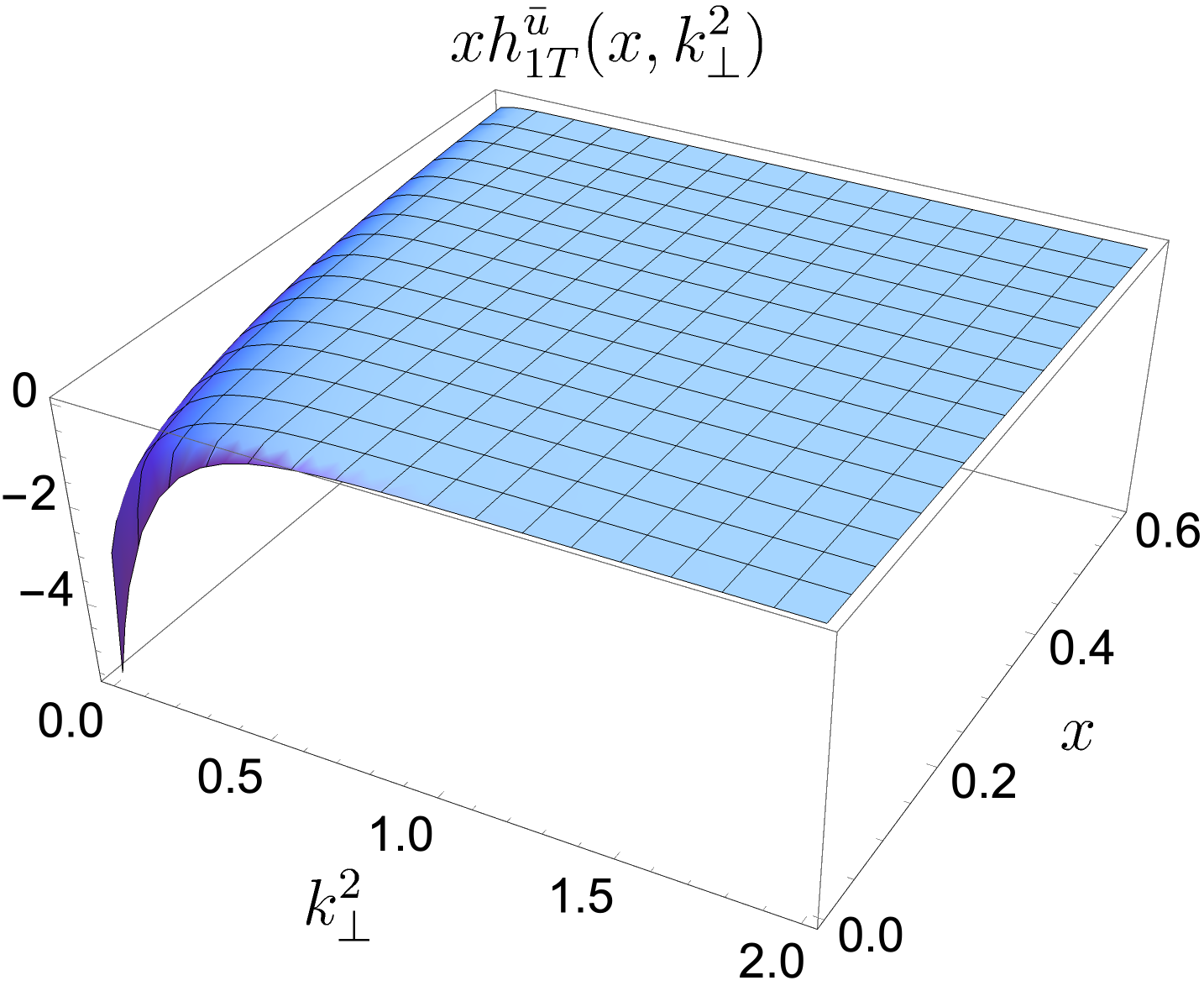}\vspace{0.4cm}\\
	\includegraphics[scale=0.22]{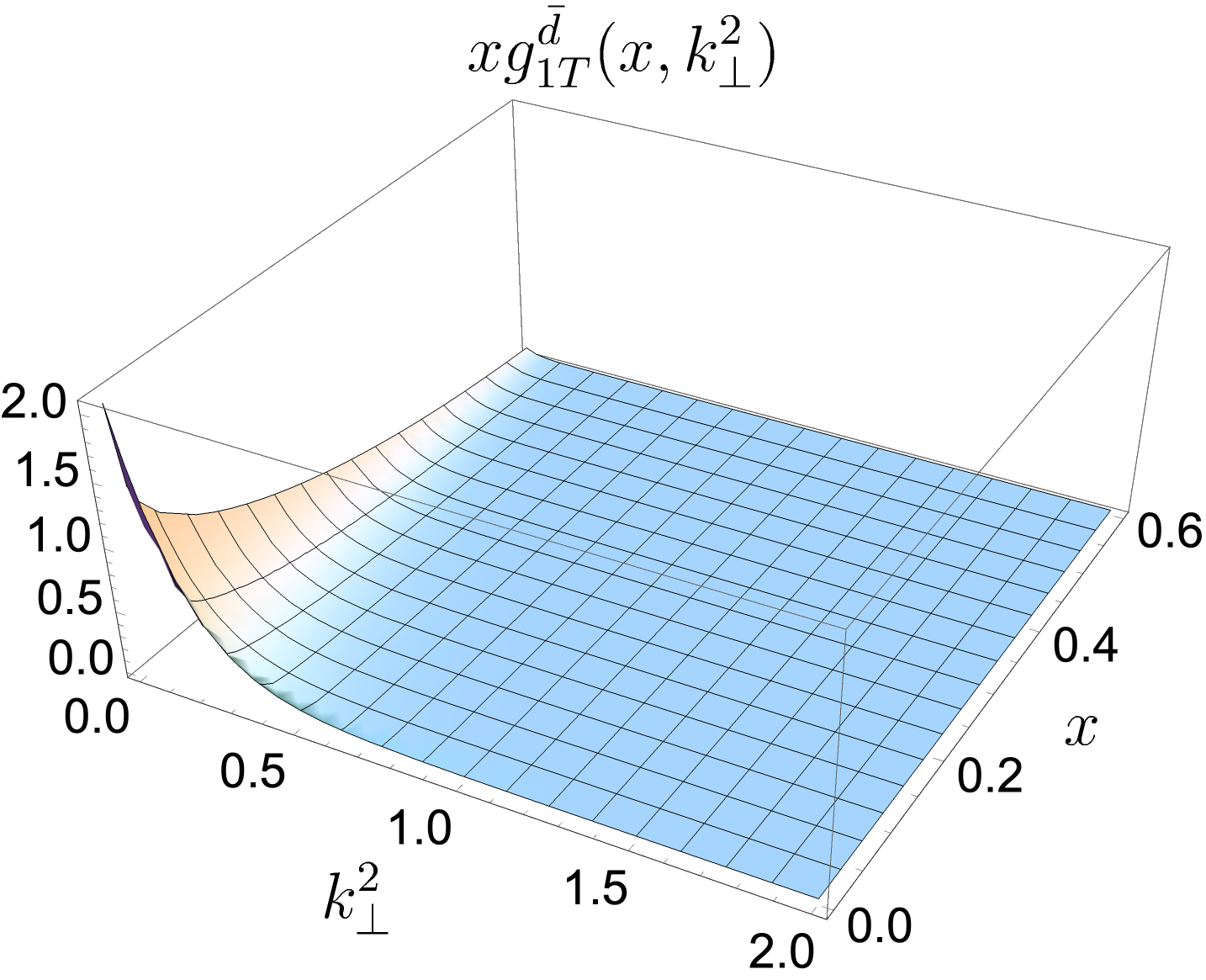}\hspace{0.4cm} 
	\includegraphics[scale=0.22]{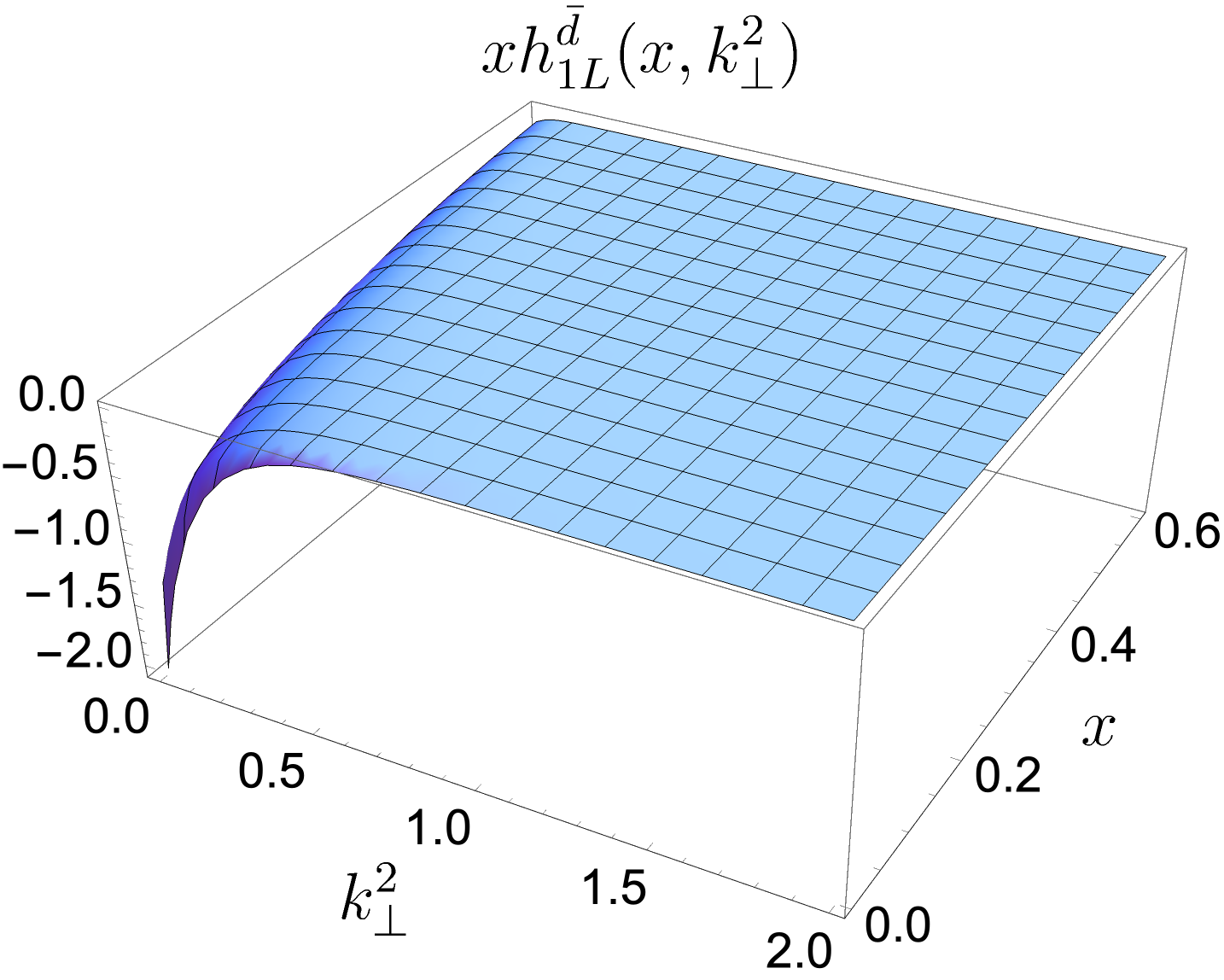}\hspace{0.4cm}
	\includegraphics[scale=0.22]{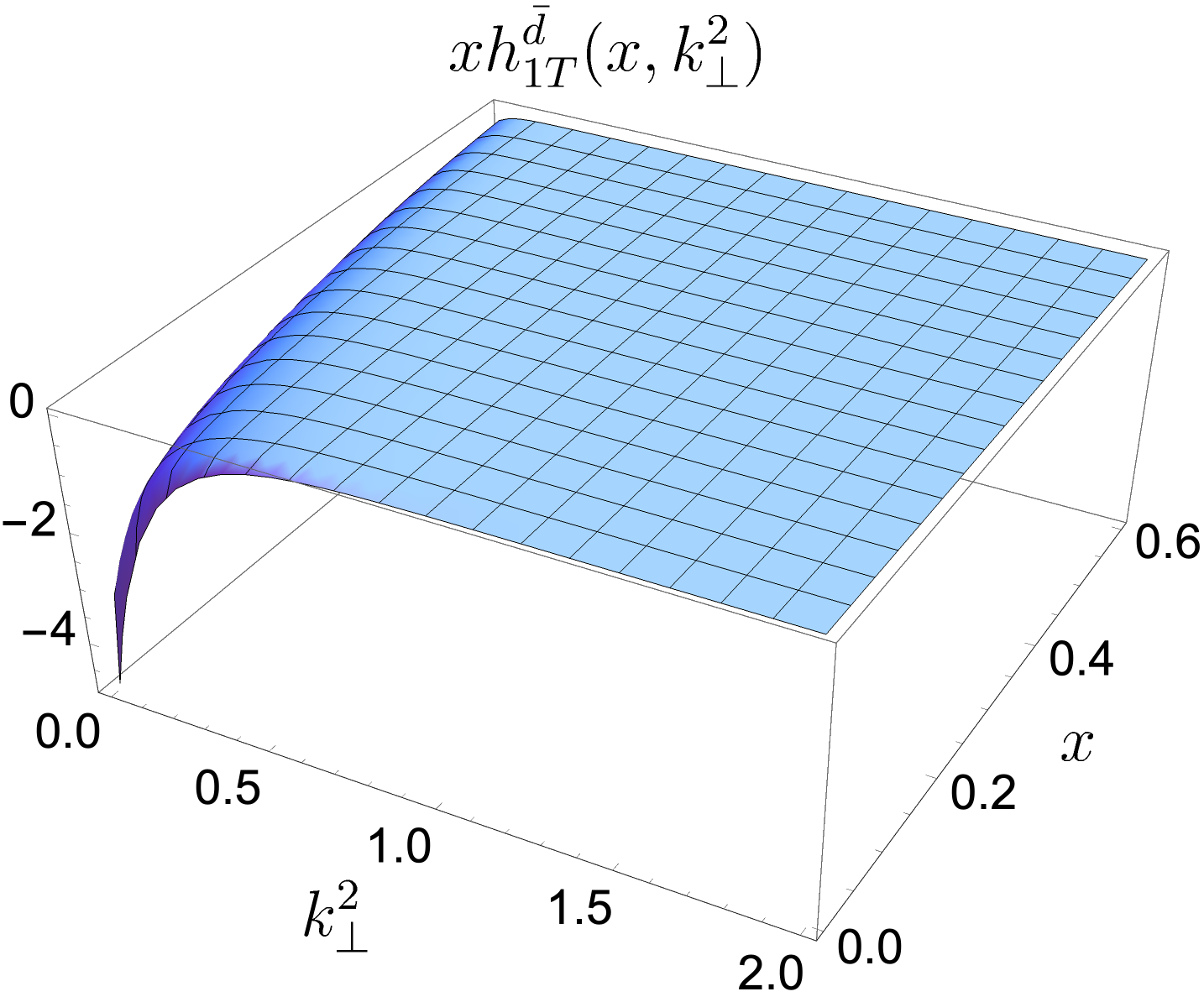}\vspace{0.4cm} \\
	\caption{The sea quark T-even TMDs, $ x{g}^{\bar{q} }_{1T}(x,\bfk^2)$, $x {h}^{\bar{q}  }_{1L}(x,\bfk^2)$, and $ xh_{1T}^{\bar{q}}(x,\bfk^2)$ calculated in the range $0.001<x<0.6$ and momentum transfer $0< \bfk^2<2 $ GeV$^2$ in the proton within our model. The upper panel is for the $\bar{u}$ quark and the lower panel represents the distributions for the $\bar{d}$ quark. $\bfk^2$ is in units of GeV$^2$.}
	\label{3DGTMDs2}
\end{figure}

The 3D structures of the six leading-twist sea quark TMDs in our model are shown in Figs.~\ref{3DGTMDs1} and \ref{3DGTMDs2}. We present the three T-even TMDs: $f_1^{\bar{q}}(x,\bfk^{2})$,  $g_{1L}^{\bar{q}}(x,\bfk^{2})$, and $h_1^{\bar{q}}(x,\bfk^{2})$, which have their colinear limits, shown in Fig.~\ref{model fit}, \ref{helicity}, and \ref{Transversity}, respectively. The other three TMDs: $g_{1T}^{\bar{q}}(x,\bfk^{2})$,  $h_{1L}^{\bar{q}}(x,\bfk^{2})$, and $h_{1T}^{\bar{q}}(x,\bfk^{2})$ are illustrated in Fig.~\ref{3DGTMDs2}. It is clearly seen there that all six nonzero TMDs have peaks near $(x,\bfk^{2})\to (0,0)$. These peaks designate the dominant probability is to find a light sea quark in the proton in each of the different polarization configurations. 
%Accept $h_{1L}^{\bar{q}}$ and $h_{1T}^{\bar{q}}$, 
The  $xf_1^{\bar{q}}$,  $xg_{1L}^{\bar{q}}$,  $xh_1^{\bar{q}}$, and $xg_{1T}^{\bar{q}}$  show positive distributions and  they appear very similar by eye, the peak in the longitudinal direction runs higher
with decreasing $\bfk^{2}$ but falls rapidly to zero after 
$\bfk^{2}$ increases above $1$ GeV$^2$. Meanwhile, the peak in the transverse direction falls to zero when $x$ increases above $0.6$. We also notice that $xg_{1L}^{\bar{q}}(x,\bfk^{2})$ shows slightly negative distribution for the region $0.2~{\rm GeV}^2<\bfk^2<1.0~{\rm GeV}^2$ at very low-$x$. This TMD after integrating out the momenta quantifies the amount of spin contribution of the sea quarks in the proton. In this particular model, the TMDs corresponding to transversely polarized quarks within a longitudinally polarized proton, denoted by ${h}^{q \perp }_{1L}(x,\bfk^2) $ and a transversely polarized proton, denoted by ${h}^{q \perp }_{1T}(x,\bfk^2) $ are negative for both the anti-up and anti-down quarks.The negative trend of ${h}^{\bar{q} \perp }_{1L}$ for both the anti-up and the anti-down quarks is similar to the valence up and down quarks~\cite{Maji:2017bcz}, while the $ {h}^{\bar{q} \perp }_{1T}$ distribution for the sea quarks doesn't follow the same trend as in the case of valence quarks. Our model predicts negative pretzelosity for both the anti-up and the anti-down quarks.

\begin{figure}
	\centering
	\includegraphics[scale=0.22]{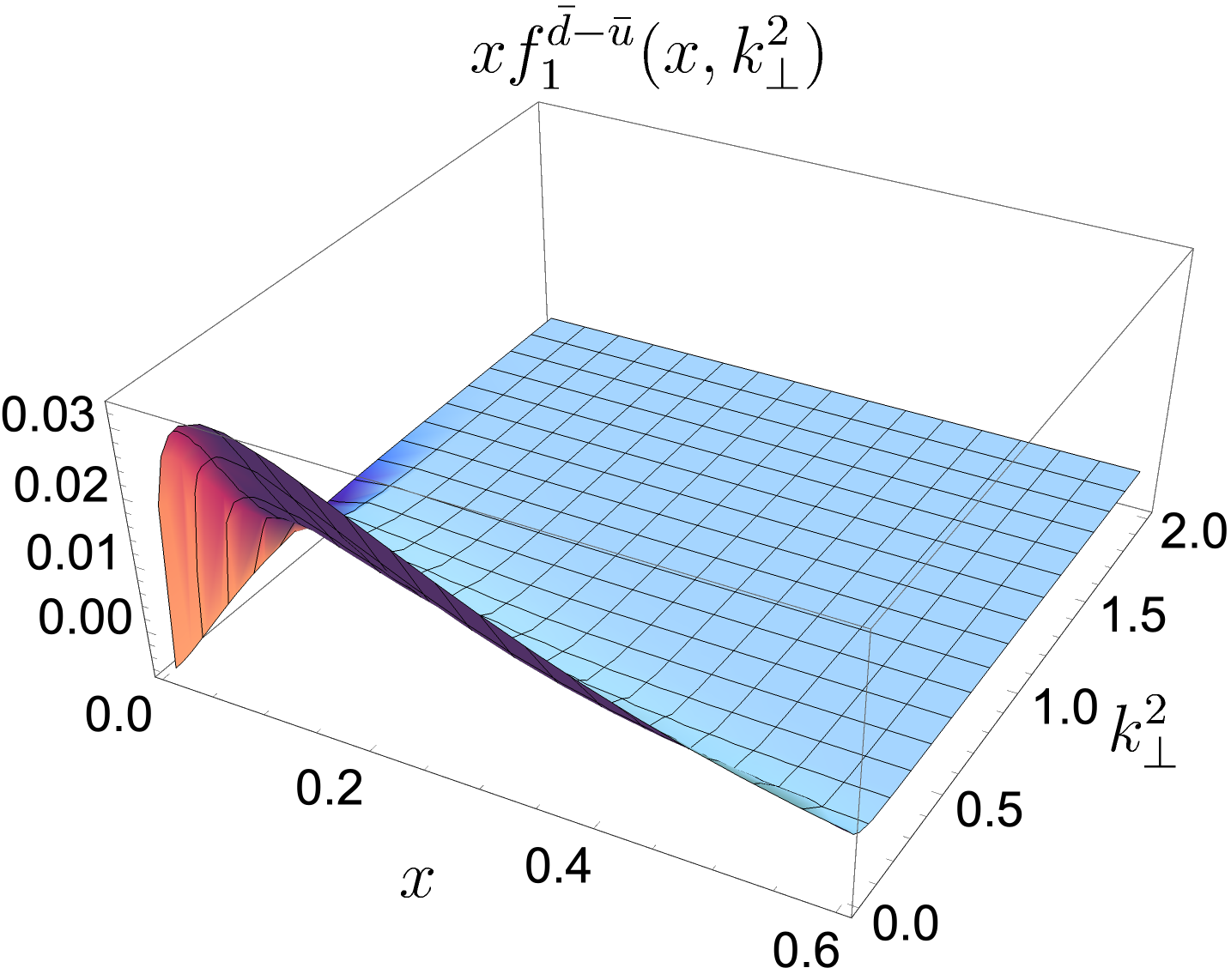}\hspace{0.4cm}
	\includegraphics[scale=0.22]{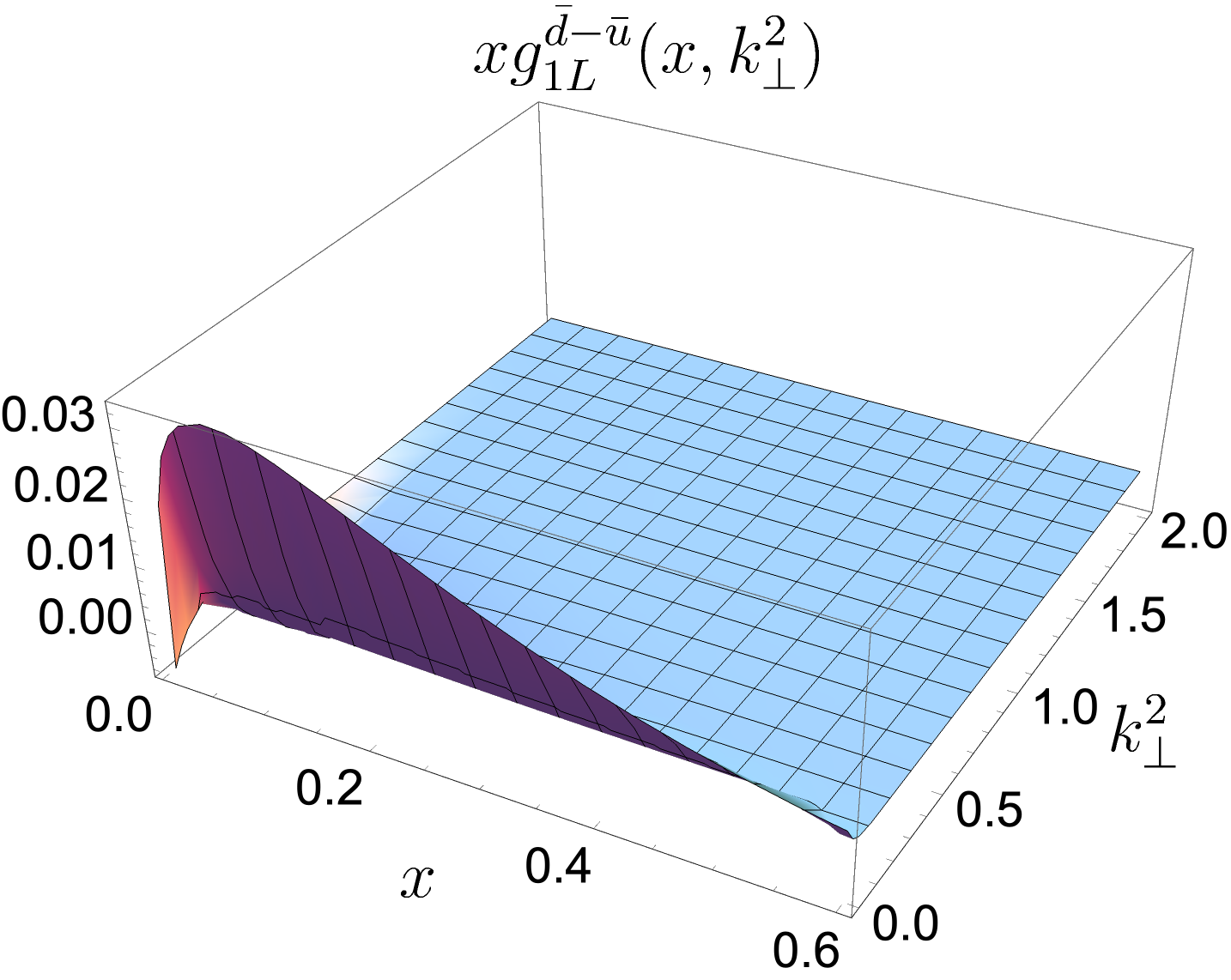}\hspace{0.4cm} 
	\includegraphics[scale=0.22]{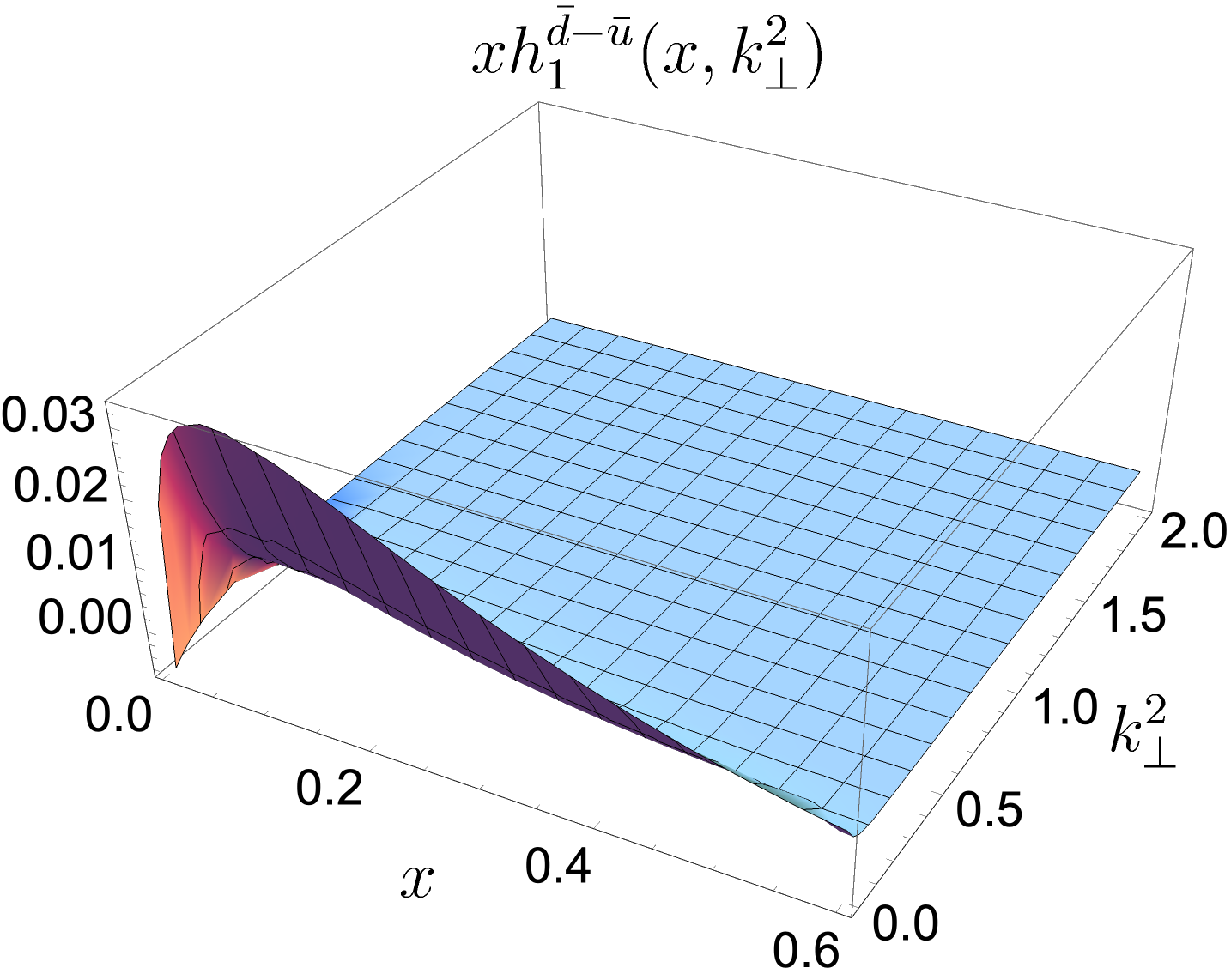}\vspace{0.4cm}\\
	\caption{TMD asymmetries of light sea quarks in  the proton, $xf_1^{\bar{d}-\bar{u}}$ (left plot),   $xg_{1L}^{\bar{d}-\bar{u}}$ (middle plot), and $xh_1^{\bar{d}-\bar{u}}$ (right plot) calculated in the range $0.005<x<0.6$ and momentum transfer $0< \bfk^2<2 $ GeV$^2$ in the proton within our model. $\bfk^2$ is in units of GeV$^2$.}
	\label{TMDasyms}
\end{figure}

The features of all the anti-up quark TMDs are very similar to those for the anti-down quark. However, we also find that there exists the flavor asymmetries in the transverse momentum plane, as can be seen from Fig.~\ref{TMDasyms}. The asymmetry distributions $xf_1^{\bar{d}-\bar{u}}$,   $xg_{1L}^{\bar{d}-\bar{u}}$, and $xh_1^{\bar{d}-\bar{u}}$ are found to be very similar in our model having a peak around $(x,\bfk^2)=(0.03,0)$ but falls rapidly to zero with increasing $\bfk^2$. Note that the ($\bar{d}-\bar{u}$) TMD asymmetries are negative at low-$x$ independent of choice of $\bfk^2$, indicating the density of anti-up quark is  larger than that of the anti-down quark at very small-$x$ domain, which is also observed in the PDFs and the GPDs asymmetries in Figs.~\ref{pdf asym} and ~\ref{2DGPDs}, respectively.
%================================
%=================================
%==================================

%===================================
\begin{figure}
	\centering
        \includegraphics[scale=0.32]{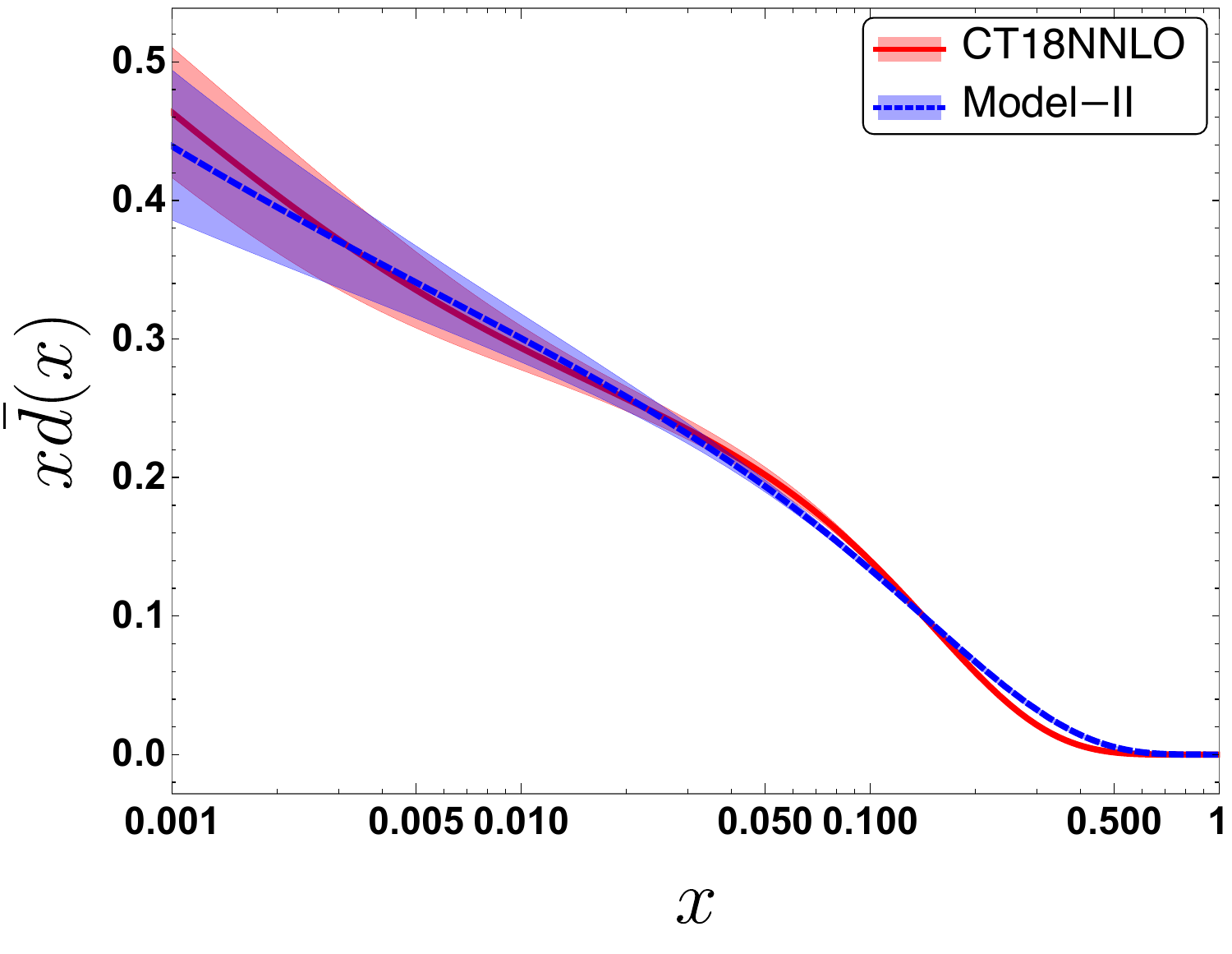}\hspace{0.5cm}
	\includegraphics[scale=0.32]{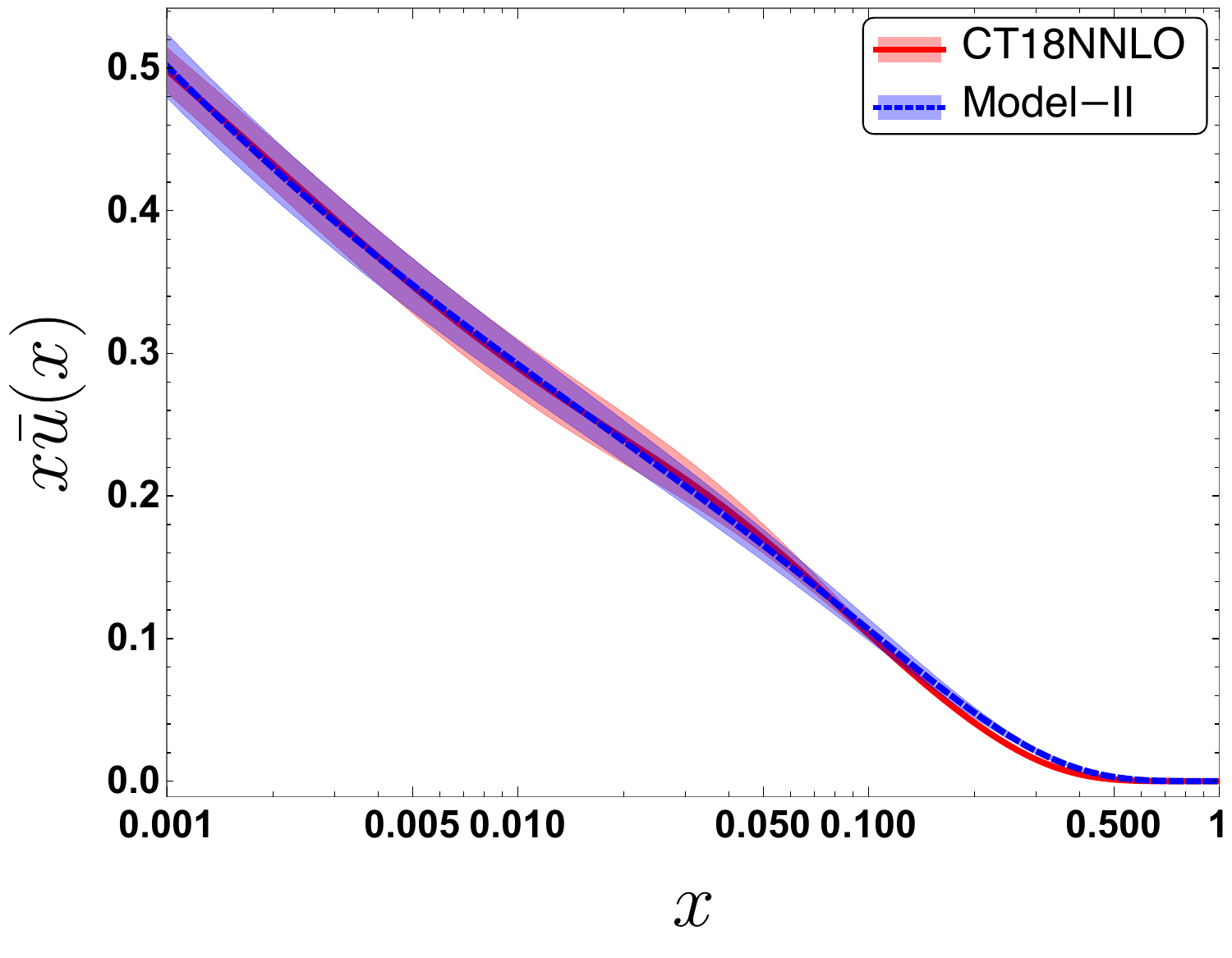}\vspace{0.5cm} \\
	\caption{The sea quark unpolarized distributions in the proton, left panel:  $ x {\bar d}(x)$ and right panel: $ x {\bar u}(x)$ as a function of longitudinal momentum fraction $x$ in the kinematics region $0.001 \leq x \leq 1 $ at $Q_0=2$ GeV. Our model results (dashed-blue lines with blue bands) are compared with the CTEQ18 NNLO dataset (solid-red lines with red bands)~\cite{Hou:2019efy}}.
	\label{model2 fit}
\end{figure}
%====================================
%==================================
 \begin{figure}
	\centering
	\includegraphics[scale=0.32]{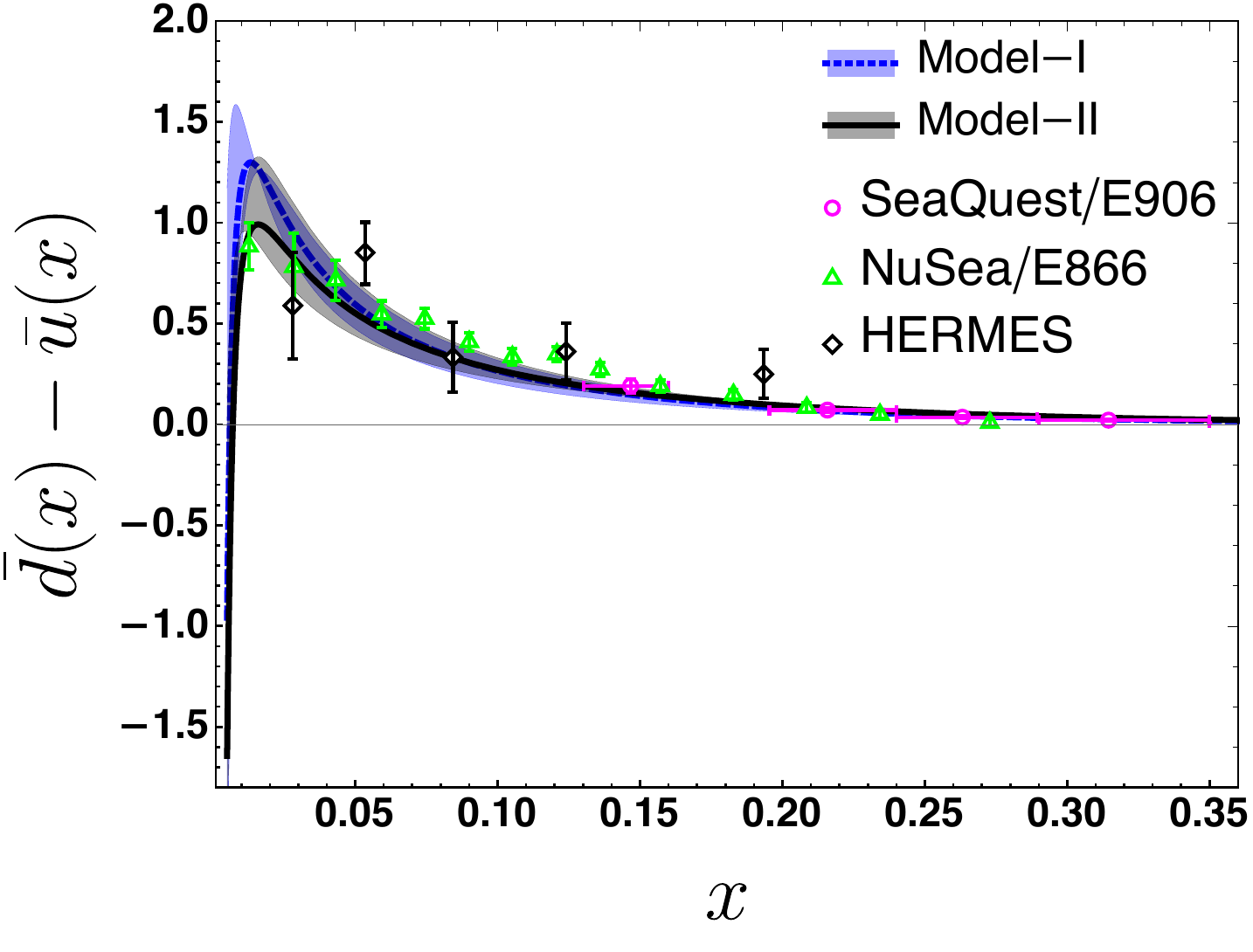}\hspace{0.5cm}
	\includegraphics[scale=0.32]{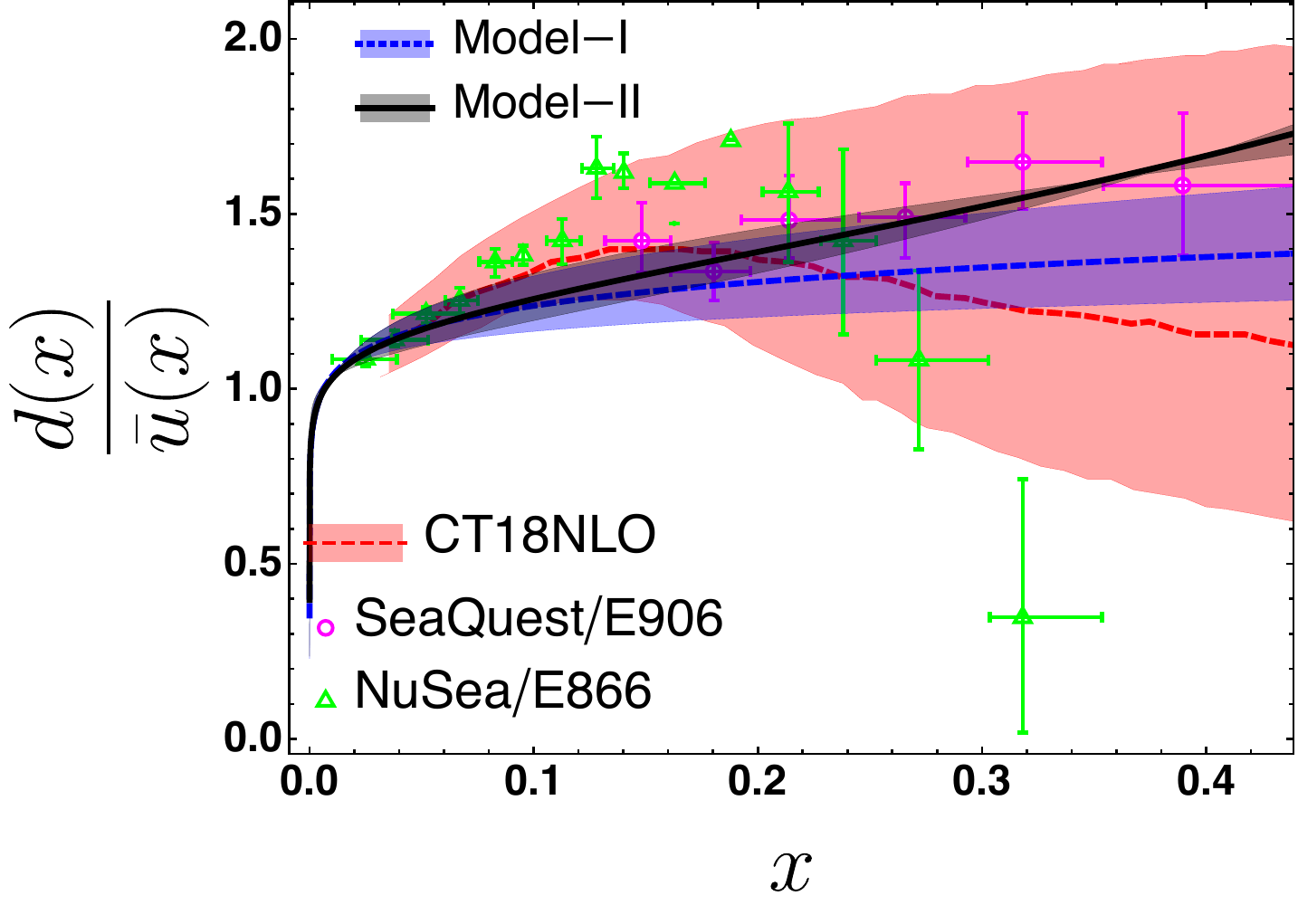}\vspace{0.5cm} \\
        \caption{The comparison of light flavor sea asymmetries in the proton between the Model-I (blue lines with blue bands) and Model-II (black lines with black bands), left panel: $\bar{d}(x)-\bar{u}(x)$ and right panel: $\bar{d}(x)/\bar{u}(x)$ as a function of $x$. Our results
 are compared with the experimental data from the SeaQuest/E906~\cite{SeaQuest:2022vwp} (magenta points), NuSea/E866~\cite{NuSea:2001idv} (green points), and HERMES~\cite{HERMES:1998uvc}(black points) Collaborations. In the right panel, we also include the CTEQ18 NLO global analysis~\cite{Hou:2019efy} (blue line with blue band) for the comparison.} 
	\label{pdf asym2}
\end{figure}
%==============================
\section{Model-II}\label{sec5}
%==============================
For further investigation of the model dependency in the flavor asymmetry,
instead of choosing Eq.~\eqref{WaveFun} in our light-front spectator model, we use a different form of the wave function adopted from a phenomenological model discussed in Ref.~\cite{Brodsky:2022kef},
\eq
\varphi(x,\bfk^2)=\sqrt{A} \frac{4 \pi }{\kappa}\sqrt{\frac{\log[1/(1-x)]}{x}}x^{\alpha}(1-x)^{3+\beta} \exp{\bigg[-\frac{\bfk^{2}}{2\kappa^{2}}\bigg]}.
\label{WaveFun2}
\en
We refer this model as Model-II.
%===================================
 \begin{table}
\caption{Fitted parameters for Model-II at $Q_0=2$ GeV.}
	\centering
	\begin{tabular}{ |c|c|c|c|c|c|c|c|} 
		\hline\hline
		& A  & $\alpha $  & $\beta$   \\
		\hline
		%	\multirow{3}{4em}{Multiple row} & cell2 & cell3 \\ 
		$\overline{d}$ \quad \quad	&  \quad $0.162_{0.015}^{-0.012}$ &  \quad $-0.572_{0.015}^{-0.014}$& \quad $-0.253_{0.164}^{-0.133}$   \\
		$\overline{u}$	\quad  \quad &  \quad  $0.116_{-0.01}^{0.009}$&  \quad  $-0.606_{-0.003}^{0.002}$ & \quad  $-0.022_{-0.004}^{0.004}$  \\
		\hline \hline
	\end{tabular}
	\label{Table-2}
\end{table}
%==================================

%================
\subsection{PDFs}
%================
In this particular model, the unpolarized sea quark PDF reads
\eq 
f_{1}^{\bar{q}}(x) &=&
\int \frac{\text {d}^2\bfk}{16\pi^3} \,
\biggl[ |\psi^{\uparrow}_{\bar{q}; + \frac{1}{2}}(x,\bfk)|^2
+ |\psi^{\uparrow}_{\bar{q};- \frac{1}{2}}(x,\bfk)|^2
\biggr] \nonumber\\
&=& 2 A x^{2 \alpha-1}(1-x)^{6+2 \beta} \log \left (\frac{1}{1-x}\right),
\en
while the helicity PDF  $ g_{1L}(x)$ vanishes. 
We fit $f_{1}^{\bar{q}}(x)$  with the CT18NNLO data at $Q_0=2$ GeV
and obtain a different set of the model parameters, $\alpha$, $\beta$, and $A$ given in Table~\ref{Table-2}.  The results of the fits for  $x{\bar{q}}(x)$ are shown in Fig.~\ref{model2 fit}. The value of $\chi^2$/d.o.f. for the fits of the $ x {\bar d}(x)$ and $ x {\bar u}(x)$ are $ 0.2530$ and $ 0.0234$, respectively in the region $0.001<x<0.1$. 

Figure~\ref{pdf asym2} compares the flavor asymmetries between between the Model-I (using Eq.~\eqref{WaveFun} and parameters listed in Table~\ref{Table-1}) and the Model-II. We find that the  ${\bar d}(x)-{\bar u}(x)$ asymmetry in both models is almost identical. However, for the ${\bar d}(x)/{\bar u}(x)$ asymmetry, the Model-II describes the growth trend toward higher $x$ of the latest experimental data from the SeaQuest/E906 Collaboration~ \cite{SeaQuest:2022vwp} better than the Model-I. 
%=====================================
%==============
\subsection{GPDs and OAM}
%================
The GPDs can be evaluated following the similar procedure as described in Eqs.~\eqref{GPDHq} to \eqref{GPDHtildeq} for the Model-I. Using the light-front wave functions, the explicit expressions of the sea quark GPDs at zero skewness in the Model-II read
\eq
H^{\bar{q}}(x, 0, -\Delta_\perp^2)&=&  2 A x^{2 \alpha-1}(1-x)^{6+2 \beta} \log \left (\frac{1}{1-x}\right)\left(1 -\frac{\Delta_\perp^2 (1-x)^2}{4 \kappa^2}\right )  \exp{\bigg[-\frac{(1-x)^2}{4\kappa^{2}}\Delta_\perp^2\bigg]}\,,\\
E^{\bar{q}}(x, 0, -\Delta_\perp^2) &=& \frac{ M_N}{\kappa} 2 A x^{2 \alpha-1}(1-x)^{7+2 \beta} \exp{\bigg[-\frac{(1-x)^2}{4\kappa^{2}}\Delta_\perp^2\bigg]}\,,\\
\tilde{H}^q(x, 0, -\Delta_\perp^2) &= &  0\,.
\en

  \begin{figure}
	\centering
	\includegraphics[scale=0.32]{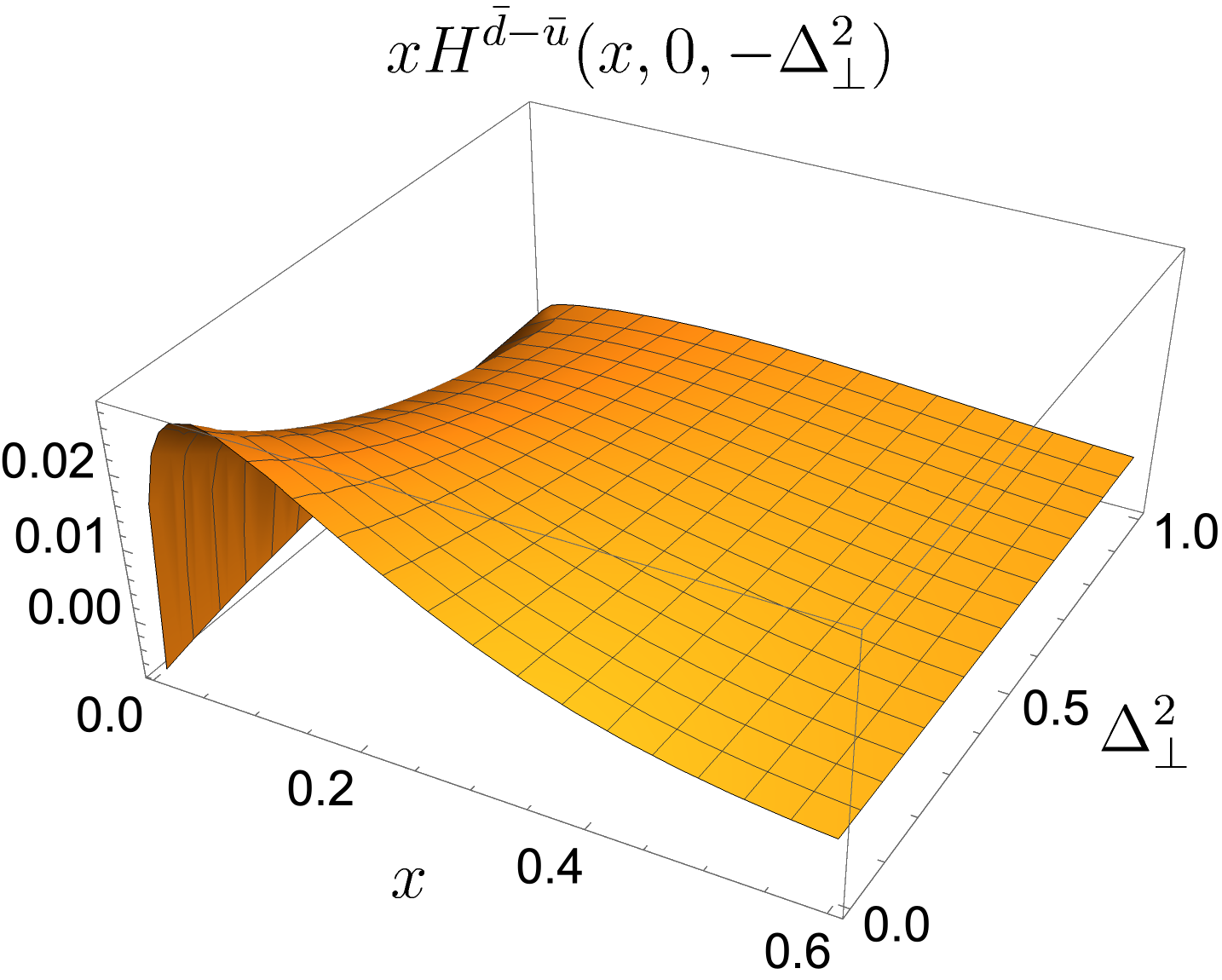}\hspace{0.5cm}
	\includegraphics[scale=0.32]{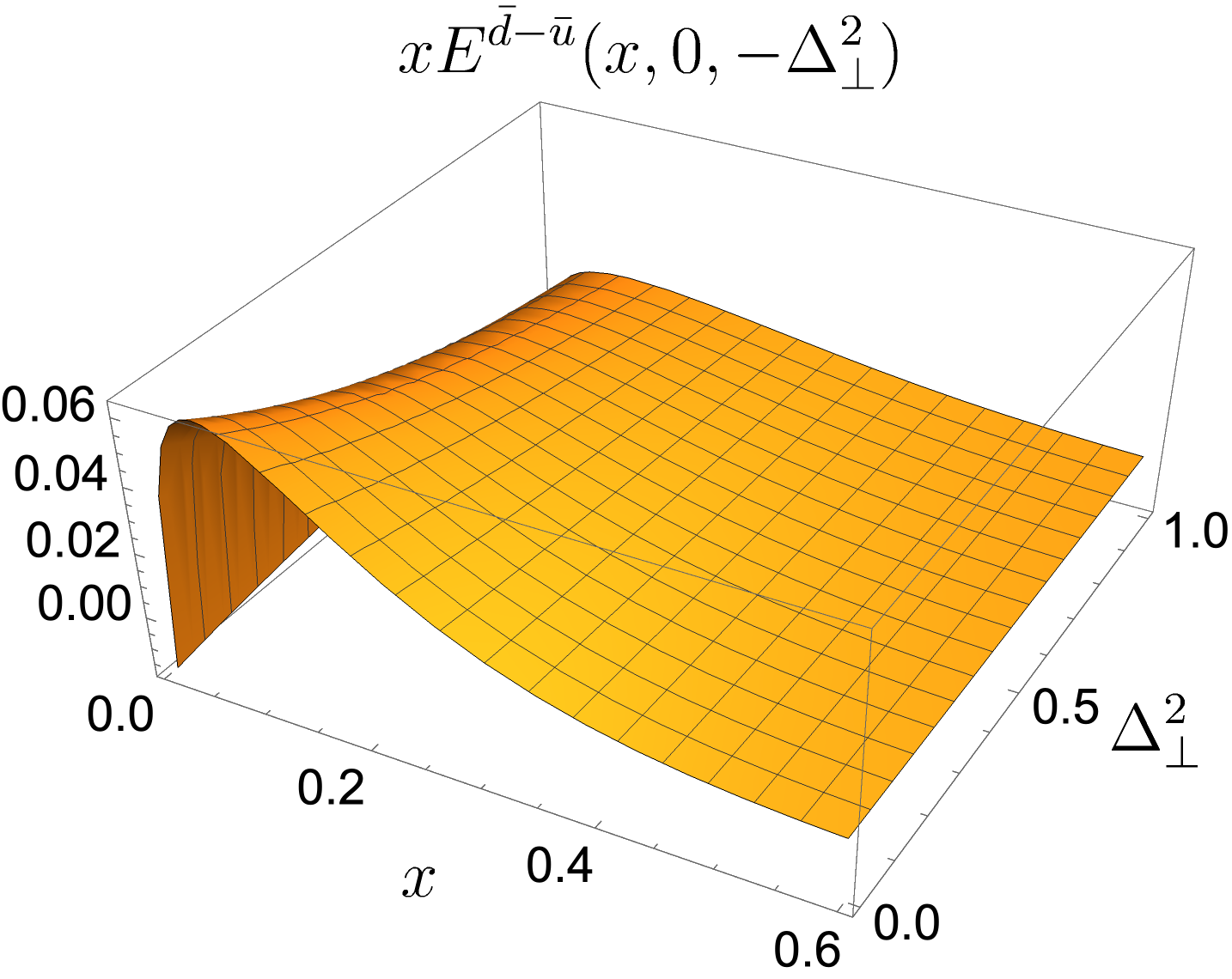}\vspace{0.4cm}
	\caption{The electric $ x H^{\bar{d}-\bar{u}} $ (left plot) and magnetic $ x E^{\bar{d}-\bar{u}} $ (right plot) GPDs asymmetry in the range $0.005<x<0.6$ and the momentum transfer $0<\Delta_{\perp}^2<1$ from Model-II.}
	\label{GPDasym3d2}
\end{figure}  
%=====================================
\begin{figure}
	\centering
	\includegraphics[scale=0.3]{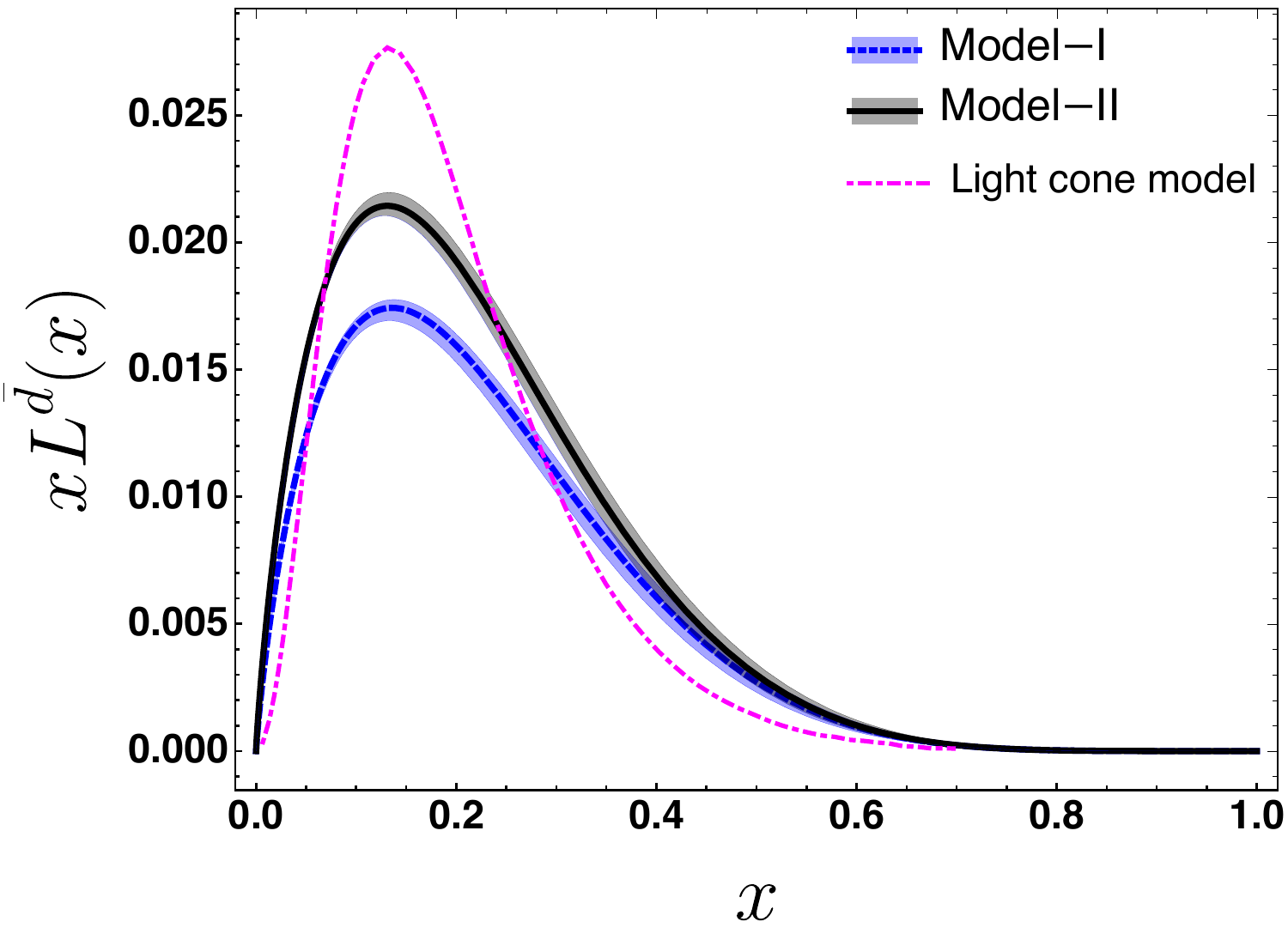} \hspace{0.5cm}
     \includegraphics[scale=0.3]{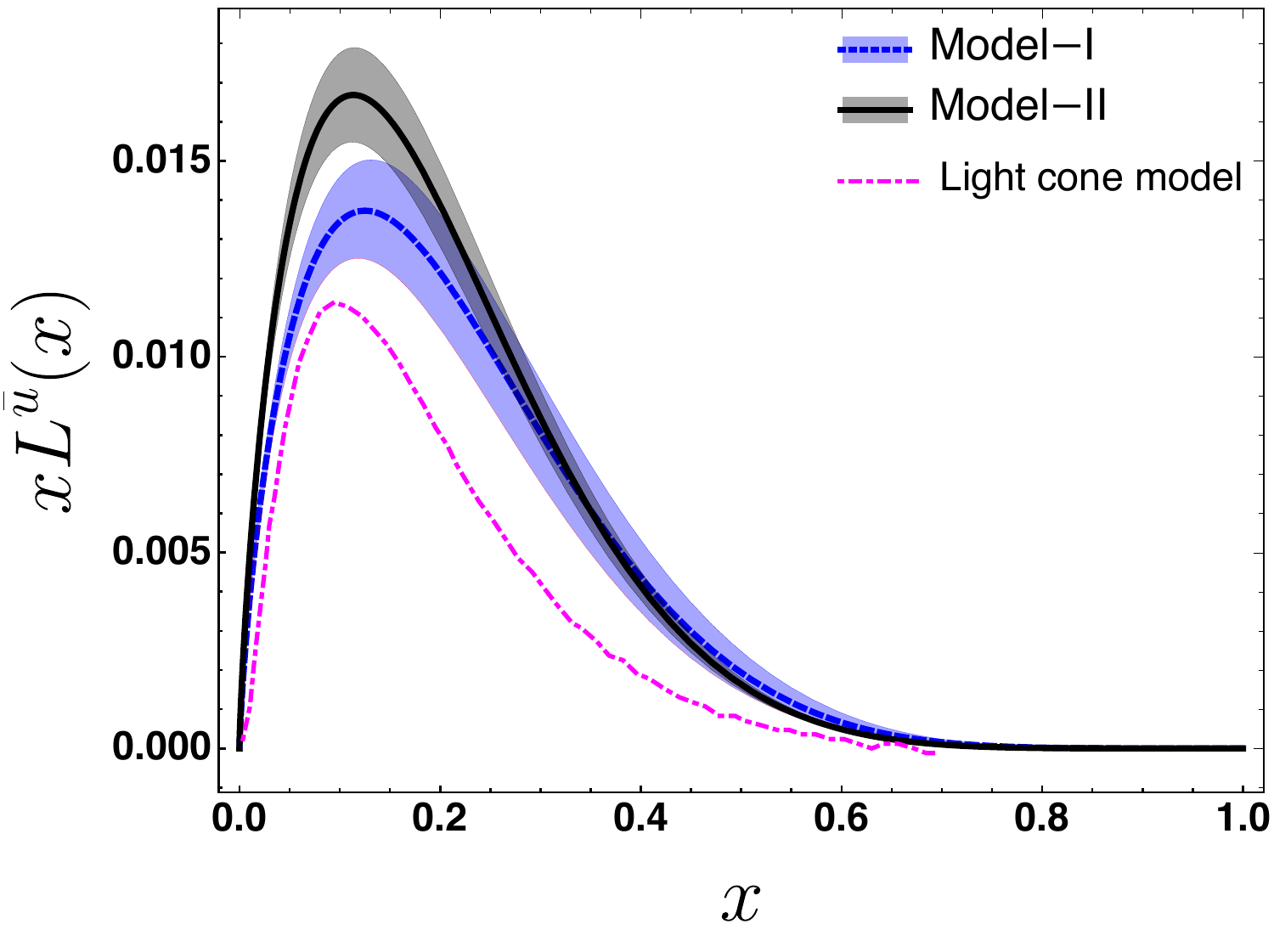}
	\caption{Comparison of the sea quark OAM distributions, $xL^{\bar{q}}(x)$, among the Model-I, the Model-II, and a light-cone model~\cite{Luan:2023lmt}. The left plot is for the $\bar{d}$ quark and the right plot is for the $\bar{u}$ quark. }
\label{OAM}
\end{figure}

Figure~\ref{GPDasym3d2} presents the qualitative behavior of $ x H^{\bar{d}-\bar{u}}(x,0,-\Delta_{\perp}^2) $ and $ x E^{\bar{d}-\bar{u}}(x,0,-\Delta_{\perp}^2) $ computed using the  Model-II. The distributions are plotted in the region $0.005<x<0.6 $ and $0<\Delta_{\perp}^2<1.0$ GeV$^2$.  We observe that there is not much difference in magnitude of the peaks compared to that in the Model-I (Fig.~\ref{GPDasym}) but with increasing $x$ and $\Delta_{\perp}^2$, the behaviors are quite different between these two models. The asymmetries in the Model-I fall faster than the Model-II. We find that Model-II provides compatible behaviour of the distributions $ x H^{\bar{d}-\bar{u}}(x,0,-\Delta_{\perp}^2) $ and $ x E^{\bar{d}-\bar{u}}(x,0,-\Delta_{\perp}^2) $ as observed  in non-local covariant chiral effective theory~\cite{He:2022leb}.

%=================================
 \begin{table}
	%\label{Tab:modelparameters}
  \caption{The Kinetic OAM, helicity and total angular momentum of up and down sea quarks in Model-I, Model-II and a Light cone model \cite{Luan:2023lmt}. }
	\centering
	\begin{tabular}{ |c|c|c|c|c|c|c|c|} 
		\hline\hline
		&$ L^{\bar {u}} $  & $L^{\bar {d}} $  & $ \Delta \Sigma^{\bar{u}}/2$ & $ \Delta \Sigma^{\bar{d}}/2$ & $ J^{\bar{u}}=L^{\bar {u}}+\Delta \Sigma^{\bar{u}}/2$ & $ J^{\bar{d}}=L^{\bar {d}}+\Delta \Sigma^{\bar{d}}/2$ \\
		\hline
		Model-I \quad \quad	&  \quad $0.040 \pm 0.003 $ &  \quad  $ 0.048\pm 0.002$ & \quad $0.013\pm 0.001$ & \quad $ 0.015\pm 0.001 $ & \quad $ 0.053\pm 0.004$ & \quad $0.063\pm 0.003$  \\
		Model-II	\quad  \quad &  \quad  $0.048\pm 0.003$&  \quad  $0.059\pm 0.002$ & \quad 0.00 & \quad 0.00 & \quad  $0.048\pm 0.003$&  \quad  $0.059\pm 0.002$  \\
           Light cone model \cite{Luan:2023lmt}	\quad  \quad &  \quad  $0.025 $&  \quad  $0.046$ & \quad 0.00 & \quad 0.00 & \quad  $0.025$&  \quad  $0.046$  \\
		\hline \hline
	\end{tabular}
	\label{TableOAM}
\end{table}
%================================

Figure~\ref{OAM} compares the  distribution $xL^{\bar{q}}(x)$ between light cone model \cite{Luan:2023lmt}, the Model-I  and the Model-II for both the down and up sea quarks. The qualitative nature of the distribution is more or less same in these three models except the magnitude in the light-cone model.
Table \ref{TableOAM} displays our predictions for the up and down sea quarks' OAM, spin, and total angular momentum. These results are evaluated within the region $0.001<x<1$ for the Model-I and Model-II. We observe that the down sea quark contributes more to the proton spin than the up sea quark. The spin contribution vanishes  in the Model-II and the light-cone model \cite{Luan:2023lmt}. The total angular momentum in the Model-I is higher than the Model-II and the light-one model for both sea quarks. While the down quark's kinetic OAM is comparable in the Model-I and the light cone model, the up quark's kinetic OAM is comparable in the Model-I and the Model-II.
%========================
\subsection{TMDs}
%========================
Employing the light-front wave functions, we also calculate the sea quark TMDs in the Model-II. The unpolarized TMD in this particular model is expressed as 
\eq
{f}^{q  }_1(x,\bfk^2)&=& \frac{1}{\pi \kappa^2}A x^{2 \alpha-1}(1-x)^{6+2 \beta} \log(1/1-x) \left(1+\frac{\bfk^2}{\kappa^2} \right) \exp{\left[-\frac{\bfk^{2}}{\kappa^{2}}\right]}\,.
\en

\begin{figure}
	\centering
 \includegraphics[scale=0.32]{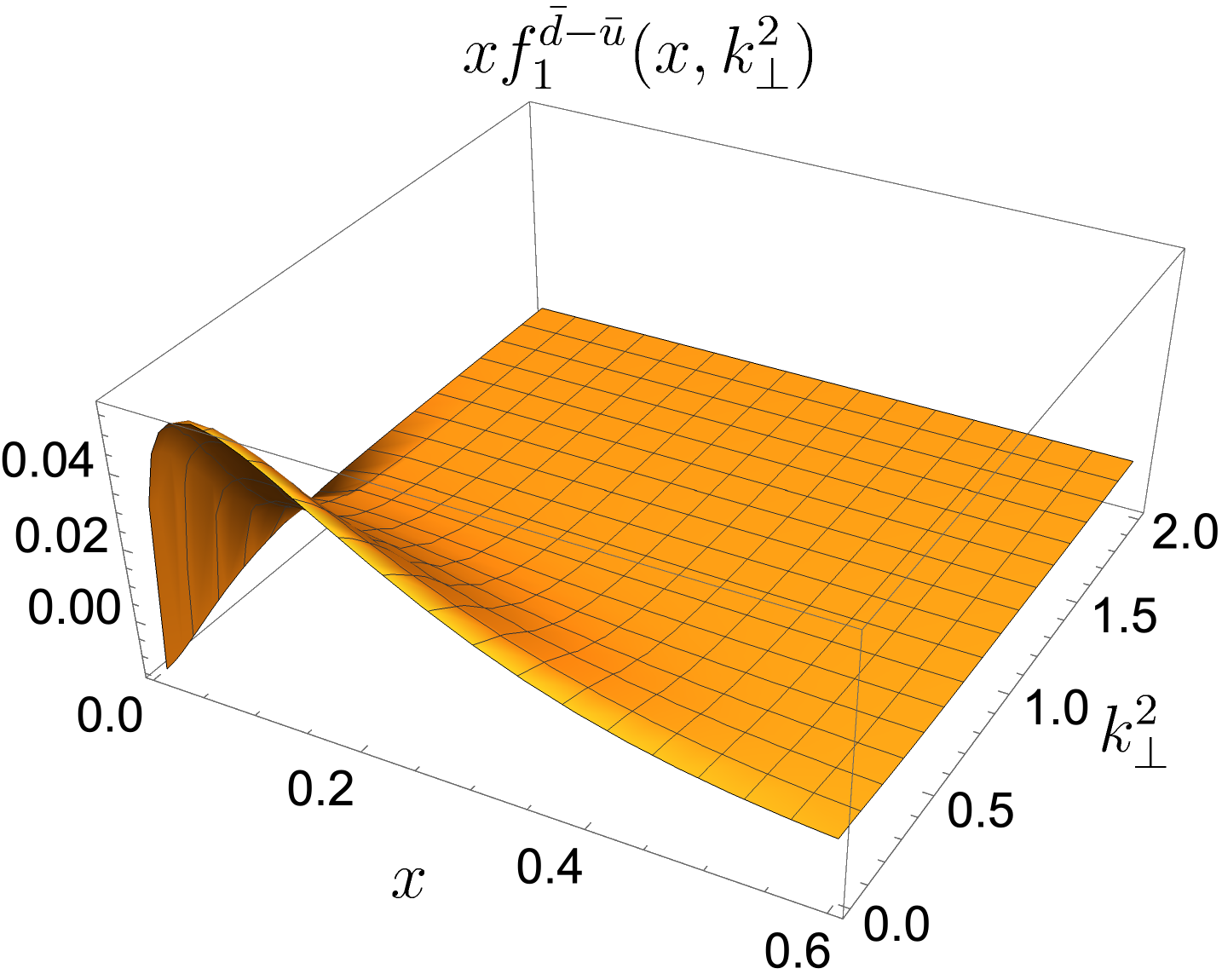}
	\caption{Flavor asymmetry in the TMD, $xf_1^{\bar{d}-\bar{u}}(x,\bfk^2)$ calculated in the range $0.005<x<0.6$ and momentum transfer $0< \bfk^2<2 $ GeV$^2$ in the proton within Model-II.}
	\label{TMD:M2}
\end{figure}

The flavor asymmetry in TMD within the Model-II is shown in Fig.~\ref{TMD:M2}. Here again, we notice that the qualitative nature of $xf_1^{\bar{d}-\bar{u}}(x,\bfk^2)$ in the Model-II is more or less same as observed in the Model-I (Fig.~\ref{TMDasyms}). The magnitude of the peak is comparable to that in Model-I but with increasing $\bfk^2$, the asymmetries in the Model-II fall much slower than the Model-I.
%=============================
%============================
\section{Summary and Outlook}\label{SUMMARY}
%============================
In this work, we have studied the flavor asymmetries of the light sea quarks within the proton by employing
a light-front spectator model, where the sea quarks are viewed as active partons and the remaining
components of the proton are considered as the spectator. We have formulated this model based on the light-front wave function modelled by the soft wall AdS/QCD and  parameterized by fitting the unpolarized parton distribution functions (PDFs) of light sea quarks from the CTEQ18 NNLO global analysis at the scale $Q_0=2$ GeV. We have calculated the sea quark leading-twist PDFs, generalized parton distributions (GPDs), and transverse momentum dependent parton distribution functions (TMDs) and their flavor asymmetries. We have observed that at small-$x$ ($<0.01$), ${\bar{u}}(x)$ is larger than ${\bar{d}}(x)$ in our model. The ${\bar{d}}(x)-{\bar{u}}(x)$ asymmetry agrees well with the available experimental data from the NuSea/E866, HERMES, and SeaQuest/E906 Collaborations~\cite{NuSea:2001idv,HERMES:1998uvc,SeaQuest:2022vwp}. From the comparison between the model and the experiments, we have found that our model describes well the trend of the latest experimental data of the ${\bar{d}}(x)/{\bar{u}}(x)$ asymmetry from the the SeaQuest/E906 Collaboration~\cite{SeaQuest:2022vwp}, however fails to explain the overall behavior shown by NuSea/E866 data~\cite{NuSea:2001idv}. Similar results are also observed  in the pion model~\cite{Choudhary:2023bap,Alberg:2021nmu} and the chiral light-front perturbation theory~\cite{Alberg:2017ijg}. Our model leads to more or less satisfactory results for the sea quark helicity PDFs when we compare with the experimental data from the COMPASS \cite{COMPASS:2010hwr} and the HERMES \cite{HERMES:2003gbu}. It should be noted that  the signs of sea quark helicity PDFs are not fixed by the experiments. 

We have presented results for the sea quark unpolarized and helicity-dependent GPDs of the proton in  momentum space for zero skewness and we found that the qualitative
behavior of the GPD asymmetries in our model bears similarities to other phenomenological models~\cite{He:2022leb,Luan:2023lmt}. We have employed these GPDs to study the orbital angular momentum (OAM) distribution at the density level.
Within our model, we have found that the $\bar{d}$ quark distribution dominates over the $\bar{u}$ quark distribution and their qualitative features are in 
accord with the behavior observed in a light-front baryon-meson fluctuation model~\cite{Luan:2023lmt}. We have obtained that the sea quark OAMs, $L^{\bar {u}}=0.040\pm 0.003$ and $L^{\bar {d}}=0.048\pm 0.002$ strongly dominate over their spin contributions, $\Delta \Sigma^{\bar{u}}/2=0.013\pm0.001$ and $\Delta \Sigma^{\bar{d}}/2=0.015\pm0.001$.

We have further investigated the leading-twist TMDs of the sea quarks within the proton. In this
study, the gauge link has been set to unity, which leaves us
six nonzero T-even TMDs out of the eight leading-twist TMDs. We have observed that the properties of all nonzero TMDs for both the $\bar{u}$ and $\bar{d}$ are are very similar. They have peaks at $(x,\bfk^{2})\to (0,0)$ and fall rapidly as $x$ and $\bfk^{2}$ increase. It is not surprising since the sea quark density is much higher at small-$x$ domain. The sea asymmetry in the TMDs has been observed only when $\bfk^{2}$ is small ($<0.2$ GeV$^2$) for all values of $x$. For further investigation of the model dependency of the sea quark properties, we have considered a different form of the light-front wave function (Model-II) and studied those asymmetries in this model. We have noticed that this model shows more or less similar characteristics of the sea quarks as observed in the Model-I.

 The proposed light-front spectator models can be further employed to compute the other properties of the sea quarks, such as the chiral-odd GPDs, T-odd TMDs, and Wigner distributions, etc., in the proton as well as to study the role of sea quarks in the single and double spin asymmetries. The present calculation can be straightforwardly
extended to investigate the strange and the charm flavor asymmetries in the proton.

\section*{Acknowledgements}
CM is supported by new faculty start up funding by the Institute of Modern Physics, Chinese Academy of Sciences, Grant No. E129952YR0. CM also thanks the Chinese Academy of Sciences Presidents International Fellowship Initiative for the support via Grants No. 2021PM0023.

\bibliography{Ref.bib}

\end{document}